# Single-Molecule Magnets: Jahn-Teller Isomerism and the Origin of Two Magnetization Relaxation Processes in $Mn_{12}$ Complexes

**Sheila M. J. Aubin,[1a] Ziming Sun,[1a] Hilary J. Eppley,[1b] Evan M. Rumberger,[1a] Ilia A. Guzei,[1c] Kirsten Folting,[1b] Peter K. Gantzel,[1a] Arnold L. Rheingold,[1c] George Christou,[1b,*] and David N. Hendrickson[1a,*]**

*Contribution from the Department of Chemistry and Biochemistry-0358, University of California at San Diego, La Jolla, California 92093, the Department of Chemistry and Molecular Structure Center, Indiana University, Bloomington, Indiana 47405-4001, and the Department of Chemistry and Biochemistry, University of Delaware, Newark, Delaware 19716.*

## Abstract

Several single-molecule magnets with the composition $[Mn_{12}O_{12}(O_2CR)_{16}(H_2O)_x]$ (x = 3 or 4) exhibit two out-of-phase ac magnetic susceptibility signals, one in the 4-7 K region and the other in the 2-3 K region. New $Mn_{12}$ complexes were prepared and structurally characterized and the origin of the two magnetization relaxation processes was systematically examined. Different crystallographic forms of a $Mn_{12}$ complex with a given R substituent exist, where the two forms have different compositions of solvent molecules of crystallization and this results in two different arrangements of bound $H_2O$ and carboxylate ligands for the two crystallographically different forms with the same R substituent. The X-ray structure of cubic crystals of $[Mn_{12}O_{12}(O_2CEt)_{16}(H_2O)_3] \cdot 4H_2O$ (space group $P\bar{1}$) (complex **2a**) has been reported previously. The more prevalent needle-form of $[Mn_{12}O_{12}(O_2CEt)_{16}(H_2O)_3]$ (complex **2b**) crystallizes in the monoclinic space group $P2_1/c$, which at –170 °C has *a* = 16.462(7) Å, *b* =

22.401(9) Å, $c$ = 20.766(9) Å, β = 103.85(2)°, and Z = 4. The arrangements of $H_2O$ and carboxylate ligands on the $Mn_{12}$ molecule are different in the two crystal forms. The complex $[Mn_{12}O_{12}(O_2CC_6H_4\text{-}p\text{-}Cl)_{16}(H_2O)_4] \cdot 8CH_2Cl_2$ (**5**) crystallizes in the monoclinic space group *C*2/c, which at –172 °C has $a$ = 29.697(9) Å, $b$ = 17.708(4) Å, $c$ = 30.204(8) Å, β = 102.12(2)° and Z = 4. The ac susceptibility data for complex **5** show that it has out-of-phase signals in both the 2-3 K and the 4-7 K ranges. X-ray structures are also reported for two isomeric forms of the *p*-methylbenzoate complex. $[Mn_{12}O_{12}(O_2CC_6H_4\text{-}p\text{-}Me)_{16}(H_2O)_4] \cdot (HO_2CC_6H_4\text{-}p\text{-}Me)$ (**6**) crystallizes in the monoclinic space group *C*2/c, which at 193 K has $a$ = 40.4589(5) Å, $b$ = 18.2288(2) Å, $c$ = 26.5882(4) Å, β = 125.8359(2)° , and Z = 4. $[Mn_{12}O_{12}(O_2CC_6H_4\text{-}p\text{-}Me)_{16}(H_2O)_4] \cdot 3(H_2O)$ (**7**) crystallizes in the monoclinic space group *I*2/a, which at 223 K has $a$ = 29.2794(4) Å, $b$ = 32.2371(4) Å, $c$ = 29.8738(6) Å, β = 99.2650(10)° , and Z = 8. The $Mn_{12}$ molecules in complexes **6** and **7** differ in their arrangements of the four bound $H_2O$ ligands. Complex **6** exhibits an out-of-phase ac peak ($\chi_M$'') in the 2-3 K region, whereas the hydrate complex **7** has a $\chi_M$'' signal in the 4-7 K region. In addition, however, in complex **6**, one $Mn^{III}$ ion has an abnormal Jahn-Teller distortion axis oriented at an oxide ion, and thus **6** and **7** are Jahn-Teller isomers. This reduces the symmetry of the core of complex **6** compared with complex **7**. Thus, complex **6** likely has a larger tunneling matrix element and this explains why this complex shows a $\chi_M$'' signal in the 2-3 K region, whereas complex **7** has its $\chi_M$'' peak in the 4-7 K, i.e., the rate of tunneling of magnetization is greater in complex **6** than complex **7**. Detailed $^1$H NMR experiments (2-D COSY and TOCSY) lead to the assignment of all proton resonances for the benzoate and *p*-methyl-benzoate $Mn_{12}$ complexes and confirm structural integrity of the $(Mn_{12}O_{12})$ complexes upon dissolution. In solution there is rapid ligand exchange and no evidence for the different isomeric forms of $Mn_{12}$ complexes seen in the solid state.

## Introduction

Nanomagnets on the scale of 1 to 100 nm are expected to exhibit unusual properties and are therefore the subject of much current interest.[2,3] The most common preparative method for nanomagnets involves fragmenting bulk ferro- or ferri-magnetic materials, although the protein ferritin has also been found[4] to be an effective vehicle for the synthesis and study of such nanomagnets. Very recently molecules have been found that function as nanomagnets, and they have thus been termed single-molecule magnets (SMM's).[5-41] The magnetic "response" comes not from domains in crystals where the spins on a large number of metal sites are correlated as in normal magnets, but from individual non-interacting molecules, where each SMM has a large enough magnetic moment and magnetic anisotropy to function as a magnet. In response to an external magnetic field, the magnetic moment of the SSM can be magnetized with its spin either "up" or "down" along the axial magnetic anisotropy axis. After the magnetic moment of this molecular species is oriented in an external field and the external field then removed, the moment (spin) of the molecule will only very slowly reorient if the temperature is low. The most thoroughly studied SMM is $[Mn_{12}O_{12}(O_2CMe)_{16}(H_2O)_4] \cdot 2(HO_2CMe) \cdot 4(H_2O)$ (**1**), commonly called "$Mn_{12}$-acetate" or even simply "$Mn_{12}$".

There are at least two major reasons for studying nanomagnets composed of single molecules. One goal is to construct the ultimate high-density memory device; such a molecular memory[42] device could conceivably store two or more bits per molecule. A second goal is the elucidation of how quantum-mechanical behavior at the macroscopic scale underlies classical behavior at the macroscopic scale.[43] In a large sample of nanomagnets (*i.e.*, a classical system) it is possible to see quantum mechanical tunneling of the magnetization underlying the classical magnetization hysteresis response. Friedman *et al.*[24,25] reported for the first time the observation of resonant magnetization tunneling for $Mn_{12}$ molecules when they observed steps at regular intervals of magnetic field in the magnetization hysteresis loop for oriented crystals of $Mn_{12}$-acetate (**1**). The presence of these steps has been confirmed for this same complex by two other groups.[26,27] The observed steps in the hysteresis loop correspond to an increase in the rate of decrease of magnetization occurring when there is an alignment of energy levels in the two parts of the double-well potential for a single-molecule magnet.

In short, a SMM is magnetizable because it has a potential-energy barrier between its "spin up" and "spin down" states. There are two requirements: the SMM must have a large spin (S) ground state and there has to be a large *negative* magnetic anisotropy (i.e., easy axis type anisotropy) resulting from zero-field splitting (ZFS) in the ground state with $D < 0$ (D is the axial ZFS parameter). The "$Mn_{12}$-acetate" complex has a $S = 10$ ground state that experiences axial zero-field splitting with $D = -0.50\ cm^{-1}$. The barrier height for magnetization reversal scales as $S^2|D|$ (= 50 $cm^{-1}$) and at 2 K the $Mn_{12}$-acetate complex has a magnetization relaxation half-life in the order of 2 months.

There are several experimental manifestations of the fact that a molecule has a significant barrier for magnetization reversal and thus is functioning as a SMM at low temperatures: (i) there will be a divergence between the zero-field-cooled and field-cooled magnetization at some "blocking" temperature; (ii) perhaps the most classic indication is the observation of a hysteresis loop in the magnetization versus external magnetic field response; and (iii) there will be a frequency-dependent out-of-phase ac magnetic susceptibility signal because at low temperatures the magnetization of a single-molecule magnet will not be able to keep in phase with an oscillating magnetic field.

It has often been observed[13] that samples of certain $Mn_{12}$ SMM's exhibit not one but two out-of-phase ac signals. One peak occurs in the 4-7 K region and the other in the 2-3 K region, and both are frequency dependent. In this paper we systematically examine the origin of the two out-of-phase ac signals. The characterizations of new single-molecule $Mn_{12}$ magnets are also reported. It has been determined that isomeric forms of a given complex can be obtained that differ in the placement of the four $H_2O$ ligands and/or exhibit the recently identified phenomenon of Jahn-Teller isomerism. It is concluded that these isomeric differences are the origin of the differing magnetic behaviors.

## Experimental Section

**Compound Preparation**. All chemicals and solvents were used as received. All preparations and manipulations were performed under aerobic conditions. $(NBu_4^n)[MnO_4]$ was prepared as described in ref 44. *WARNING*: Organic permanganates should be handled with extreme caution. Detonation of some organic permanganates have been reported while drying at high temperatures. $[Mn_{12}O_{12}(O_2CEt)_{16}(H_2O)_3]$ (**2**) and $[Mn_{12}O_{12}(O_2CPh)_{16}(H_2O)_4]$ (**3**) were prepared as previously described.[13]

**$[Mn_{12}O_{12}(O_2CC_6H_4$-*p*-$Cl)_{16}(H_2O)_4]\cdot 8CH_2Cl_2$ (Complex 5)**. $Mn(ClO_4)_2$ (4.00 g, 11.0 mol) was dissolved in ethanol (40 mL) followed by the addition of *p*-chlorobenzoic acid (8.65 g, 55.2 mmol). Additional ethanol (290 mL) was added in order to dissolve the acid. Solid $(NBu_4^n)[MnO_4]$ (1.55 g, 4.28 mmol) was slowly added resulting in a reddish-brown solution that was filtered and left uncapped and undisturbed for 2 weeks. The resulting brown microcrystals of $[Mn_{12}O_{12}(O_2CC_6H_4$-*p*-$Cl)_{16}(H_2O)_4]\cdot 8CH_2Cl_2$ (**5**) (9% yield based on total available Mn) were collected on a frit and washed with ethanol. These microcrystals were recrystallized by slow evaporation of a $CH_2Cl_2$ solution resulting in black cubic crystals that are suitable for structural analysis by X-ray crystallography. Selected IR data (KBr, $cm^{-1}$): 3400 (w, b), 1593 (s), 1542 (s, b), 1412 (vs), 1174 (s), 1091 (s), 1010 (s), 859 (s), 771 (s), 739 (m), 661 (m), 610 (m), 551 (m). Anal. Calcd (Found) for $C_{112}H_{72}O_{48}Cl_{16}Mn_{12}$ : C, 39.42 (39.49); H, 2.12 (1.89).

**$[Mn_{12}O_{12}(O_2CC_6H_4$-*p*-$Me)_{16}(H_2O)_4] \cdot HO_2CC_6H_4$-*p*-Me (Complex 6)**. $Mn(ClO_4)_2$ (4.00 g, 11.0 mmol) was dissolved in 100% ethanol (20 mL) followed by the addition of *p*-methylbenzoic acid (19.13 g, 140.5 mmol) and additional ethanol (180 mL). Solid $(NBu_4^n)[MnO_4]$ (1.55 g, 4.28 mmol) was slowly added followed by filtration. The reddish-brown filtrate was left uncapped and undisturbed for 2 weeks. The resulting brown solid $[Mn_{12}O_{12}(O_2CC_6H_4$-*p*-$Me)_{16}(H_2O)_4]$ (7% yield based on Mn) was collected on a frit, washed with 100% ethanol, and recrystallized from $CH_2Cl_2$ hexanes, yielding black needles. Later it was discovered that higher yields (20%) were obtained if the synthesis was done in 2% $H_2O$:98% ethanol. The resulting microcrystals were washed with 100% ethanol and dried under vacuum. Recrystalization from anhydrous $CH_2Cl_2$:hexanes lead to black microcrystals. Anal. Calcd (Found) for $C_{136}H_{128}O_{50}Mn_{12}$ : C, 50.70 (50.78); H, 4.00 (3.94).

**$[Mn_{12}O_{12}(O_2CC_6H_4$-*p*-$Me)_{16}(H_2O)_4] \cdot 3H_2O$ (Complex 7)**. This complex was prepared in an analogous fashion as for complex **6** except instead of 100% ethanol, a 20% $H_2O$/80% ethanol solution (260 mL) was used. The yield based on total available Mn was 9%. Anal. Calcd (Found) for $C_{130}H_{132}O_{54}Mn_{12}$ : C, 49.0 (48.9); H, 4.04 (4.12).

**$[Mn_{12}O_{12}(O_2CC_6H_4$-*p*-$Bu^t)_{16}(H_2O)_4] \cdot CH_2Cl_2$ (Complex 8)**. $Mn(ClO_4)_2$ (2.15 g, 5.53 mmol) was dissolved in ethanol (20 mL) followed by the addition of *p*-tert-butylbenzoic acid (12.52 g, 70.2 mmol) and enough ethanol (100 mL) to result in a solution. Solid $(NBu_4^n)[MnO_4]$ (1.55 g, 4.28 mmol) was slowly added to the clear solution. The resulting reddish-brown solution was filtered and hexane (125 mL) was stirred into the filtrate. The solution was left uncapped and undisturbed for 2 weeks. The resulting brown microcrystals of $[Mn_{12}O_{12}(O_2CC_6H_4$-*p*-$Bu^t)_{16}(H_2O)_4] \cdot CH_2Cl_2$ (**8**) (8% yield based on Mn) were collected on a frit, washed with ethanol, and recrystallized from $CH_2Cl_2$/hexane resulting in small black cubic crystals. Anal. Calcd (Found) for $Mn_{12}O_{48}C_{177}H_{218}Cl_2$: C, 55.3 (55.2); H, 5.72 (5.69); Mn, 17.15 (16.7).

**X-ray Crystallography**

**Complexes 2b and 5**. Data were collected for $[Mn_{12}O_{12}(O_2CEt)_{16}(H_2O)_3]$ (**2b**) and $[Mn_{12}O_{12}(O_2CC_6H_4\text{-}p\text{-}Cl)_{16}(H_2O)_4] \cdot 8(CH_2Cl_2)$ (complex **5**) with a modified four-circle Picker diffractometer; details of the diffractometer, low temperature facilities and computation procedures employed by the Molecular Structure Center are available elsewhere.[45] For each complex, a suitable single crystal was selected, affixed to a glass fiber with silicone grease and transferred to the goniostat where it was cooled for characterization and data collection. Crystallographic data are shown in Table 1. For both complexes, the X-ray structures were solved using a combination of direct methods (MULTAN 78) and Fourier methods and refined by a full-matrix least squares.

For complex **5**, an automated search for peaks followed by analysis using the programs DIRAX[46] and TRACER[47] led to a C centered monoclinic unit cell. After complete data collection the systematic extinction of *h01* for $l=2n+1$ limited the choice of possible space groups to Cc or C2/c. The choice of the centrosymmetric space group C2/c was confirmed by the solution and refinement of the structure. After correction for absorption, data processing produced a set of 10,080 unique reflections and $R_{av} = 0.051$ for the averaging of 3469 reflections measured more than once. Four standard reflections measured every 300 reflections showed no significant trends. Unit cell dimensions were obtained by an unrestrained least-squares fit of the setting angle for 58 carefully center reflections having 2θ values between 22 and 30 degrees.

The structure was solved by using a combination of direct methods (SHELXS-86) and Fourier techniques. The positions of the seven Mn atoms, Mn(1) through Mn(7), in the asymmetric unit were obtained from the initial solution. The remaining non-hydrogen atoms were obtained from iteration of least-squares refinement and difference Fourier calculations. In addition to one half of the $Mn_{12}$-complex, the asymmetric unit was found to contain eight partially occupied molecules of dichloromethane which had been used as solvent. Two of the *p*-chlorobenzoate ligands were disordered (50%). In the final cycles of refinement, the positions of the non-hydrogen atoms in the molecule were refined using anisotropic thermal parameters, except for the disordered atoms in the benzoate ligands which were kept isotropic. Atoms Cl(96), C(97), Cl(98), Cl(99) and Cl(101) in the solvent molecules were refined with anisotropic thermal parameters and the remainder of the solvent atoms were refined using isotropic thermal parameters. The $Mn_{12}$-complex is located on a crystallographic two-fold axis (0,y,0.25) in the space group C2/c, position e. Mn(3) and Mn(7) are located on the two-fold axis.

For complex **2b**, a preliminary search for peaks, and analysis with the programs DIRAX and TRACER revealed a primitive monoclinic unit cell. Unit cell dimensions were determined by an unrestrained least-squares fit of the setting angles for 74 reflections located between 10 and 21 degrees in 2θ. The upper limit for data collection was set at 40° because weak diffraction was present as high 2ϕ angles. When data collection was complete, an attempt was made to collect data between 40° and 45°, but the crystal was lost due to ice accumulation. Systematic extinction of *h01* for *1* = 2n + 1 and of *0k0* for *k* = 2*n* + 1 uniquely identified the space group as $P2_1/c$. The Mn atoms were refined with anisotropic thermal parameters, and the remaining atoms were refined isotropically. Analysis of these data resulted in a final R(F)[$R_w$(F)] of 0.09 (0.08). No hydrogen atoms were located, but they were introduced in fluxed calculated positions with isotropic thermal parameters. Data having $F < 3.0\sigma$ were given zero weight. The final difference map was essentially featureless, several peaks of 1.3 e/$Å^3$ were located near Mn or O atoms and a hole of -1.4 e/$Å^3$ was observed.

**Complexes 6 and 7**. Data were collected on a Siemens P4/CCD diffractometer. The systematic absences in the diffraction data were consistent for the reported space groups. In both cases [complex **6** (**7**)], either of the monoclinic space groups *Cc (Ia)* or *C*2/*c* (*I*2/*a*) was indicated, but only the latter centrosymmetric space group *C*2/*c* (*I*2/*a*) was preferred based on the chemically reasonable and computationally stable results of the refinement. Crystallographic data are given in Table 2. The structures of complexes **6** and **7** were solved using direct methods, completed by subsequent difference Fourier synthesis and refined by block-matrix least-squares procedures in the case of **7**, and by full-matrix least-squares procedures in the case of **6**. Empirical absorption corrections for **6** and **7** were applied by using the program DIFABS described by Walker, N. and Stuart, D. in *Acta Cryst*. **1983**, *A39*, 158. The non-hydrogen atoms in the complexes were refined with anisotropic displacement coefficients. The non-hydrogen atoms in the solvate molecules were refined isotropically. Except as noted all hydrogen atoms were treated as idealized contributions. In complex **7** three solvate molecules of water were located and were modeled and refined as oxygen atoms. Hydrogen atoms on these water molecules as well as on the four aqua ligands in the complex were ignored. In the case of complex **6**, the molecule of the complex resides on a two-fold axis. The hydrogen atoms on the aqua ligands were ignored. There is one solvate molecule of *p*-toulic acid present in the asymmetric unit. Owing to severe disorder, this solvate molecule was modeled as carbon atoms and refined isotropically. All software and sources of the scattering factors are contained in the SHELXTL (version 5.03) program library (G. Sheldrick, Siemens XRD, Madison, WI).

**Physical Measurements**

Alternating current (ac) magnetic susceptibility measurements were collected on an MPMS2 Quantum Design SQUID magnetometer equipped with 1T magnet and capable of achieving temperatures of 1.7 to 400 K. The ac field range is $1 \times 10^{-4}$ to 5 G, oscillating in a frequency range of $5 \times 10^{-4}$ Hz to 1512 Hz. Pascal's constants[48] were used to approximate the diamagnetic molar susceptibility contribution for each complex and were subtracted from the experimental molar susceptibility data to give the paramagnetic molar susceptibility data. Ac magnetic susceptibility data were collected on microcrystalline and frozen solution samples in an ac field of 1G, oscillating at frequencies of 50, 250 or 1000 Hz and a dc field of 0G. The frozen solution ac magnetic susceptibility experiments were performed on a sample of 2.4 mg of **2b** dissolved in 0.5 mL $CD_2Cl_2$ and 0.5 mL toluene-$d_8$ in a sealed quartz tube. The diamagnetism of just the solvent and quartz were measured and subtracted from the above data to determine the paramagnetism due to complex **2b**. $^1$H NMR spectra were collected on a 300 MHz Varian Gemini 2000 NMR spectrometer or a 500 MHz Varian$^{UNITY}$ Inova NMR spectrometer.

**Results and Discussion**

**Synthesis and Strategy**. The main goal of this research was to determine the origin of the two different out-of-phase ac magnetic susceptibility signals seen for $Mn_{12}$ complexes. This involved preparing new $Mn_{12}$ complexes, particularly complexes that exhibit only the 2-3 K range out-of-phase ac signal or only the 4-7 K range signal. The key to the puzzle was found in the content of solvate of crystallization. Several $Mn_{12}$ complexes crystallize with extra acid molecules or solvent molecules in the crystal in addition to the $Mn_{12}$ complexes. This leads to $Mn_{12}$ complexes that are geometrical isomers and/or possess unusual Jahn-Teller distortions at certain $Mn^{III}$ ions.

There are basically two different synthetic procedures available for making new $[Mn_{12}O_{12}(O_2CR)_{16}(H_2O)_x]$ (x = 3 or 4) complexes. In one case, the acetate groups on the R = $CH_3$ (complex **1**) molecule can be substituted in a toluene solution by other carboxylates. This substitution process is driven by the greater acidity of the incoming carboxylic acid and/or the removal by distillation of the azeotrope of acetic acid and toluene. Several treatments with the new carboxylate are sometimes needed to replace all of the acetate groups.

The second synthetic approach involves the direct reaction of $Mn(ClO_4)_2$ in ethanol with $(NBu_4^n)[MnO_4]$ in the presence of the desired carboxylic acid. Yields of product are generally less than those obtained with the ligand substitution method. The new $Mn_{12}$ complexes reported in this paper (**5-8**) were prepared with the second approach. The complexes to be discussed are **2a** and **2b**, **5**, **6**, **7**, and **8** and they are listed below:

$[Mn_{12}O_{12}(O_2CEt)_{16}(H_2O)_3] \cdot 4(H_2O) \cdot ½(C_6H_5CH_3)$ (**complex 2a**)

$[Mn_{12}O_{12}(O_2CEt)_{16}(H_2O)_3]$ **(complex 2b)**

$[Mn_{12}O_{12}(O_2CC_6H_4\text{-}p\text{-}Cl)_{16}(H_2O)_4] \cdot 8(CH_2Cl_2)$ **(complex 5)**

$[Mn_{12}O_{12}(O_2CC_6H_4\text{-}p\text{-}Me)_{16}(H_2O)_4]\cdot(HO_2CC_6H_4\text{-}p\text{-}Me)$ **(complex 6)**

$[Mn_{12}O_{12}(O_2CC_6H_4\text{-}p\text{-}Me)_{16}(H_2O)_4] \cdot 3(H_2O)$ **(complex 7)**

$[Mn_{12}O_{12}(O_2CC_6H_4\text{-}p\text{-}Bu^t)_{16}(H_2O)_4] \cdot CH_2Cl_2$ **(complex 8)**

The X-ray structure of the cubic crystals (space group $P\bar{1}$) of complex **2a** was reported earlier.[13] It was discovered in this work that the needle crystals of complex **2b** are the most prevalent form of the propionate $Mn_{12}$ complex. The needle form crystallizes in the $P2_1/c$ space group (*vide infra*). Obviously, the change in solvation between **2a** and **2b** correlates with a change in space groups.

It was found that the solvation composition of $Mn_{12}$ complexes can be controlled by changes in the solvent used in the direct reaction approach. Brown microcrystalline samples of complexes **5**, **6**, and **8** were obtained by adding solid $(NBu_4^n)[MnO_4]$ to a 100% ethanol solution of $Mn(ClO_4)_2$ and the appropriate carboxylic acid. These microcrystalline products were recrystallized from $CH_2Cl_2$ or from $CH_2Cl_2$/hexane mixtures to give black crystals. A change in the reaction solvent from anhydrous ethanol to a 80% EtOH/20% $H_2O$ medium gives complex **7**. Complexes **6** and **7** have the same $Mn_{12}$ complex, but differ in solvate molecule compositions. The two forms of the $Mn_{12}$ propionate, complexes **2a** and **2b**, also differ in their solvate molecule compositions.

In the following sections of this paper, single-crystal X-ray structures are presented for complexes **2b** and **5**, **6** and **7**. The X-ray structures of **2a** (previously reported[13]) and **2b** show that there are two crystallographically different forms of this $Mn_{12}$ propionate complex. Single-crystal X-ray structures are also reported to show that complexes **6** and **7** are crystallographically different forms of the *p*-methylbenzoate $Mn_{12}$ complex. The structurally different crystal forms of a given $Mn_{12}$ complex are shown in this paper to have different magnetization relaxation properties.

**Alternating Current Magnetic Susceptibility**. In the ac susceptibility experiment, the magnetic field is oscillated at a particular frequency. An out-of-phase ac magnetic susceptibility signal is observed when the rate at which the magnetic moment of a molecule (or a collection of molecules) flips is close to the operating frequency of the ac magnetic field. The maximum in the out-of-phase ac signal occurs when the rate at which the magnetic moment flips is equal to the frequency of the oscillating field. The magnetic moment of paramagnetic molecules typically flips at a rate of $\sim 10^9$ Hz and, therefore, no out-of-phase component of ac magnetic susceptibility is seen in the frequency range of 50-1000 Hz. The origin of the out-of-phase peak observed for the $Mn_{12}$-acetate complex **1** has been the focus of considerable research.[9] One frequency-dependent peak is observed for the acetate complex **1** in the 4-7 K region. Detailed magnetic susceptibility and EPR studies on complex **1** have shown that it has an S = 10 ground state that experiences a large negative axial zero-field splitting, $\mathbf{H} = D\mathbf{S}^2_z$ where $D = -0.5$ cm$^{-1}$. Thus, in zero dc field there are two equivalent lowest energy states corresponding to $m_S = \pm 10$ and the highest energy state corresponds to $m_S = 0$ (Figure 1). The height of the potential energy barrier is $|D|[(10)^2-(0)^2] = 50$ cm$^{-1}$. In order for the magnetic moment of a $Mn_{12}$ molecule to flip from "up" to "down" it must either climb over the potential energy barrier in a thermally activated process or pass through the barrier by quantum tunneling. When $kT < |D|m_s^2$, the rate at which the magnetic moment of a molecule flips becomes sluggish and this relaxation phenomenon appears as an out-of-phase ac susceptibility peak within a certain range of frequencies. This leads to one out-of-phase ac signal per $Mn_{12}$ molecule or for a collection of identical $Mn_{12}$ molecules. However, it does *not* explain why a given complex would exhibit two out-of-phase ac signals.

Eppley *et al.*[13] reported in 1995 that complex **2** exhibits two frequency-dependent out-of-phase ac magnetic susceptibility peaks in the temperature region of 2-3 K and 4-7 K. The lower temperature peak was found to be ~1/2 the intensity of the higher temperature peak. In order to further investigate the origin of the two $\chi_M$''peaks, new $Mn_{12}$ molecules were prepared. We have found that not only is it possible to prepare samples that show only one peak in the 4-7 K region, but samples of $Mn_{12}$ complexes can also be prepared that show two peaks (~2-3 K and 4-7 K) and still other samples have been prepared that only exhibit one peak in the 2-3 K region.

Plotted in the upper part of Figure 2 is $\chi_M'T$ *versus* temperature for the *p*-tert-butylbenzoate complex **8** measured at frequencies of 50, 250 and 1000 Hz. The quantity $\chi_M'$ is the in-phase component of the ac magnetic susceptibility. At 50 Hz, $\chi_M'T$ remains relatively constant at a value of 44 $cm^3$ K $mol^{-1}$ between 10 and 3.2 K. Below 3.2 K, the value of $\chi_M'T$ decreases abruptly to 7.82 $cm^3$ K $mol^{-1}$ at 2 K. At higher frequencies the decrease in $\chi_M'T$ occurs at higher temperatures. If all of the $Mn_{12}$ molecules in a sample were populating only a S = 9 ground state with g = 1.98, the value of $\chi_M'T$ would plateau at 44 $cm^3$ K $mol^{-1}$, which agrees well with the experimental $\chi_M'T$ plateau value of 44 $cm^3$ K $mol^{-1}$. In the lower part of Figure 2 is a plot of $\chi_M''$ *versus* T at 50, 250 and 1000 Hz for complex **8**, where $\chi_M''$ is the out-of-phase component of the ac magnetic susceptibility. The maximum in the out-of-phase ac peak occurs at 2.4 K with a $\chi_M''$ value of 4.95 $cm^3$ $mol^{-1}$ at 50 Hz. On closer inspection a small out-of-phase peak is also found at 4.2 K with a maximum $\chi_M''$ value of 0.112 $cm^3$ $mol^{-1}$. The low temperature peak has an intensity 44 times larger than the higher temperature peak. Complex **8** is the first $Mn_{12}$ complex to show predominantly an out-of-phase ac susceptibility peak in the 2-3 K region. All of the previously reported $Mn_{12}$ complexes have predominantly an out-of-phase ac magnetic susceptibility peak in the 4-7 K region.

Several experiments were carried out in order to determine the origin of the two out-of-phase ac peaks. In the upper part of Figure 3, $\chi_M'T$ is plotted as a function of temperature in the range of 2 to 10 K for the p-chlorobenzoate complex **5**. Two plateau regions in the plot of $\chi_M'T$ *versus* T are seen, one in the 6-10 K region with a $\chi_M'T$ value of 44 $cm^3$ K $mol^{-1}$ and another in the 2.6-3.6 K region with a $\chi_M'T$ value of 23 $cm^3$ K $mol^{-1}$. The lower part of Figure 3 shows that two out-of-phase ac magnetic susceptibility peaks are observed in the same temperature regions that steps occur in the $\chi_M'T$ *versus* T plot. At 50 Hz, the maxima in the peaks are found at 4.4 K with a $\chi_M$'' value of 1.29 $cm^3\ mol^{-1}$, and at 2.0 K with a $\chi_M$'' value of 2.28 $cm^3\ mol^{-1}$. The intensity of the 2.0 K peak is 1.8 times larger than that for the 5.0 K peak. Higher operating frequencies shift the $\chi_M$'' peaks to higher temperatures. Small differences in the ratio of the two $\chi_M$'' peaks were observed for different samples of complex **5** synthesized with the direct synthesis approach. For another sample of complex **5**, for example, the ratio of the intensities of the 2 K peak to 5 K peak was found to be 1.2 (plot not shown). Factors such as reaction conditions and solvate molecules in the crystal lattice were probed as the origin(s) of the two ac signals. Unfortunately, complex **5** was not an ideal candidate for these studies because recrystallization leads to poor yields and the crystals of complex **5** readily lose solvate molecules. However, the *p*-methylbenzoate $Mn_{12}$ complex was found to be a better candidate for a systematic study.

Slight modifications (see **Experimental Section**) in the preparation of the *p*-methylbenzoate complex result in changes in the number and types of solvate molecules in the $Mn_{12}$ crystal lattice and also lead to changes in the intensity ratio of the two out-of-phase ac peaks. In Figures 4 and 5 are plotted $\chi_M$'' *vs.* T and $\chi_M$'T *vs.* T, respectively, for complex **6** (upper) and for complex **7** (lower) in the temperature range of 2-10 K at frequencies of 50, 250 or 1000 Hz. Both samples of these forms of the p-methylbenzoate $Mn_{12}$ complex have two frequency-dependent out-of-phase ac peaks, one in the 2-3 K region and the other in the 4-7 K region. However, complex **6** has predominantly a peak in the 2-3 K region (17 times more intense), whereas complex **7** (lower trace) has predominantly a peak in the 4-7 K region (25 times more intense). Several crystalline samples of $[Mn_{12}O_{12}(O_2CC_6H_4\text{-}p\text{-}Me)_{16}(H_2O)_4] \cdot x(HO_2CC_6H_4\text{-}p\text{-}Me) \cdot y(H_2O)$ were prepared and characterized. As x and y were varied, so did the intensity ratio of the two out-of-phase $\chi_M$'' peaks. Powder X-ray data collected for complex **6** and complex **7** indicate that these samples are crystallographically different. It appears that one crystallographically different form exhibits the lower temperature out-of-phase ac signal, whereas the higher temperature peak is attributable to the other crystallographic form. Single crystal structures were solved for crystals picked from each sample (**6** and **7**), and the results of these X-ray structures indicate the origins of the two ac $\chi_M''$ peaks (*vide infra*).

Ac magnetic susceptibility data were also collected for a frozen solution of $[Mn_{12}O_{12}(O_2CEt)_{16}(H_2O)_3]$ in $CD_2Cl_2/C_6D_5CD_3$ (1:1) to further examine the origin of the two out-of-phase ac signals. As can be seen in Figure 6 (lower trace), two $\chi_M''$ peaks are seen in the temperature regions of 2-3 and 4-7 K. For comparison purposes the data are also shown for a polycrystalline sample. There is a shift in peak temperatures and a change in the ratio of intensities of the two peaks when the state of this complex is changed from polycrystalline to a frozen solution. In the frozen solution the intensity ratio of the 4-7 K peak to the 2-3 K peak is 3.7, while this ratio is 12 for the microcrystalline sample. Clearly, the species responsible for the 2-3 K and 4-7 K responses are capable of existence in solution as well as the solid state.

**X-ray Structure of $[Mn_{12}O_{12}(O_2CC_6H_4\text{-}p\text{-}Cl)_{16}(H_2O)_4] \cdot 8CH_2Cl_2$ (Complex 5).** At –172 °C, this complex crystallizes in the monoclinic *C2/c* space group. An ORTEP plot of complex **5** is shown in Figure 7 and selected bond distances and angles are given in Tables 3 and 4, respectively. In addition to one half of the $Mn_{12}$ complex, the asymmetric unit was found to contain eight partially occupied $CH_2Cl_2$ solvate molecules. Two of the *p*-chlorobenzoate ligands were found to be disordered (50% in two positions). The $Mn_{12}$ complex is located on a crystallographic two-fold axis passing through Mn(3), Mn(7) and the mid-points of the Mn(1)-Mn(1a) and Mn(2)-Mn(2a) vectors. The complex possesses a $[Mn_{12}(\mu_3\text{-}O)_{12}]$ core comprising a central $[Mn^{IV}_4O_4]^{8+}$ cubane held within a non-planar ring of eight $Mn^{III}$ ions by eight $\mu_3$-$O^{2-}$ ions. Distances and angles within the core are very similar to those in previously reported[5,9,12] $[Mn_{12}O_{12}]$ complexes. The $Mn_{12}$ molecule of complex **5** has virtual $D_2$ symmetry. Overall, the structure of complex **5** is very similar to $[Mn_{12}O_{12}(O_2CPh)_{16}(H_2O)_4]$ (complex **3**).[9]

Peripheral ligation of complex **5** is provided by sixteen $\eta^2,\mu$-carboxylate groups and four $H_2O$ ligands. X-ray structures are now available for six molecules with the composition $[Mn_{12}O_{12}(O_2CR)_{16}(H_2O)_4]^n$ (n = 0 or 1-). Different forms of this molecule have been characterized based on different positionings of the $H_2O$ ligands. For all of the complexes, the eight $Mn^{III}$ ions fall into two groups of 4 $Mn^{III}$ ions. In group I, each $Mn^{III}$ ion is bonded to a *single* $Mn^{IV}$ via two oxo bridges while in group II each $Mn^{III}$ is bonded to *two* $Mn^{IV}$ via two oxo bridges. The four $H_2O$ ligands coordinate only to the 4 $Mn^{III}$ ions in group II. For the *p*-chlorobenzoate complex **5** there are two $H_2O$ ligands on Mn(5) and two on Mn(5a). This same "trans" combination of two $H_2O$/two $H_2O$ ligand positioning was also found[9] for the benzoate complex **3**. The $Mn_{12}$ acetate complex **1** has one $H_2O$ ligand on each of the four $Mn^{III}$ ions. Furthermore, the four $H_2O$ ligands of complex **1** are arranged in an alternating up, down, up, down arrangement, giving each molecule in complex **1** a 222 site symmetry. Thus, the $Mn_{12}$ acetate complex **1** has a much higher crystal site symmetry than does any other $Mn_{12}$ molecule. Finally, the anion in $(PPh_4)[Mn_{12}O_{12}(O_2CEt)_{16}(H_2O)_4]$ (complex **4**) has a two, one, one pattern of $H_2O$ ligation. Thus, one manganese atom has two $H_2O$ ligands and two manganese atoms have one $H_2O$ ligand each.

If one considers carefully the structure of $[Mn_{12}O_{12}(O_2CR)_{16}(H_2O)_4]^n$, it is possible to identify 11 different isomeric forms of a complex based on different arrangements of the four $H_2O$ ligands. This assumes that water coordination occurs only at the manganese ions in group II and only coordinates at the axial positions (up/down) on the $Mn^{III}$ ions. Of the 11 possible isomeric arrangements of four $H_2O$ ligands, X-ray structures have revealed only three of them. From one isomeric form to another there could be changes in pairwise Mn····Mn magnetic exchange interactions. This could lead to changes in the spin of the ground state or, at least, to changes in the distribution of excited spin states.

As shown in Figure 3, a polycrystalline sample of complex **5** exhibits two out-of-phase ac peaks. There may be two or more structurally different forms of complex **5**. Unfortunately, it is not possible from this single structure to identify which $\chi_M$'' peak arises from the structurally characterized *p*-chlorobenzoate complex. However, two crystallographic forms of the *p*-methylbenzoate and propionate derivative $[Mn_{12}O_{12}(O_2CEt)_{16}(H_2O)_3]$ have been structurally characterized and differences in the $Mn_{12}$ molecules were found.

**X-ray Structure of [$Mn_{12}O_{12}(O_2$ CEt$)_{16}(H_2O)_3$] (Complex 2b)**. Two crystal forms of $Mn_{12}O_{12}(O_2CEt)_{16}(H_2O)_3$ (**2a**, **2b**) have been identified. The two different crystals are readily distinguished since one has a cube morphology, while the other has a needle morphology. As characterized previously,[13] the crystal with cube morphology (complex **2a**) crystallizes in the triclinic space group $P\overline{1}$. The newly characterized crystal with needle morphology (complex **2b**) crystallizes in the monoclinic $P2_1/c$ space group, with four $Mn_{12}$ molecules in the unit cell and no observable lattice solvent molecules. Data collection parameters are shown in Table 1. Selected bond distances and angles are given in Tables 5 and 6, respectively. An ORTEP representation with the methyl groups and hydrogen atoms omitted for clarity is shown in Figure 8; and the unit cell is shown at the bottom of Figure 8. This newly characterized form of the propionate complex is a structural isomer of the previously characterized complex, differing in the positions of axial $H_2O$ molecules, and the position of some propionate ligands. No lattice solvent was found in the new structure (complex **2b**), whereas four $H_2O$ molecules and 1/2 toluene were found in the previously characterized structure (complex **2a**). As indicated above, the placement of water ligands has been found to vary among the various structurally characterized $Mn_{12}O_{12}$; however, this is the first instance of structural isomers among [$Mn_{12}O_{12}$] complexes with the *same* ligand set. Like the previously characterized [$Mn_{12}O_{12}(O_2CEt)_{16}(H_2O)_3$] complex (**2a**), some disorder was observed for complex **2b**. The atom C(92) refined as two positions at 50% occupancy each and an additional less well defined 40/60% disorder between O(25) of a coordinated $H_2O$ and the oxygen atoms O(75) of a propionate ligand was found. The cores of complex **2b** and, previously characterized complex **2a** are not superimposable and a best molecular fit is shown in Figure 9, emphasizing the nonsuperimposability of the axial water molecules. Also, the positions of some of the propionate ligands are nonsuperimposable. These structural differences on the $Mn_{12}O_{12}$ core could lead to changes in the electronic environments of the $Mn^{III}$ ions and thereby change the exchange interactions and the resulting magnetic properties. At this time, it is not known whether the low temperature (2-3 K) signal is due to the cube or needle form of crystals, or a third form in the sample that was not identified. Less uncertainty was achieved with the *p*-methylbenzoate complexes **6** and **7**.

**X-ray Structures of the p-Methylbenzoate Complexes 6 and 7**. Considerable effort was invested to obtain the X-ray structures of complexes **6** and **7**. This was important for we have established which crystal form gives the 2-3 K region $\chi_M$'' ac peak and which gives the 4-7 K region $\chi_M$'' ac peak. As is clear from Table 2, complexes **6** and **7** are crystallographically different. Complex **6** crystallizes in the *C*2/c space group and has one molecule of *p*-methylbenzoic acid as a solvate molecule. This solvate molecule was found to be disordered in the refinement of the X-ray data. Complex **7** crystallizes in the *I*2/a space group. In the refinement **7**, three $H_2O$ solvate molecules were located.

The $Mn_{12}$ complexes in **6** and **7** have the same composition of $[Mn_{12}O_{12}(O_2CC_6H_4\text{-}p\text{-}Me)_{16}(H_2O)_4]$, and in Figure 10 are given ORTEP representations of their cores (*i.e.*, without *p*-methylbenzoate ligands). The cores are quite similar to those of other $Mn_{12}$ complexes; tables of bond distances and angles are given in Tables 7-10. From Figure 10 it can be seen that the $Mn_{12}$ complexes in complexes **6** and **7** differ in the positioning of the four $H_2O$ ligands. Complex **6** has two $H_2O$ ligands on the Mn(5) atom and one $H_2O$ each on the Mn(7) and Mn(7A) atoms. Thus, it has a [1(down)-2-1(up)] pattern for the four $H_2O$ ligands. Complex **7** has two $H_2O$ ligands on the Mn(11) atom and one $H_2O$ each on the Mn(12) and Mn(9) atoms. Thus, complex **7** has a [1(down)-1(up)-2] pattern for the four $H_2O$ ligands.

Complexes **6** and **7** have one other very important difference in their structures. Each $Mn^{III}$ ion in a given $[Mn^{IV}_4Mn^{III}_8O_{12}(O_2CR)_{16}(H_2O)_4]$ complex experiences a Jahn-Teller (JT) elongation where the Mn-ligand distances to two trans-located ligands are appreciably longer than those for the other four ligands. As shown in Figure 11 (top), all of the JT elongation axes in the hydrate complex **7** are all very roughly parallel and perpendicular to the plane of the disc-like $Mn_{12}O_{12}$ core. This is as generally expected from bonding considerations, since JT elongation axes avoid the $Mn\text{-}O^{2-}$ bonds, usually the shortest and strongest bonds around the Mn atom. For complex **6** in Figure 11 (bottom), however, it can be seen that one JT axis is abnormally oriented; the situation is slightly complicated by the fact that the molecule has a crystallographic $C_2$ axis disordering the JT axis about two positions (shown with dashed lines in Figure 11) but, nevertheless, the JT axis is clearly not in its normal position but in an abnormal position containing a core $O^{2-}$ ion.

Complexes **6** and **7** thus display the new phenomenon of Jahn-Teller isomerism, as recently identified[49] and defined for $[Mn_{12}O_{12}(O_2CCH_2Bu^t)_{16}(H_2O)_4]$ *i.e.*, non-equivalent molecules differing in the relative orientation of one or more JT distortion axes. The change in the orientation of the JT axis becomes evident on considering the Mn-O bonds at the unusual Mn(6) atom in complex **6**. The elongated Mn-O bonds at Mn(9) in complex **7** [2.115(6) and 2.199(7) Å] are instead normal lengths in complex **6** [1.991(8)) Å]. Similarly, although complicated by the $C_2$ axis, the elongated Mn-$O^{2-}$ [1.985(5) Å] and Mn-O [2.036(7) Å] bonds in complex **6** are noticeably longer than in complex **7** [1.898(5)/1.899(5) and 1.939(6)/1.956(7) Å, respectively]. As described below, this difference in structure involving the JT axes is believed to be the origin of the markedly different magnetic properties between complexes **6** and **7**.

The two other complexes structurally characterized in this paper, complexes **2b** and **5**, show the more usual JT elongation axis orientation seen for complex **7**. Polycrystalline samples of these compounds show both the 2-3 K and the 4-7 K range $\chi_M$'' ac susceptibility signals. However, these polycrystalline samples are likely mixtures of different geometrical and JT distortion isomers. In the case of the propionate molecules two different isomers, complexes **2a** and **2b**, have already been structurally characterized.

**NMR Spectroscopy**. In the solid state, the propionate and *p*-methylbenzoate $Mn_{12}$ complexes each have two structural isomers. The positions of the axial $H_2O$ and some of the carboxylate ligands are the main differences between the isomers. In frozen solution ac magnetic studies of the propionate complex, two $\chi_M$'' ac peaks are seen, just as observed for a solid sample. However, the ratios of the two $\chi_M$'' peaks are not the same. In order to try to rationalize these differences, solution $^1$H NMR studies were carried out to probe the structure of $Mn_{12}$ complexes in solution. Proton peak assignments were made using a combination of 1-D and 2-D $^1$H NMR spectroscopy.

Extensive $^1$H NMR studies and peak assignment for the propionate complex **2** are published elsewhere.[13] A total of seven carboxylate proton peaks were seen in solution $^1$H NMR studies on complex **2**, and this indicates that the effective solution symmetry is higher than the solid-state symmetry ($C_1$). These results suggest *intra*molecular ligand exchange processes involving especially the axial $H_2O$ and carboxylate ligands in solution. Such exchange of ligands would result in one isomeric form of the propionate $Mn_{12}$ complex converting to another, and the amounts of different isomeric forms in solution is not, therefore, necessarily the same as for the solid state.

Although many examples of paramagnetic NMR spectra of metal acetate complexes exist in the literature, definitive assignments of resonances due to benzoate protons are much less common. A series of experiments were performed in order to correctly assign the proton resonances in the complex spectra of the dodecanuclear $Mn_{12}$ benzoate derivatives. In the past, the use of $^{19}$F NMR data for para-substituted benzoates has supported the integrity of the $Mn_{12}$ complex in solution; however, individual resonances in the proton spectra were not assigned.

Recent use of a combination of $T_1$ measurements and 2-D NMR has allowed the assignment of these spectra. The 1-D $^1$H NMR spectrum of $[Mn_{12}O_{12}(O_2CPh)_{16}(H_2O)_4]$ (**3**) shows approximately nine peaks as shown in Figure 12 and as summarized in Table 6, as well as peaks from the protio-$CD_2Cl_2$ solvent and $Et_2O$. Assuming virtual $D_2$ symmetry in solution, complex **3** should have four types of carboxylates (with four carboxylates in each): two types of equatorial carboxylates bridging $Mn^{III}\cdots Mn^{III}$ pairs, axial carboxylates bridging $Mn^{III}\cdots Mn^{III}$ pairs, and axial carboxylates bridging $Mn^{III}\cdots Mn^{IV}$ pairs. For each of these four types of carboxylates there should be three resonances due to the *o*-, *m*-, and *p*- protons, giving a total of 12 peaks. There are only nine signals, so it is expected that three of the nine peaks are doubly intense signals, and this is borne out by integrations of the peaks. All of the peaks from the complex are found within the relatively narrow paramagnetic chemical shift window of 14 ppm to -2 ppm, as might be expected from a complex with all protons significantly removed (5 bonds) from the paramagnetic center. The broadness of these peaks varies greatly, and it is expected that the ortho-protons, which are closer to the paramagnetic metal center should be broadest. On the basis of comparison with $^{1}H$ NMR data for other $Mn_{12}$ complexes substituted in various positions on the aromatic ring, and integration and $T_1$ values measured by the inversion recovery [(180°-t-90°)] method, initial assignments of the *o*-, *m*-, and *p*- protons were made. The meta-proton peaks were assigned as those at *ca*. 13(A), 10(C) and 5(F) ppm, with the last peak as a doubly intense resonance. The signals due to the para-protons are only half as intense as those from the meta-positions, because there is only one para-position on each ring, and thus the peaks at *ca*. 7.3(E), 4.2(G), and -0.8(I) ppm were assigned as due to the para-proton resonances. The ortho-proton resonances were significantly broader than the other peaks, and were easily assigned as the peaks at *ca*. 11(B), (D), and 1.2(H) ppm. Unfortunately, due to the overlap of the ortho-proton peaks with other peaks, it was difficult to get an accurate integration of these extremely broad peaks.

After these tentative assignments were made, a 2-D COSY spectrum of complex **3** was run. Although 2-D paramagnetic NMR was not a very useful diagnostic tool in the case of $[Mn_{12}O_{12}(O_2CEt)_{16}(H_2O)_3]$ (**2**), it was hoped that the significantly longer $T_1$ (and $T_2$) values of the benzoate complex would facilitate the observation of cross peaks. As can be seen in Figure 12, the COSY spectrum of $[Mn_{12}O_{12}(O_2CPh)_{16}(H_2O)_4]$ (**3**) shows a number of cross peaks between the proton resonances. Coupling can be seen between the meta- and para-protons using this method, but no cross peaks to the ortho-proton can be seen. Because peaks F and I are doubly intense, they can be assigned as due to the overlap of the meta- and para-protons respectively of the two types of $Mn^{III}\cdots Mn^{III}$ equatorial carboxylates.[12] This observation is consistent with the previously characterized NMR spectra of $Mn_{12}$ complexes. Also, because peaks C and E are the narrowest, have longer $T_1$ values, and are located closest to shift values of benzoic acid, these are assigned as due to the axial $Mn^{III}\cdots Mn^{IV}$ bridging carboxylates. In the acetate and propionate complexes, the $^{1}$H NMR resonances of the axial $Mn^{III}\cdots Mn^{IV}$ carboxylates also experience smaller isotropic shifts relative to their $Mn^{III}\cdots Mn^{III}$ counterparts.[13]

TOCSY (total correlation spectroscopy) has been used successfully to detect cross peaks of paramagnetic species that are not visible using the COSY technique.[50] As shown in Figure 13, the TOCSY spectrum of complex **3** shows an additional cross peak between peaks A and D, and allows the assignment of D as the axial $Mn^{IV}\cdots Mn^{III}$ ortho-proton. No cross peaks are seen to the other ortho-protons, probably due to their weakness and broadness, but their assignments can be made on the basis of the size of the peak (see Table 6). The observed distribution of coupled *o*-, *m*-, and *p*-protons for the $Mn^{III}\cdots Mn^{III}$ carboxylates (both axial and equatorial) show isotropic shift patterns consistent with a π-delocalization mechanism through the ring of the carboxylate (*o*- and *p*- in a direction opposite to that of the *m*-proton). However, for the equatorial carboxylates, this pattern is imposed on top of an overall upfield shift of all three resonances. The protons of the $Mn^{III}\cdots Mn^{IV}$ carboxylates show a markedly different pattern. It is probable that, in these highly anisotropic molecules, dipolar contributions to the isotropic shift of these resonances play a significant role.

Unfortunately, the $^{1}$H NMR spectrum of the p-methylbenzoate complex shows a much broader and less well-resolved NMR spectrum as shown in Figure 14. The 2-D TOCSY was not particularly useful because the prominent cross peaks in the benzoate derivative are those due to the meta/para-proton pairs. The protons of the *p*-methyl group are further removed from the meta-proton position, and hence result in no cross peaks. Some assignments of this spectrum are possible though, based on comparison with the benzoate complex. Replacement of the benzoate para-proton with a para-Me group results in some marked changes in the 1-D $^{1}$H NMR spectrum. The *o*- and *m*-proton resonances have similar chemical shifts to those of complex **3**, and hence, were assigned by comparison as shown in Table 7. For the doubly intense upfield para-Me peak I in Figure 13, replacement of the proton with a methyl group results in a striking paramagnetic shift in the opposite direction: the corresponding Me proton resonance is peak B of Figure 14. The *p*-Me resonance F was assigned as the axial $Mn^{III}\cdots Mn^{IV}$ bridging carboxylate because of its proximity to the diamagnetic region, and peak H was assigned as the axial $Mn^{III}\cdots Mn^{III}$ bridging carboxylate. It appears that the additional peaks at *ca*. 7.9, 7.3 and 2.5 ppm are present from a small amount of toluic acid in solution.

**Magnetization Hysteresis Loops**. Since complexes **6** and **7** have barriers for changing their magnetic moments from "spin up" to "spin down" (Figure 1), it is informative to examine the change in the magnetization of a sample as an external field is changed. For an oriented sample of a SMM, steps can be seen at regular intervals of magnetic field in the magnetization hysteresis loop. These steps result from a quantum mechanical tunneling of the magnetization.

Oriented samples of complexes **6** and **7** were prepared by suspending a few small crystals of either complex in fluid eicosane in the 312-318 K range. In the presence of a 5.5 T magnetic field each crystallite orients with its principal axis of magnetization parallel to the direction of the external field. The eicosane is then cooled to room temperature and this gives a wax cube with the few crystallites magnetically oriented inside.

Figure 15 shows the magnetization hysteresis data measured for complex **7**. Magnetization hysteresis loops are seen in the 1.72-2.50 K range. The coercive magnetic field and consequently the area enclosed within a hysteresis loop increase as the temperature is decreased. It is instructive to examine the features seen in one of these loops. At 1.72 K the eicosane cube with oriented crystallites is first exposed to a magnetic field of +4.0 T. In this field there is a saturation of the magnetization. In reference to Figure 1 which shows the energetics in zero external field, a field of +4.0 T leads to the $m_s$ = -10 level being considerably stabilized in energy relative to the $m_s$ = +10 level. All molecules have their moments aligned parallel ("spin up") to the external magnetic field and are in the $m_s$ = -10 state at 1.72 K. The magnetic field is then swept from 4.0 T to zero. If there was no barrier for converting from "spin up" to "spin down", then at zero external field there would be equal numbers of molecules with "spin up" and "spin down", i.e., the magnetization would go to zero at zero field. This is not the case, however, because there is a barrier, and at 1.72 K the $Mn_{12}$ molecules do not have enough thermal energy to go over the barrier. Reversal of the direction of the external field, followed by increasing the field to –4.0 T again leads to magnetization saturation. In this case the $m_s = +10$ "spin down" state is stabilized in energy and all molecules are in the $m_s = +10$ state. The field is then cycled from –4.0 T to zero, reversed and then cycled back to +4.0 T. At 1.72 K the coercive magnetic field for complex **7** is *ca*. 2 T.

Close examination of the magnetization hysteresis loops shown in Figure 15 for complex **7** shows that each hysteresis loop is not smooth; steps are seen at regular intervals of the external field. These steps are due to tunneling of the magnetization. That is, a $Mn_{12}$ molecule in the "spin up" state can either be thermally activated over the barrier to the "spin down" state or it reverses its direction of magnetization by tunneling through the barrier. The steps are clear evidence of tunneling of magnetization[24-27], which can be understood in the following manner. After saturation in a +4.0 T field, the external field about the crystallites of complex **7** is reduced to zero (Figure 1) and at this instant all of the molecules are in the $m_s = -10$ state. Depending on the rate of sweep of the external field, some of the molecules may tunnel from the $m_s = -10$ to the $m_s = +10$ state, or more generally from the $m_s = -n$ to the $m_s = +n$ state (n = 10, 9, 8, …, 1). Thus, at zero field we see the first step. Reversal of the external field, followed by changing the field from zero to ~ -0.48 T leads to the appearance of a second step at H = -0.48 T. At this external field the $m_s = -10$ state has the same energy as the $m_s = +9$ state. This alignment of energy levels leads to resonant magnetization tunneling. In this way as the field is swept from zero to –4.0 T steps are seen at regular intervals of ~0.48 T. No steps are seen as the field is swept from +4.0 T toward zero. Only when the field is zero do we see the first step on the reverse sweep. The second field reversal leads to regular steps as the field is swept from zero to +4.0 T.

Magnetization hysteresis loops were also measured for an oriented eicosane cube of complex **6** at the temperatures of 1.72, 2.20, 2.00, 1.90 and 1.80 K (Figure 16). The hysteresis loops for complex **6** look quite different than those for complex **7**. When the external field is reduced from +4.4 T to zero, the magnetization falls off dramatically, and the coercive fields are considerably less for complex **6** than for complex **7**. Thus, these two *p*-methylbenzoate $Mn_{12}$ complexes experience quite different kinetic barriers for reversal of magnetization. It must be emphasized that the sweep rate for all the loops was 25 Oe/s.

First derivative plots were calculated for each of the hysteresis loops (lower plots in Figures 15 and 16). For complex **7** on the sweep from +4.0 to –4.0 T the first step is seen at zero field, followed by steps at –0.467, -0.910, -1.42, and –1.85 T. On the sweep from –4.0 T to +4.0 T the first step is seen at zero field, followed by steps at +0.467, +0.910, +1.42, and +1.85 T. On the sweep from –4.0 T to +4.0 T for complex **6**, steps are seen at zero field, +0.501, +0.992, +1.42 and +1.87 T.

A careful analysis of the hysteresis data for complex **6** shows that the external field value at each step does shift slightly with temperature. Friedman *et al.*[24,25] confirmed that the small shift is due to the fact that each $Mn_{12}$ molecule does not just experience the applied field (H) but rather it experiences a magnetic induction (B) due to combination of the external field and dipolar fields from neighboring molecules in the crystal. The magnetic induction B is given as $B = H + 4\pi M$ and is invariant for each step. Figure 17 gives a plot of magnetic induction *vs*. the temperature for the five hysteresis loops measured for complex **6**. It can be seen that the steps determined at different temperatures have essentially the same magnitude of increment in magnetic induction. In this way the increment was found to be 0.47 T for complex **6**, and 0.48 T for complex **7.**

From the hysteresis loop data it is clear that the *p*-methylbenzoate complex **6** has an appreciably greater rate of magnetization relaxation than does isomeric complex **7**. This can be quantified by analyzing the frequency dependencies of the $\chi_M''$ signals for the two complexes shown in Figure 4. Ac susceptibility data were collected at 8 different frequencies from 1.0 Hz to 1512 Hz for complex **7**. From the peaks in the $\chi_M''$ *vs*. temperature plots, values of the magnetization relaxation time $\tau$ were determined at each temperature. Figure 18 gives an Arrhenius plot of $\ln(1/\tau)$ *vs*. the inverse absolute temperature (1/T) for complex **7**. The data were least-squares fit to the Arrhenius eqn (1) to give the values of $\tau_0 = 7.7 \times 10^{-9}$ s and $U_{eff} = 64$ K.

$$\tau = \tau_0 \exp(U_{eff} / kT) \qquad \textbf{(1)}$$

A similar analysis of the frequency dependence of the dominant low temperature $\chi_M$'' peak in the ac data for complex **6** gives $\tau_0 = 2.0 \times 10^{-10}$ s and $U_{eff}$ = 38 K. The activation energy ($U_{eff}$) for reversal of the direction of the magnetization for complex **6** ($U_{eff}$ = 38 K) is considerably less than that ($U_{eff}$ = 64 K) for the isomeric complex **7**. The $Mn_{12}$-acetate complex **1** has been reported[10,11] to have a $U_{eff}$ value of 62 K, very close to the value for complex **7**.

**Origin of Two Out-of-Phase AC Susceptibility Peaks**. The presence of two different crystallographic forms of the $Mn_{12}$ *p*-methylbenzoate complex is the key to the origin of two out-of-phase ac peaks. This is a kinetic phenomenon, not just based on thermodynamics summarized in the potential-energy double well given in Figure 1. The $Mn_{12}$-acetate complex **1** has an S = 10 ground state that experiences an axial zero-field splitting ($DS_z^2$) where D = -0.50 $cm^{-1}$. The potential-energy barrier U for complex **1** can be simply calculated to be U = 50 $cm^{-1}$ = 70 K, whereas the activation energy for reversal of magnetization for complex **1** was experimentally determined[10,11] to the $U_{eff}$ = 62 K. In a simple sense $U_{eff}$ is less than U for complex **1** because this $Mn_{12}$ complex reverses its direction of magnetization not only by going over the barrier in Figure 1, but also by tunneling through the barrier.

Chudnovsky *et al.*[51] have discussed the mechanism of resonant magnetization tunneling for a SMM such as complex **1**. It was assumed that the magnetization tunneling occurs as a result of a transverse magnetic field and the rates of tunneling between pairs of $+m_S$ and $-m_S$ states were calculated. It was concluded that the rate of tunneling for the $m_S$ = -10 to $m_S$ = +10 conversion of a S = 10 molecule occurs with a lifetime longer than the universe when the transverse magnetic field is small. The rate calculated for the $m_S$ = -3 to $m_S$ = +3 tunneling (see Figure 1) was found to be close to the experimental value. It was suggested that at low temperatures where steps are seen on hysteresis loops that an individual molecule is excited by a phonon in an Orbach process from the $m_S$ = -10 level successively to the $m_S$ = -9., -8, -7, -6, -5, -4, and finally the $m_S$ = -3 level. After it is excited to the $m_S$ = -3 level, the $Mn_{12}$ molecule then tunnels to the $m_S$ = +3 level and then it quickly relaxes to the $m_s$ = +10 level. A single tunneling channel ($m_S$ = -3 to $m_S$ = +3) would be opened up and this gives the first step at zero external field in the hysteresis loop.

In addition to a transverse magnetic field, it has now been shown[52] that other interactions, such as a transverse quartic zero-field interaction, are probably also important in influencing the rate of magnetization tunneling. For each $Mn_{12}$ molecule the spin Hamiltonian given in eq (2) applies:

$$\hat{H} = \hat{H}_A + \hat{H}_Z + \hat{H}_{sp} + \hat{H}_T \quad \textbf{(2)}$$

The first term $\hat{H}_A$ is for the axial (longitudinal) zero-field interactions, the leading terms of which are given as

$$\hat{H}_A = D\hat{S}_Z^2 - B\hat{S}_Z^4 \quad \textbf{(3)}$$

The parameter D is considerably larger than B and gauges the second-order axial zero-field splitting. The second term in eqn (2), $\hat{H}_Z$, is just the Zeeman term, which in its simplest form is given in eq (4).

$$\hat{H}_Z = g\mu_B \hat{H}_Z \cdot \hat{S}_Z \quad \textbf{(4)}$$

The term $\hat{H}_{sp}$ represents the spin-phonon coupling, where a given $Mn_{12}$ complex interacts with phonons in the crystal. The last term $\hat{H}_T$, representing transverse interactions, is the most important in terms of the rate of magnetization tunneling. Some of the larger terms in $\hat{H}_T$ are given in eq (5):

$$\hat{H}_T = g\mu_B \hat{H}_x \cdot \hat{S}_x + E(\hat{S}_x^2 - \hat{S}_y^2) - B_4(\hat{S}_+^4 + \hat{S}_-^4) \quad \textbf{(5)}$$

The raising and lowering operators are given as $\hat{S}_\pm = \hat{S}_x \pm i\hat{S}_y$. The transverse magnetic field $\hat{H}_x$, the rhombic zero-field operator $(\hat{S}_x^2 - \hat{S}_y^2)$ and the quartic zero-field operator $(\hat{S}_+^4 + \hat{S}_-^4)$ mix together the $m_S$ wavefunctions and this facilitates tunneling of the magnetization. There is still considerable research needed to understand this tunneling phenomenon.[53]

The $Mn_{12}$-acetate complex **1** is excited by phonons to an $m_S$ = -3 tunneling channel. Tunneling of the magnetization in the ground state levels ($m_S$ = ±10) has not been observed for complex **1**. Tunneling from the lowest energy level has been observed for two other single-molecule magnets. The complex $[Mn_4O_3Cl(O_2CMe)_3(dbm)_3]$ (**9**) where $dbm^-$ is he monoanion of dibenzoylmethane, has an S = 9/2 ground state.[54] Since complex **9** shows frequency-dependent out-of-phase ac susceptibility signals and magnetization hysteresis loops below 0.90 K, this complex is a SMM. Steps are seen on each hysteresis loop. An Arrhenius plot of the magnetization relaxation data for complex **9** indicates a thermally activated region between 2.0 and 0.70 K and a temperature-independent region at temperatures below 0.70 K. A fit of the data in the temperature-dependent region gives $U_{eff}$ = 11.8 K and $\tau_0 = 3.6 \times 10^{-7}$ s. With the D-value obtained from high-frequency EPR (HFEPR) data for this S = 9/2 complex **9**, U can be calculated as 15.2 K ( = 10.6 $cm^{-1}$). It was concluded[54] that the temperature-independent magnetization relaxation must correspond to magnetization tunneling between the lowest degenerate levels, the $m_S$ = 9/2 and -9/2 levels for the S = 9/2 complex **9**. A temperature-independent magnetization relaxation below 0.35 K has also been reported for a $Fe^{III}_8$ complex (**10**) that has an S = 10 ground state.[55] Complex **10** has been found to have $U_{eff}$ = 24.5 K and zero-field interaction parameters of D = -0.27 K and E = -0.046 K. Tunneling in the lowest energy $m_s = \pm10$ levels is seen for this $Fe^{III}_8$ complex.

It is important to note that $Mn_4$ complex **9** and $Fe_8$ complex **10** show larger rates of tunneling in the lowest-energy level than does the $Mn_{12}$-acetate complex **1** because they possess relatively large transverse interactions, *i.e.*, $\hat{H}_T$ terms in eq (5). Due to its crystal site symmetry the $Mn_{12}$-acetate complex **1** has no rhombic zero-field splitting (E = 0). Complexes **9** and **10** probably show tunneling in the lowest-energy levels because each of these complexes is of lower symmetry and this gives a non-zero E value. Tunneling is not totally due to rhombic zero-field interactions for it has been shown that transverse magnetic fields are also important. The magnetic field can result from an external magnetic field or there could be an internal field in the crystal from neighboring molecules. In fact, a transverse component of the magnetic field created by the nuclear spins within the molecule is also important.[54,56,57]

The question at hand is why is the magnetization relaxation in *p*-methylbenzoate $Mn_{12}$ complex **6** so much different than that for the isomeric complex **7**. There are two separate parts to this question: (i) What is (are) the structural difference(s) between complexes **6** and **7** that represent(s) the origin of the difference in their magnetic properties? and (ii) what property or factor contributing to the observed magnetic behavior is affected by the structural difference(s)? The answer to part (i) is that there are two types of structural difference in the solid-state structures of **6** and **7**, the relative dispositions of the $H_2O$ and $RCO_2^-$ ligands, and the Jahn-Teller isomerism involving relative orientation of the JT axes. For the forms varying in $H_2O/RCO_2^-$ disposition, there could be resulting small changes in the $Mn^{III}..Mn^{III}$ pairwise exchange interactions, but these are, in our opinion, unlikely to have a major influence on the properties of the complexes. In fact, preliminary work has identified $Mn_{12}$ complexes that show the same geometric isomers, but different ac susceptibility characteristics. In contrast, the reorientation of one JT axis could have a more significant effect. The JT axis defines the local z axis at each $Mn^{III}$ ion, and thus the reorientation of a JT axis reorients the singly-occupied $d_{z^2}$ orbital. This will affect the pairwise $Mn^{III}..Mn^{III}$ and $Mn^{III}..Mn^{IV}$ exchange interactions involving the unique $Mn^{III}$ ion. Thus, it is reasonable that, even if the ground state spin S is not changed, the distribution of excited spin states may differ between the two complexes **6** and **7**. Also, since the molecular D value is a vector projection of the single-ion values, a reorientation of one $Mn^{III}$ anisotropy could lead to a small change to the D value between **6** and **7**, and thus affect the barrier U. Finally, the molecular symmetry of the molecule is decreased when one of the JT axes is reoriented into an equatorial position. All the described differences between **6** and **7** resulting from the JT isomerism could be contributing to some extent to the observed magnetic differences.

Complex **6** shows its out-of-phase ac susceptibility signals at essentially one-half the temperature for the peaks for complex **7**. For complex **6** it is found that $U_{eff}$ = 38 K, whereas complex **7** has been evaluated to have $U_{eff}$ = 64 K. Complex **7** behaves similarly to the high-symmetry $Mn_{12}$-acetate complex **1**. As described above, complex **7** has all of its $Mn^{III}$ JT distortion axes oriented nearly parallel, as in complex **1**. In contrast, the JT distortion axis at one $Mn^{III}$ ion in complex **6** is found to be tipped *ca*. 90° from the other axes and complex **6** thus has a lower symmetry than complex **7**. It is likely that the rhombic zero-field interactions in complex **6** are significantly larger than those in complex **7** and this means that the tunneling matrix elements for complex **6** are larger than those for complex **7**. As a consequence, complex **6** may well have a different tunneling channel than complex **7**. Complex **7** behaves similarly to complex **1** and therefore has an $m_S$ = -3 to $m_S$ = +3 tunneling channel. It could be suggested that complex **6** has a lower energy tunneling channel, such as an $m_S$ = -5 to $m_S$ = +5 channel, or at least a much faster rate of magnetization tunneling.

It is unlikely that the difference in magnetization relaxation rates between complexes **6** and **7** is due to very different potential-energy barriers, *i.e.*, values of U. The axial zero-field splitting parameters (D values) are similar for the two complexes, as indicated by the steps in the hysteresis loops; the increments between steps are essentially the same for the two isomers. The analysis of variable-field dc magnetization data for complex **6** indicates that it might have an S = 9 ground state, whereas complex **7** has an S = 10 ground state. This would have to be confirmed by HFEPR data. A change in ground state spin from S = 10 to S = 9 without a change in D would only lead to a decrease of 19% in the height of the potential-energy barrier shown in Figure 1. It is suspected that the rate of magnetization tunneling in complex **6** is appreciably greater than in complex **7** primarily because the rhombic zero-field interactions are much greater in complex **6**. Very detailed HFEPR or inelastic neutron scattering experiments are now required to evaluate the precise magnitude of the rhombic zero-field interactions in these complexes.

**Concluding Comments**

X-ray structures have been reported for four new $Mn_{12}$ single-molecule magnets. The structures of two different isomeric forms of the p-methylbenzoate $Mn_{12}$ complex were reported: $[Mn_{12}O_{12}(O_2CC_6H_4\text{-p-Me})_{16}(H_2O)_4]$ · $(HO_2CC_6H_4\text{-}p\text{-Me})$ (**6**) and $[Mn_{12}O_{12}(O_2CC_6H_4\text{-}p\text{-Me})_{16}(H_2O)_4]$ · $3(H_2O)$ (**7**). The $Mn_{12}$ molecules in complexes **6** and **7** were found to differ not only as geometric isomers involving different positioning of the $H_2O$ and carboxylate ligands, but in the orientations of $Mn^{III}$ JT elongation axes; complex **6** has one $Mn^{III}$ ion with an unusually oriented JT axis. Complex **6** is also unusual in having its out-of-phase ac susceptibility signal at relatively low temperatures, and low activation energy for magnetization relaxation. Consideration of the crystal structures of **6** and **7** has identified two main structural differences, geometrical isomerism involving the peripheral $H_2O/RCO_2^-$ ligands and Jahn-Teller isomerism. It is considered unlikely that the former has a significant effect on the magnetic properties. In contrast, the latter is believed to be the source of the magnetic differences, affecting the symmetry of the molecule as well as the anisotropy and spin state energy distribution. It is believed that complex **6** has appreciably larger rhombic zero-field interactions $[E(\hat{S}_x^2 - \hat{S}_y^2)]$ than complex **7**, i.e., the lower structural symmetry of complex **6** leads to the enhanced rhombicity of zero-field interactions and this will lead to a relatively fast rate of quantum mechanical tunneling of magnetization. Thus, it is concluded that the faster magnetization relaxation rate of **6** is primarily the result of larger quantum tunneling rates compared with **7**.

**Acknowledgments**. This work was supported by the National Science Foundation (G.C. and D.N.H.). The ac magnetic susceptibility measurements were performed with a MPMS2 SQUID magnetometer provided by the Center for Interface and Material Science, funded by the W. A. Keck Foundation.

**Supporting Material Available**: Tables giving full crystallographic details for complexes **2b**, **5**, **6** and **7** (00 pages). See any current masthead page for ordering or internet access instructions.

**Table 1**. Crystallographic Data for $[Mn_{12}O_{12}(O_2CC_6H_4\text{-}p\text{-}Cl)_{16}(H_2O)_4] \cdot 8CH_2Cl_2$ (Complex **5**) and $[Mn_{12}O_{12}(O_2CEt)_{16}(H_2O)_3]$ (Complex **2b**).

| parameter | Complex **5** | Complex **2b** |
|---|---|---|
| Formula | $C_{112}H_{72}Cl_{16}Mn_{12}O_{48} \cdot 8CH_2Cl_2$ | $C_{48}H_{86}Mn_{12}O_{47}$ |
| Formula Wt. g/mol[a] | 4091.73 | 2074.44 |
| Crystal System | Monoclinic | Monoclinic |
| Space group | $C2/c$ | $P2_1/c$ |
| a, Å | 29.697(9) | 16.462(7) |
| b, Å | 17.708(4) | 22.401(9) |
| c, Å | 30.204(8) | 20.766(9) |
| β,° | 102.12(2) | 103.85(2) |
| V, $Å^3$ | 15529.05 | 7435.13 |
| *Z* | 4 | 4 |
| Crystal dim. mm | 0.016 x 0.022 x 0.034 | 0.30 x 0.05 x 0.04 |
| T, °C | -172 | -170 |
| Radiation λ, Å[b] | 0.71069 | 0.71069 |
| $\rho_{calc}$ g/cm$^3$ | 1.750 | 1.853 |
| μ, cm$^{-1}$ | 15.684 | 20.641 |
| Octants | +h, +k, ±l | +h, +k, ±l |
| ScanSpeed, °/min | 6.0 | 6.0 |
| Scan Width, ° | 1.5 | 1.4 + dispersion |
| Data collected | 6 2θ 45 | 6 2θ 40 |
| Total Data | 14252 | 7780 |
| Unique Data | 10080 | 6963 |
| $R_{merge}$ | 0.051 | 0.104 |
| Obsd. Data[c] | 6257 | 3951 |
| R($R_w$)[d,e] | 0.109(0.109) | 0.902(0.811) |
| Goodness of fit[f] | 2.035 | 1.225 |

[a]including solvent molecules. [b]graphite monochromator cF > 3σ(F).
[d]$R = \sum(\|F_0| - |F_c\|)\sum|F_0|$. [e]$R_w = [\sum w(|F_0| - |F_c|)^2 / w|F_0|^2]^{1/2}$ where $w = 1/\sigma^2(|F_0|)$.
[f]$gof = [\sum w(|F_0| - |F_c|)^2]/(n-p)]^{1/2}$; n is the number of observed reflections, p is the number of refined parameters.

**Table 2**. Crystallographic Data for Complexes **6** and **7**.

| parameter | Complex **7** | Complex **6** |
|---|---|---|
| Formula | $C_{128}H_{126}Mn_{12}O_{51}$ | $C_{136}H_{128}Mn_{12}O_{50}$ |
| Formula Wt. g/mol[a] | 3139.46 | 3221.66 |
| Space group | *I*2/*a* | *C*2/*c* |
| a, Å | 29.2794(4) | 40.4589(5) |
| b, Å | 32.3271(4) | 18.2288(2) |
| c, Å | 29.8738(6) | 26.5882(4) |
| β,° | 99.2650(10) | 125.8359(2) |
| V, $Å^3$ | 27907.2(8) | 15897.1(4) |
| *Z* | 8 | 4 |
| Crystal color, habit | black block | black block |
| *D*(calc), g $cm^3$ | 1.488 | 1.346 |
| μ(MoKα), $cm^{-1}$ | 11.31 | 9.94 |
| temp, K | 223(2) | 193(2) |
| absorption correction | empirical | empirical |
| T(max)/T(min) | 1.000/0.390 | 1.000/0.382 |
| diffractometer | Siemens P4/CCD | |
| radiation | MoKα (λ= 0.71071 Å) | |
| *R(F)*, %[a] | 8.80 | 10.21 |
| *R(wF²)*, %[a] | 21.58 | 24.88 |

[a]Quantity minimized =

$$R(wF^2) = \sum[w(|F_0|^2 - |F_c|^2)^2 / \sum[(wF_0{}^2)^2]^{1/2}; R = \sum(||F_0| - |F_c||) / \sum|F_0|$$

**Table 3**. Selected[a] Bond Distances (Å) for $[Mn_{12}O_{12}(O_2CC_6H_4\text{-}p\text{-}Cl)_{16}(H_2O)_4] \cdot 8CH_2Cl_2$ (Complex **5**).

| A | B | distance (Å) | A | B | distance (Å) |
|---|---|---|---|---|---|
| Mn(1) | Mn(1)a | 2.836(5) | Mn(4) | O(10) | 1.885(13) |
| Mn(1) | Mn(2) | 2.816(4) | Mn(4) | O(11) | 1.9082(15) |
| Mn(1) | Mn(2)a | 2.971(3) | Mn(4) | O(18) | 2.211(18) |
| Mn(1) | Mn(4) | 2.790(3) | Mn(4) | O(38) | 1.931(3) |
| Mn(1) | O(8) | 1.9357(14) | Mn(4) | O(48) | 2.181(16) |
| Mn(1) | O(9)a | 1.889(15) | Mn(4) | O(56) | 1.932(13) |
| Mn(1) | O(9) | 1.933(12) | Mn(5) | O(11) | 1.878(13) |
| Mn(1) | O(10) | 1.8619(13) | Mn(5) | O(12) | 1.873(9) |
| Mn(1) | O(11) | 1.883(14) | Mn(5) | O(14) | 2.189(18) |
| Mn(1)□ | O(16) | 1.917(16) | Mn(5) | O(15) | 2.186(14) |
| Mn(2) | Mn(2)a | 2.819(8) | Mn(5) | O(58) | 1.950(9) |
| Mn(2)□ | Mn(6) | 2.790(3) | Mn(5) | O(66) | 1.957(14) |
| Mn(2) | O(8) | 1.898(17) | Mn(6) | O(12) | 1.9020(17) |
| Mn(2) | O(8)A | 1.926(13) | Mn(6) | O(13) | 1.907(13) |
| Mn(2) | O(9)A | 1.9231(13) | Mn(6) | O(28) | 2.221(17) |
| Mn(2) | O(12) | 1.869(12) | Mn(6) | O(68) | 1.966(14) |
| Mn(2) | O(13) | 1.8576(10) | Mn(6) | O(76) | 2.194(17) |
| Mn(2) | O(26) | 1.892(17) | Mn(6) | O(86) | 1.9125(28) |
| Mn(3) | O(10) | 1.925(7) | Mn(7) | O(13) | 1.898(13) |
| Mn(3) | O(10)A | 1.925(7) | Mn(7) | O(78) | 2.134(12) |
| Mn(3) | O(36) | 1.989(7) | Mn(7) | O(88) | 1.974(12) |
| Mn(3) | O(46) | 2.068(20) | | | |

[a]For the $[Mn_{12}O_{12}]$ core and $H_2O$ molecules only; a full listing is available in the supplementary material.

**Table 4**. Selected[a] Bond Angles (°) for $[Mn_{12}O_{12}(O_2CC_6H_4\text{-}p\text{-}Cl)_{16}(H_2O)_4] \cdot 8CH_2Cl_2$ (Complex **5**).

| A | B | C | angle (°) | A | B | C | angle (°) |
|---|---|---|---|---|---|---|---|
| O(8) | Mn(1) | O(9) | 79.1(4) | O(36) | Mn(3) | O(46) | 91.1(6) |
| O(8) | Mn(1) | O(9)a | 84.0(5) | O(36) | Mn(3) | O(46)a | 84.4(5) |
| O(8) | Mn(1) | O(10) | 174.8(5) | O(46) | Mn(3) | O(46)a | 174.0(5) |
| O(8) | Mn(1) | O(11) | 98.9(4) | O(10) | Mn(4) | O(11) | 83.1(4) |
| O(8) | Mn(1) | O(16) | 91.3(5) | O(10) | Mn(4) | O(18) | 85.4(6) |
| O(9)a | Mn(1) | O(9) | 83.2(7) | O(10) | Mn(4) | O(38) | 96.9(4) |
| O(9) | Mn(1) | O(10) | 97.1(4) | O(10) | Mn(4) | O(48) | 91.9(6) |
| O(9)a | Mn(1) | O(10) | 92.1(5) | O(10) | Mn(4) | O(56) | 178.8(3) |
| O(9) | Mn(1) | O(11) | 171.4(7) | O(11) | Mn(4) | O(18) | 85.6(5) |
| O(9)a | Mn(1) | O(11) | 88.3(6) | O(11) | Mn(4) | O(38) | 175.5(7) |
| O(9) | Mn(1) | O(16) | 95.5(6) | O(11) | Mn(4) | O(48) | 91.5(5) |
| O(9)a | Mn(1) | O(16) | 175.2(13) | O(11) | Mn(4) | O(56) | 96.0(4) |
| O(10) | Mn(1) | O(11) | 84.4(4) | O(18) | Mn(4) | O(38) | 89.9(5) |
| O(10) | Mn(1) | O(16) | 92.6(4) | O(18) | Mn(4) | O(48) | 176.2(3) |
| O(11) | Mn(1) | O(16) | 92.9(6) | O(18) | Mn(4) | O(56) | 93.8(6) |
| O(8) | Mn(2) | O(8)a | 84.1(5) | O(38) | Mn(4) | O(48) | 93.0(5) |
| O(8)a | Mn(2) | O(9)a | 79.6(4) | O(38) | Mn(4) | O(56) | 84.0(4) |
| O(8) | Mn(2) | O(9)a | 84.1(5) | O(48) | Mn(4) | O(56) | 88.8(6) |
| O(8)a | Mn(2) | O(12) | 172.8(7) | O(11) | Mn(5) | O(12) | 92.0(5) |
| O(8) | Mn(2) | O(12) | 88.8(6) | O(11) | Mn(5) | O(14) | 93.2(5) |
| O(8)a | Mn(2) | O(13) | 96.2(4) | O(11) | Mn(5) | O(15) | 92.1(6) |
| O(8) | Mn(2) | O(13) | 91.8(5) | O(11) | Mn(5) | O(58) | 92.7(5) |
| O(8)a | Mn(2) | O(26) | 95.1(6) | O(11) | Mn(5) | O(66) | 174.3(3) |
| O(8) | Mn(2) | O(26) | 176.1(6) | O(12) | Mn(5) | O(14) | 89.9(5) |
| O(9)a | Mn(2) | O(12) | 98.7(3) | O(12) | Mn(5) | O(15) | 94.5(5) |
| O(9)a | Mn(2) | O(13) | 174.4(7) | O(12) | Mn(5) | O(58) | 175.2(6) |
| O(9)a | Mn(2) | O(26) | 92.0(5) | O(12) | Mn(5) | O(66) | 93.2(5) |
| O(12) | Mn(2) | O(13) | 85.0(4) | O(14) | Mn(5) | O(15) | 173.0(5) |
| O(12) | Mn(2) | O(26) | 91.9(6) | O(14) | Mn(5) | O(58) | 89.6(5) |
| O(13) | Mn(2) | O(26) | 92.1(5) | O(14) | Mn(5) | O(66) | 84.8(6) |
| O(10)a | Mn(3) | O(10) | 92.6(4) | O(15) | Mn(5) | O(58) | 85.5(5) |
| O(10) | Mn(3) | O(36) | 92.5(3) | O(15) | Mn(5) | O(66) | 89.6(5) |
| O(10) | Mn(3) | O(36)a | 174.6(4) | O(58) | Mn(5) | O(66) | 82.0(5) |
| O(10)a | Mn(3) | O(36)a | 92.5(3) | O(12) | Mn(6) | O(13) | 82.8(4) |
| O(10) | Mn(3) | O(46)a | 90.3(6) | O(12) | Mn(6) | O(28) | 85.0(5) |
| O(10) | Mn(3) | O(46) | 93.8(6) | O(12) | Mn(6) | O(68) | 96.1(4) |
| O(10)a | Mn(3) | O(46) | 90.3(6) | O(12) | Mn(6) | O(76) | 90.6(5) |
| O(36)a | Mn(3) | O(36) | 82.4(4) | O(12) | Mn(6) | O(86) | 176.0(7) |
| O(36) | Mn(3) | O(46)a | 84.4(6) | O(13) | Mn(6) | O(28) | 84.9(6) |

[a]For the $[Mn_{12}O_{12}(H_2O)_4]$ core only; a full listing is available in the supplementary material.

**Table 5**. Selected Bond Distances (Å) for [$Mn_{12}O_{12}(O_2CEt)_{16}(H_2O)_3$ ] (Complex **2b**).

| | | | |
|---|---|---|---|
| Mn(1)...Mn(2) | 2.854(5) | Mn(1)...Mn(3) | 2.897(5) |
| Mn(1)...Mn(4) | 2.814(2) | Mn(1)...Mn(5) | 2.766(5) |
| Mn(1)...Mn(6) | 3.433(5) | Mn(1)...Mn(12) | 3.452(5) |
| Mn(2)...Mn(3) | 2.813(5) | Mn(2)...Mn(4) | 2.889(5) |
| Mn(2)...Mn(6) | 3.412(5) | Mn(2)...Mn(7) | 2.740(5) |
| Mn(2)...Mn(8) | 3.448(5) | Mn(3)...Mn(4) | 2.814(4) |
| Mn(3)...Mn(8) | 3.419(5) | Mn(3)...Mn(9) | 2.748(5) |
| Mn(3)...Mn(10) | 3.459(5) | Mn(4)...Mn(10) | 3.405(5) |
| Mn(4)...Mn(11) | 2.773(4) | Mn(4)...Mn(12) | 3.437(5) |
| Mn(5)...Mn(6) | 3.333(5) | Mn(5)...Mn(12) | 3.310(5) |
| Mn(6)...Mn(7) | 3.400(5) | Mn(7)...Mn(8) | 3.371(5) |
| Mn(8)...Mn(9) | 3.451(5) | Mn(9)...Mn(10) | 3.352(5) |
| Mn(10)...Mn(11) | 3.459(5) | Mn(11)...Mn(12) | 3.323(5) |
| Mn(1)–O(13) | 1.904(17) | Mn(1)–O(14) | 1.903(9) |
| Mn(1)–O(16) | 1.884(10) | Mn(1)–O(17) | 1.876(17) |
| Mn(1)–O(18) | 1.868(11) | Mn(1)–O(28) | 1.900(8) |
| Mn(2)–O(13) | 1.921(11) | Mn(2)–O(14) | 1.933(17) |
| Mn(2)–O(15) | 1.909(11) | Mn(2)–O(19) | 1.834(11) |
| Mn(2)–O(20) | 1.899(18) | Mn(2)–O(33) | 1.928(10) |
| Mn(3)–O(13) | 1.891(11) | Mn(3)–O(15) | 1.876(9) |
| Mn(3)–O(16) | 1.921(16) | Mn(3)–O(21) | 1.860(17) |
| Mn(3)–O(22) | 1.894(10) | Mn(3)–O(38) | 1.907(7) |
| Mn(4)–O(14) | 1.920(11) | Mn(4)–O(15) | 1.905(17) |
| Mn(4)–O(16) | 1.929(7) | Mn(4)–O(23) | 1.860(10) |
| Mn(4)–O(24) | 1.835(17) | Mn(4)–O(43) | 1.907(9) |
| Mn(5)–O(17) | 1.937(10) | Mn(5)–O(18) | 1.895(17) |
| Mn(5)–O(30) | 2.204(10) | Mn(5)–O(48) | 1.941(11) |
| Mn(5)–O(53) | 2.112(10) | Mn(5)–O(58) | 1.939(18) |
| Mn(6)–O(18) | 1.915(15) | Mn(6)–O(19) | 1.866(14) |
| Mn(6)–O(25) | 2.286(18) | Mn(6)–O(50) | 1.919(15) |
| Mn(6)–O(55) | 2.068(11) | Mn(6)–O(63) | 1.916(17) |
| Mn(7)–O(19) | 1.912(18) | Mn(7)–O(20) | 1.904(12) |
| Mn(7)–O(35) | 2.158(8) | Mn(7)–O(65) | 1.932(15) |
| Mn(7)–O(68) | 1.926(19) | Mn(7)–O(73) | 2.135(10) |
| Mn(8)–O(20) | 1.874(15) | Mn(8)–O(21) | 1.884(15) |
| Mn(8)–O(21) | 1.884(15) | Mn(8)–O(26) | 2.272(7) |
| Mn(8)–O(70) | 1.976(16) | Mn(8)–O(75) | 2.162(7) |
| Mn(8)–O(78) | 1.969(15) | Mn(9)–O(21) | 1.880(11) |
| Mn(9)–O(22) | 1.884(17) | Mn(9)–O(40) | 2.251(7) |
| Mn(9)–O(80) | 1.966(18) | Mn(9)–O(83) | 1.933(12) |
| Mn(9)–O(88) | 2.134(8) | Mn(10)–O(22) | 1.890(9) |
| Mn(10)–O(23) | 1.870(13) | Mn(10)–O(27) | 2.200(14) |
| Mn(10)–O(85) | 1.952(13) | Mn(10)–O(90) | 2.141(14) |
| Mn(10)–O(93) | 1.960(8) | Mn(11)–O(23) | 1.930(15) |
| Mn(11)–O(24) | 1.885(9) | Mn(11)–O(45) | 2.167(14) |
| Mn(11)–O(95) | 1.957(9) | Mn(11)–O(98) | 1.990(16) |
| Mn(11)–O(103) | 2.127(13) | Mn(12)–O(17) | 1.892(9) |
| Mn(12)–O(24) | 1.888(14) | Mn(12)–O(60) | 1.932(16) |
| Mn(12)–O(100) | 1.943(9) | Mn(12)–O(105) | 2.086(14) |

**Table 6**. Selected Bond Angles (°) for [$Mn_{12}O_{12}(O_2CEt)_{16}(H_2O)_3$] (Complex **2b**).

| | | | |
|---|---|---|---|
| O(13)–Mn(1)–O(14) | 83.7(6) | O(13)–Mn(1)–O(16) | 80.9(6) |
| O(13)–Mn(1)–O(17) | 174.4(2) | O(13)–Mn(1)–O(18) | 97.9(6) |
| O(13)–Mn(1)–O(28) | 90.2(5) | O(14)–Mn(1)–O(16) | 85.2(4) |
| O(14)–Mn(1)–O(18) | 90.1(4) | O(14)–Mn(1)–O(28) | 173.8(7) |
| O(16)–Mn(1)–O(17) | 94.2(6) | O(16)–Mn(1)–O(18) | 175.3(5) |
| O(16)–Mn(1)–O(28) | 92.5(4) | O(17)–Mn(1)–O(18) | 86.8(6) |
| O(17)–Mn(1)–O(28) | 92.9(6) | O(18)–Mn(1)–O(28) | 92.1(4) |
| O(18)–Mn(1)–O(28) | 92.1(4) | O(13)–Mn(2)–O(14) | 82.5(6) |
| O(13)–Mn(2)–O(15) | 82.8(4) | O(13)–Mn(2)–O(19) | 91.5(4) |
| O(13)–Mn(2)–O(20) | 90.2(6) | O(13)–Mn(2)–O(33) | 175.8(6) |
| O(14)–Mn(2)–O(15) | 81.8(6) | O(14)–Mn(2)–O(19) | 92.3(7) |
| O(14)–Mn(2)–O(20) | 172.7(4) | O(14)–Mn(2)–O(33) | 94.7(6) |
| O(15)–Mn(2)–O(19) | 172.2(7) | O(15)–Mn(2)–O(20) | 97.8(6) |
| O(15)–Mn(2)–O(33) | 93.8(4) | O(19)–Mn(2)–O(20) | 87.4(6) |
| O(19)–Mn(2)–O(33) | 91.8(5) | O(20)–Mn(2)–O(33) | 92.6(6) |
| O(20)–Mn(2)–O(33) | 92.6(6) | O(13)–Mn(3)–O(15) | 84.4(4) |
| O(13)–Mn(3)–O(16) | 80.3(6) | O(13)–Mn(3)–O(21) | 98.8(6) |
| O(13)–Mn(3)–O(22) | 173.9(6) | O(13)–Mn(3)–O(38) | 87.6(4) |
| O(15)–Mn(3)–O(16) | 84.8(6) | O(15)–Mn(3)–O(21) | 87.0(6) |
| O(15)–Mn(3)–O(22) | 91.5(4) | O(15)–Mn(3)–O(38) | 171.2(5) |
| O(16)–Mn(3)–O(21) | 171.8(4) | O(16)–Mn(3)–O(22) | 94.8(6) |
| O(16)–Mn(3)–O(38) | 97.7(5) | O(21)–Mn(3)–O(22) | 85.5(6) |
| O(21)–Mn(3)–O(38) | 90.4(5) | O(22)–Mn(3)–O(38) | 96.7(4) |
| O(14)–Mn(4)–O(15) | 82.2(6) | O(14)–Mn(4)–O(16) | 83.6(4) |
| O(14)–Mn(4)–O(23) | 173.3(4) | O(14)–Mn(4)–O(24) | 93.6(6) |
| O(14)–Mn(4)–O(43) | 91.7(4) | O(15)–Mn(4)–O(16) | 83.8(5) |
| O(15)–Mn(4)–O(23) | 98.9(6) | O(15)–Mn(4)–O(24) | 175.0(2) |
| O(15)–Mn(4)–O(43) | 89.5(6) | O(16)–Mn(4)–O(23) | 90.0(4) |
| O(16)–Mn(4)–O(23) | 90.0(4) | O(16)–Mn(4)–O(24) | 93.0(6) |
| O(16)–Mn(4)–O(43) | 172.2(6) | O(23)–Mn(4)–O(24) | 84.9(6) |
| O(23)–Mn(4)–O(43) | 94.9(4) | O(24)–Mn(4)–O(43) | 93.5(6) |
| O(17)–Mn(5)–O(18) | 84.3(6) | O(17)–Mn(5)–O(30) | 87.0(4) |
| O(17)–Mn(5)–O(48) | 177.4(4) | O(17)–Mn(5)–O(53) | 84.5(4) |
| O(17)–Mn(5)–O(58) | 95.6(6) | O(18)–Mn(5)–O(30) | 84.4(5) |
| O(18)–Mn(5)–O(48) | 95.3(6) | O(18)–Mn(5)–O(53) | 93.5(6) |
| O(18)–Mn(5)–O(58) | 174.9(4) | O(30)–Mn(5)–O(48) | 90.4(4) |
| O(30)–Mn(5)–O(48) | 90.4(4) | O(30)–Mn(5)–O(53) | 171.4(5) |
| O(30)–Mn(5)–O(58) | 90.5(5) | O(48)–Mn(5)–O(53) | 98.0(5) |
| O(48)–Mn(5)–O(58) | 84.6(6) | O(53)–Mn(5)–O(58) | 91.6(6) |
| O(18)–Mn(6)–O(19) | 94.4(6) | O(18)–Mn(6)–O(25) | 90.1(9) |
| O(18)–Mn(6)–O(50) | 91.6(7) | O(18)–Mn(6)–O(55) | 92.6(6) |
| O(18)–Mn(6)–O(63) | 172.8(2) | O(19)–Mn(6)–O(25) | 91.9(8) |
| O(19)–Mn(6)–O(63) | 90.6(7) | O(25)–Mn(6)–O(50) | 77.1(8) |
| O(25)–Mn(6)–O(55) | 172.6(8) | O(25)–Mn(6)–O(63) | 84.6(9) |
| O(50)–Mn(6)–O(55) | 95.8(6) | O(50)–Mn(6)–O(63) | 82.5(7) |
| O(55)–Mn(6)–O(63) | 92.1(6) | O(19)–Mn(7)–O(20) | 85.1(7) |
| O(19)–Mn(7)–O(35) | 85.8(6) | O(19)–Mn(7)–O(65) | 94.0(7) |
| O(19)–Mn(7)–O(68) | 178.50(19) | O(19)–Mn(7)–O(73) | 90.6(6) |
| O(20)–Mn(7)–O(35) | 88.2(4) | O(20)–Mn(7)–O(65) | 173.6(3) |
| O(20)–Mn(7)–O(68) | 93.9(7) | O(20)–Mn(7)–O(73) | 90.9(5) |
| O(35)–Mn(7)–O(65) | 85.4(4) | O(35)–Mn(7)–O(68) | 93.0(6) |
| O(35)–Mn(7)–O(73) | 176.4(8) | O(65)–Mn(7)–O(68) | 86.9(8) |
| O(65)–Mn(7)–O(73) | 95.4(5) | O(68)–Mn(7)–O(73) | 90.5(6) |
| O(20)–Mn(8)–O(21) | 93.4(7) | O(20)–Mn(8)–O(26) | 91.9(4) |
| O(20)–Mn(8)–O(70) | 93.1(6) | O(20)–Mn(8)–O(75) | 91.5(5) |
| O(20)–Mn(8)–O(78) | 175.2(7) | O(21)–Mn(8)–O(26) | 90.9(4) |
| O(21)–Mn(8)–O(70) | 173.3(6) | O(21)–Mn(8)–O(75) | 89.6(5) |
| O(21)–Mn(8)–O(78) | 90.9(6) | O(26)–Mn(8)–O(70) | 87.1(4) |
| O(26)–Mn(8)–O(75) | 176.5(6) | O(26)–Mn(8)–O(78) | 85.6(4) |
| O(70)–Mn(8)–O(75) | 92.0(5) | O(70)–Mn(8)–O(78) | 82.6(6) |
| O(75)–Mn(8)–O(78) | 90.9(5) | O(21)–Mn(9)–O(22) | 85.2(6) |
| O(21)–Mn(9)–O(40) | 84.8(4) | O(21)–Mn(9)–O(80) | 92.5(6) |
| O(21)–Mn(9)–O(83) | 174.6(4) | O(21)–Mn(9)–O(88) | 89.3(4) |
| O(22)–Mn(9)–O(40) | 86.1(5) | O(22)–Mn(9)–O(80) | 174.4(2) |
| O(22)–Mn(9)–O(83) | 93.7(6) | O(22)–Mn(9)–O(88) | 93.2(5) |
| O(40)–Mn(9)–O(80) | 88.6(5) | O(40)–Mn(9)–O(83) | 89.8(4) |
| O(40)–Mn(9)–O(88) | 174.0(4) | O(80)–Mn(9)–O(83) | 88.1(6) |
| O(80)–Mn(9)–O(88) | 91.8(5) | O(83)–Mn(9)–O(88) | 96.1(4) |
| O(22)–Mn(10)–O(23) | 95.8(5) | O(22)–Mn(10)–O(27) | 93.1(5) |
| O(22)–Mn(10)–O(85) | 89.4(5) | O(22)–Mn(10)–O(90) | 92.5(5) |
| O(22)–Mn(10)–O(93) | 173.6(7) | O(23)–Mn(10)–O(27) | 90.8(5) |
| O(23)–Mn(10)–O(85) | 172.4(7) | O(23)–Mn(10)–O(90) | 93.1(5) |
| O(23)–Mn(10)–O(93) | 90.1(5) | O(27)–Mn(10)–O(85) | 83.4(5) |
| O(27)–Mn(10)–O(90) | 172.8(4) | O(27)–Mn(10)–O(93) | 89.4(5) |
| O(85)–Mn(10)–O(90) | 92.2(5) | O(85)–Mn(10)–O(93) | 84.9(5) |
| O(90)–Mn(10)–O(93) | 84.6(5) | O(23)–Mn(11)–O(24) | 81.6(5) |
| O(23)–Mn(11)–O(45) | 89.6(6) | O(23)–Mn(11)–O(95) | 93.0(6) |
| O(23)–Mn(11)–O(98) | 171.1(6) | O(23)–Mn(11)–O(103) | 97.1(5) |
| O(24)–Mn(11)–O(45) | 87.2(4) | O(24)–Mn(11)–O(95) | 172.7(8) |
| O(24)–Mn(11)–O(98) | 92.4(5) | O(24)–Mn(11)–O(103) | 94.2(4) |
| O(45)–Mn(11)–O(95) | 87.8(5) | O(45)–Mn(11)–O(98) | 83.5(6) |
| O(45)–Mn(11)–O(103) | 173.2(6) | O(95)–Mn(11)–O(98) | 92.3(6) |
| O(95)–Mn(11)–O(103) | 91.4(5) | O(98)–Mn(11)–O(103) | 89.8(6) |
| O(17)–Mn(12)–O(24) | 93.6(5) | O(17)–Mn(12)–O(60) | 94.6(5) |
| O(17)–Mn(12)–O(100) | 162.7(6) | O(17)–Mn(12)–O(105) | 102.9(5) |
| O(24)–Mn(12)–O(60) | 169.9(4) | O(24)–Mn(12)–O(100) | 87.1(5) |
| O(24)–Mn(12)–O(105) | 97.4(6) | O(60)–Mn(12)–O(100) | 83.3(5) |
| O(60)–Mn(12)–O(105) | 86.4(6) | O(100)–Mn(12)–O(105) | 94.1(5) |
| Mn(1)–O(13)–Mn(2) | 96.5(5) | Mn(1)–O(13)–Mn(3) | 99.5(7) |
| Mn(2)–O(13)–Mn(3) | 95.1(3) | Mn(1)–O(14)–Mn(2) | 96.1(5) |
| Mn(1)–O(14)–Mn(4) | 94.8(6) | Mn(2)–O(14)–Mn(4) | 97.2(6) |
| Mn(2)–O(15)–Mn(3) | 96.0(1) | Mn(2)–O(15)–Mn(4) | 98.5(7) |
| Mn(3)–O(15)–Mn(4) | 96.2(7) | Mn(1)–O(16)–Mn(3) | 99.1(6) |
| Mn(1)–O(16)–Mn(4) | 95.1(6) | Mn(3)–O(16)–Mn(4) | 93.9(6) |
| Mn(1)–O(17)–Mn(5) | 93.0(6) | Mn(1)–O(17)–Mn(12) | 132.7(5) |
| Mn(5)–O(17)–Mn(12) | 119.7(8) | Mn(1)–O(18)–Mn(5) | 94.6(6) |
| Mn(1)–O(18)–Mn(6) | 130.4(7) | Mn(5)–O(18)–Mn(6) | 122.0(7) |
| Mn(2)–O(19)–Mn(6) | 134.5(1) | Mn(2)–O(19)–Mn(7) | 94.0(7) |
| Mn(6)–O(19)–Mn(7) | 128.3(5) | Mn(2)–O(20)–Mn(7) | 92.2(7) |
| Mn(2)–O(20)–Mn(8) | 132.1(9) | Mn(7)–O(20)–Mn(8) | 126.4(6) |
| Mn(3)–O(21)–Mn(8) | 131.9(7) | Mn(3)–O(21)–Mn(9) | 94.6(7) |
| Mn(8)–O(21)–Mn(9) | 133.0(9) | Mn(3)–O(22)–Mn(9) | 93.3(6) |
| Mn(3)–O(22)–Mn(10) | 132.2(8) | Mn(9)–O(22)–Mn(10) | 125.3(5) |
| Mn(4)–O(23)–Mn(10) | 131.8(6) | Mn(4)–O(23)–Mn(11) | 94.0(6) |
| Mn(4)–O(24)–Mn(12) | 134.8(3) | Mn(11)–O(24)–Mn(12) | 123.5(9) |

**Table 7**. Selected[a] Bond Distances (Å) for $[Mn_{12}O_{12}(O_2CC_6H_4\text{-}p\text{-}Me)_{16}(H_2O)_4] \cdot (HO_2CC_6H_4\text{-}p\text{-}Me)$ (Complex **6**).

| A | B | distance (Å) | A | B | distance (Å) |
|---|---|---|---|---|---|
| Mn(1) | O(101) | 1.867(5) | Mn(1) | O(102) | 1.895(6) |
| Mn(1) | O(62) | 1.925(6) | Mn(1) | O(52) | 1.950(6) |
| Mn(1) | O(71) | 2.233(5) | Mn(1) | O(42) | 2.236(6) |
| Mn(1) | Mn(2) | 2.792(2) | Mn(2) | O(101) | 1.842(6) |
| Mn(2) | O(102) | 1.897(5) | Mn(2) | O(104)#1 | 1.911(5) |
| Mn(2) | O(104) | 1.940(6) | Mn(2) | O(103) | 1.942(5) |
| Mn(2) | O(72) | 1.945(5) | Mn(2) | Mn(2)#1 | 2.823(3) |
| Mn(2) | Mn(3)#1 | 2.850(2) | Mn(2) | Mn(3) | 2.951(2) |
| Mn(3) | O(105) | 1.857(5) | Mn(3) | O(106) | 1.861(6) |
| Mn(3) | O(103) | 1.911(6) | Mn(3) | O(11) | 1.912(6) |
| Mn(3) | O(104) | 1.941(5) | Mn(3) | O(103)#1 | 1.942(5) |
| Mn(3) | Mn(4) | 2.786(2) | Mn(3) | Mn(2)#1 | 2.850(2) |
| Mn(3) | Mn(3)#1 | 2.855(3) | Mn(4) | O(105) | 1.857(6) |
| Mn(4) | O(31) | 1.930(7) | Mn(4) | O(106) | 1.940(5) |
| Mn(4) | O(22) | 1.965(6) | Mn(4) | O(1)#1 | 2.196(6) |
| Mn(4) | O(12) | 2.227(6) | Mn(5) | O(102) | 1.899(6) |
| Mn(5) | O(102)#1 | 1.899(6) | Mn(5) | O(51)#1 | 1.969(6) |
| Mn(5) | O(51) | 1.969(6) | Mn(5) | O(108) | 2.275(6) |
| Mn(5) | O(108)#1 | 2.276(6) | Mn(6) | O(105)#1 | 1.985(6) |
| Mn(6) | O(105) | 1.985(6) | Mn(6) | O(2)#1 | 1.991(8) |
| Mn(6) | O(2) | 1.991(8) | Mn(6) | O(21)#1 | 2.036(7) |
| Mn(6) | O(21) | 2.036(7) | Mn(7) | O(106)#1 | 1.898(6) |
| Mn(7) | O(101) | 1.934(5) | Mn(7) | O(32)#1 | 1.966(7) |
| Mn(7) | O(61) | 1.970(7) | Mn(7) | O(41) | 2.125(6) |
| Mn(7) | O(107) | 2.343(9) | O(103) | Mn(3)#1 | 1.942(5) |
| O(1) | Mn(4)#1 | 2.196(6) | O(106) | Mn(7)#1 | 1.898(6) |

| | | |
|---|---|---|
| O(32) | Mn(7)#1 | 1.966(7) |
| O(104) | Mn(2)#1 | 1.911(5) |

[a]For the $[Mn_{12}O_{12}]$ core and $H_2O$ molecules only; a full listing is available in the supporting material.

**Table 8**. Selected[a] Bond Angles (°) for $[Mn_{12}O_{12}(O_2CC_6H_4\text{-}p\text{-}Me)_{16}(H_2O)_4] \cdot (HO_2CC_6H_4\text{-}p\text{-}Me)$ (Complex **6**).

| A | B | C | angle (°) | A | B | C | angle (°) |
|---|---|---|---|---|---|---|---|
| O(101) | Mn(1) | O(102) | 83.0(2) | O(101) | Mn(1) | O(62) | 96.8(2) |
| O(102) | Mn(1) | O(62) | 172.9(2) | O(101) | Mn(1) | O(52) | 177.3(3) |
| O(102) | Mn(1) | O(52) | 94.3(2) | O(62) | Mn(1) | O(52) | 85.9(3) |
| O(102) | Mn(1) | O(71) | 86.3(2) | O(102) | Mn(1) | O(71) | 84.1(2) |
| O(62) | Mn(1) | O(71) | 88.8(2) | O(52) | Mn(1) | O(71) | 93.3(2) |
| O(101) | Mn(1) | O(42) | 91.2(2) | O(102) | Mn(1) | O(42) | 93.4(2) |
| O(62) | Mn(1) | O(42) | 93.7(2) | O(52) | Mn(1) | O(42) | 89.1(2) |
| O(71) | Mn(1) | O(42) | 176.6(2) | O(102) | Mn(1) | Mn(2) | 42.692) |
| O(101) | Mn(2) | O(102) | 83.6(2) | O(101) | Mn(2) | O(104)#1 | 90.9(2) |
| O(102) | Mn(2) | O(104)#1 | 87.9(2) | O(101) | Mn(2) | O(104) | 174.4(2) |
| O(102) | Mn(2) | O(104) | 99.9(2) | O(104)#1 | Mn(2) | O(104) | 84.9(2) |
| O(101) | Mn(2) | O(103) | 95.9(2) | O(102) | Mn(2) | O(103) | 172.5(2) |
| O(104)#1 | Mn(2) | O(103) | 84.7(2) | O(104) | Mn(2) | O(103) | 80.1(2) |
| O(101) | Mn(2) | O(72) | 93.4(3) | O(102) | Mn(2) | O(72) | 89.5(2) |
| O(104)#1 | Mn(2) | O(72) | 174.7(2) | O(104) | Mn(2) | O(72) | 91.0(2) |
| O(103) | Mn(2) | O(72) | 97.9(2) | O(105) | Mn(3) | O(106) | 84.0(3) |
| O(105) | Mn(3) | O(11) | 96.8(3) | O(106) | Mn(3) | O(103) | 174.3(2) |
| O(103) | Mn(3) | O(11) | 92.3(2) | O(106) | Mn(3) | O(11) | 94.8(3) |
| O(03) | Mn(3) | O(11) | 90.8(3) | O(105) | Mn(3) | O(104) | 176.9(2) |
| O(106) | Mn(3) | O(104) | 98.2(2) | O(103) | Mn(3) | O(104) | 80.9(2) |
| O(11) | Mn(3) | O(104) | 89.8(2) | O(105) | Mn(3) | O(103)#1 | 93.8(2) |
| O(106) | Mn(3) | O(103)#1 | 90.8(2) | O(103) | Mn(3) | O(103)#1 | 83.6(30 |
| O(11) | Mn(3) | O(103)#1 | 172.2(2) | O(104) | Mn(3) | O(103)#1 | 83.9(2) |
| O(105) | Mn(4) | O(31) | 176.2(2) | O(105) | Mn(4) | O(106) | 81.8(2) |
| O(31) | Mn(4) | O(106) | 96.4(3) | O(105) | Mn(4) | O(22) | 95.9(3) |
| O(31) | Mn(4) | O(22) | 86.3(3) | O(106) | Mn(4) | O(22) | 172.3(3) |
| O(105) | Mn(4) | O(1)#1 | 90.7(3) | O(31) | Mn(4) | O(1)#1 | 86.193) |
| O(106) | Mn(4) | O(1)#1 | 95.0(2) | O(22) | Mn(4) | O(1)#1 | 92.4(3) |
| O(105) | Mn(4) | O(12) | 89.8(2) | O(31) | Mn(4) | O(12) | 93.493) |
| O(106) | Mn(4) | O(12) | 85.9(2) | O(22) | Mn(4) | O(12) | 86.8(3) |
| O(1)#1 | Mn(4) | O(12) | 179.0(2) | O(102) | Mn(5) | O(102)#1 | 95.7(3) |
| O(102) | Mn(5) | O(51)#1 | 172.593) | O(102)#1 | Mn(5) | O(51)#1 | 90.9(3) |
| O(102) | Mn(5) | O(51) | 90.9(3) | O(102)#1 | Mn(5) | O(5) | 172.5(2) |
| O(51)#1 | Mn(5) | O(51) | 82.4(4) | O(102) | Mn(5) | O(108) | 91.292) |
| O(102)#1 | Mn(5) | O(108) | 86.4(2) | O(51)#1 | Mn(5) | O(108) | 92.8(2) |
| O(51) | Mn(5) | O(108) | 89.9(3) | O(102) | Mn(5) | O(108)#1 | 86.4(2) |
| O(102)#1 | Mn(5) | O(108)#1 | 91.2(2) | O(51)#1 | Mn(5) | O(108)#1 | 89.9(3) |
| O(51) | Mn(5) | O(108)#1 | 92.8(2) | O(108) | Mn(5) | O(108)#1 | 176.4(3) |
| O(105)#1 | Mn(6) | O(105) | 94.0(3) | O(105)#1 | Mn(6) | O(2)#1 | 89.2(3) |
| O(105) | Mn(6) | O(2)#1 | 94.0(3) | O(105)#1 | Mn(6) | O(2) | 94.0(3) |
| O(105) | Mn(6) | O(2) | 89.2(3) | O(2)#1 | Mn(6) | O(21)#1 | 175.5(3) |
| O(105)#1 | Mn(6) | O(21)#1 | 90.5(3) | O(105) | Mn(6) | O(21)#1 | 175.5(3) |
| O(2)#1 | Mn(6) | O(21)#1 | 85.9(4) | O(2) | Mn(6) | O(2)#1 | 175.3(5) |
| O(105)#1 | Mn(6) | O(21) | 175.5(3) | O(2) | Mn(6) | O(21) | 90.6(4) |
| O(2)#1 | Mn(6) | O(21) | 90.6(4) | O(2) | Mn(6) | O(21) | 85.9(4) |
| O(21)#1 | Mn(6) | O(21) | 85.0(4) | O(106)#1 | Mn(7) | O(101) | 93.6(2) |
| O(106)#1 | Mn(7) | O(32)#1 | 93.4(3) | O(101) | Mn(7) | O(32)#1 | 164.9(3) |
| O(106)#1 | Mn(7) | O(61) | 174.8(2) | O(101) | Mn(7) | O(61) | 90.7(3) |
| O(32)#1 | Mn(7) | O(61) | 81.7(3) | O(106)#1 | Mn(7) | O(41) | 89.7(2) |
| O(101) | Mn(7) | O(41) | 94.2(2) | O(32)#1 | Mn(7) | O(41) | 99.2(3) |
| O(61) | Mn(7) | O(41) | 93.0(3) | O(106)#1 | Mn(7) | O(107) | 94.0(3) |
| O(61) | Mn(7) | O(41) | 93.0(3) | O(32)#1 | Mn(7) | O(107) | 86.3(4) |
| O(101) | Mn(7) | O(107) | 80.0(3) | O(41) | Mn(7) | O(107) | 173.3(3) |
| O(61) | Mn(7) | O(107) | 83.9(3) | Mn(2) | O(101) | Mn(1) | 97.7(2) |
| Mn(1) | O(102) | Mn(5) | 131.4(3) | Mn(1) | O(102) | Mn(2) | 94.8(3) |
| Mn(3) | O(103) | Mn(3)#1 | 95.6(3) | Mn(2) | O(102) | Mn(5) | 130.7(3) |
| Mn(3)#1 | O(103) | Mn(2) | 94.4(2) | Mn(3) | O(103) | Mn(2) | 100.0(2) |
| Mn(2)#1 | O(104) | Mn(3) | 95.4(2) | Mn(2)#1 | O(104) | Mn(2) | 94.3(2) |
| Mn(4) | O(105) | Mn(3) | 97.2(3) | Mn(2) | O(104) | Mn(3) | 99.0(2) |
| Mn(3) | O(105) | Mn(6) | 132.2(3) | Mn(4) | O(105) | Mn(6) | 123.6(3) |
| Mn(3) | O(106) | Mn(4) | 94.2(3) | Mn(3) | O(106) | Mn(7)#1 | 133.1(3) |
| Mn(1) | O(101) | Mn(7) | 122.8(3) | Mn(7)#1 | O(106) | Mn(4) | 127.8(3) |
| Mn(2) | O(101) | Mn(7) | 133.8(3) | | | | |

[a]For the $[Mn_{12}O_{12}(H_2O)_4]$ core only; a full listing is available in the supporting material. Atoms labeled with a #1 suffix are the symmetry partners related by the $C_2$ axis.

**Table 9**. Selected[a] Bond Distances (Å) for $[Mn_{12}O_{12}(O_2CC_6H_4\text{-}p\text{-}Me)_{16}(H_2O)_4] \cdot 3H_2O$ (Complex **7**).

| A | B | distance (Å) | A | B | distance (Å) |
|---|---|---|---|---|---|
| Mn(1) | O(5) | 1.870(5) | Mn(1) | O(12) | 1.871(5) |
| Mn(1) | O(1) | 1.908(5) | Mn(1) | O(3) | 1.931(5) |
| Mn(1) | O(43) | 1.935(5) | Mn(1) | O(2) | 1.945(5) |
| Mn(1) | Mn(5) | 2.802(2) | Mn(1) | Mn(2) | 2.818(2) |
| Mn(1) | Mn(3) | 2.850(2) | Mn(1) | Mn(4) | 2.966(2) |
| Mn(2) | O(10) | 1.869(5) | Mn(2) | O(11) | 1.887(5) |
| Mn(2) | O(3) | 1.902(5) | Mn(2) | O(4) | 1.907(5) |
| Mn(2) | O(1) | 1.922(5) | Mn(2) | O(36) | 1.923(5) |
| Mn(2) | Mn(8) | 2.797(2) | Mn(2) | Mn(4) | 2.845(2) |
| Mn(2) | Mn(3) | 2.953(2) | Mn(3) | O(7) | 1.845(5) |
| Mn(3) | O(6) | 1.868(5) | Mn(3) | O(2) | 1.914(5) |
| Mn(3) | O(1) | 1.929(5) | Mn(3) | O(20) | 1.932(5) |
| Mn(3) | O(4) | 1.946(5) | Mn(3) | Mn(6) | 2.796(2) |
| Mn(3) | Mn(4) | 2.8513(14) | Mn(4) | O(8) | 1.874(5) |
| Mn(4) | O(9) | 1.897(5) | Mn(4) | O(4) | 1.910(5) |
| Mn(4) | O(2) | 1.931(4) | Mn(4) | O(3) | 1.949(5) |
| Mn(4) | O(29) | 1.959(5) | Mn(4) | Mn(7) | 2.806(2) |
| Mn(5) | O(5) | 1.900(5) | Mn(5) | O(12) | 1.923(4) |
| Mn(5) | O(15) | 1.945(5) | Mn(5) | O(42) | 1.966(5) |
| Mn(5) | O(13) | 2.178(5) | Mn(5) | O(44) | 2.217(5) |
| Mn(6) | O(7) | 1.900(5) | Mn(6) | O(6) | 1.920(5) |
| Mn(6) | O(23) | 1.948(6) | Mn(6) | O(18) | 1.950(5) |
| Mn(6) | O(21) | 2.163(5) | Mn(6) | O(19) | 2.179(5) |
| Mn(7) | O(8) | 1.912(4) | Mn(7) | O(9) | 1.922(5) |
| Mn(7) | O(31) | 1.956(5) | Mn(7) | O(26) | 1.962(6) |
| Mn(7) | O(28) | 2.154(6) | Mn(7) | O(30) | 2.182(5) |
| Mn(8) | O(10) | 1.900(6) | Mn(8) | O(11) | 1.900(5) |
| Mn(8) | O(34) | 1.946(5) | Mn(8) | O(39) | 1.957(6) |
| Mn(8) | O(37) | 2.176(5) | Mn(8) | O(35) | 2.240(6) |
| Mn(9) | O(6) | 1.898(5) | Mn(9) | O(5) | 1.899(5) |
| Mn(9) | O(16) | 1.939(6) | Mn(9) | O(17) | 1.956(7) |
| Mn(9) | O(14) | 2.115(6) | Mn(9) | O(45) | 2.199(7) |
| Mn(10) | O(8) | 1.873(5) | Mn(10) | O(7) | 1.923(5) |
| Mn(10) | O(24) | 1.947(6) | Mn(10) | O(25) | 1.947(5) |
| Mn(10) | O(22) | 2.143(6) | Mn(10) | O(27) | 2.177(6) |
| Mn(11) | O(10) | 1.891(5) | Mn(11) | O(33) | 1.906(5) |
| Mn(11) | O(32) | 1.936(6) | Mn(11) | O(33) | 1.977(6) |
| Mn(11) | O(47) | 2.202(6) | Mn(11) | O(46) | 2.229(7) |
| Mn(12) | O(12) | 1.874(5) | Mn(12) | O(11) | 1.884(5) |
| Mn(12) | O(41) | 1.948(6) | Mn(12) | O(40) | 1.975(6) |
| Mn(12) | O(38) | 2.143(6) | Mn(12) | O(48) | 2.252(6) |

[a]For the [$Mn_{12}O_{12}$] core and $H_2O$ molecules only; a full listing is available in the supporting material.

**Table 10**. Selected[a] Bond Angles (°) for $[Mn_{12}O_{12}(O_2CC_6H_4\text{-p-Me})_{16}(H_2O)_4] \cdot 3H_2O$ (Complex **7**).

| A | B | C | angle (°) | A | B | C | angle (°) |
|---|---|---|---|---|---|---|---|
| O(5) | Mn(1) | O912) | 84.9(2) | O(5) | Mn(1) | O(1) | 92.1(2) |
| O(12) | Mn(1) | O(1) | 88.5(2) | O(5) | Mn(1) | O(3) | 174.9(2) |
| O(12) | Mn(1) | O(3) | 98.5(2) | O(1) | Mn(1) | O(3) | 84.2(2) |
| O(5) | Mn(1) | O(43) | 93.2(2) | O(12) | Mn(1) | O(43) | 92.4(2) |
| O(1) | Mn(1) | O(43) | 174.7(2) | O(3) | Mn(1) | O(43) | 90.5(2) |
| O(5) | Mn(1) | O(2) | 95.8(2) | O(12) | Mn(1) | O(2) | 172.0(2) |
| O(1) | Mn(1) | O(2) | 83.5(2) | O(3) | Mn(1) | O(2) | 80.3(2) |
| O(43) | Mn(1) | O(2) | 95.6(2) | O(10) | Mn(2) | O(11) | 84.1(2) |
| O(10) | Mn(2) | O(3) | 89.4(2) | O(11) | Mn(2) | O(3) | 93.3(2) |
| O(1) | Mn(2) | O(4) | 98.9(2) | O(11) | Mn(2) | O(4) | 176.1(2) |
| O(3) | Mn(2) | O(4) | 84.1(2) | O(10) | Mn(2) | O(1) | 174.1(2) |
| O(11) | Mn(2) | O(1) | 96.4(2) | O(3) | Mn(2) | O(1) | 84.6(2) |
| O(4) | Mn(2) | O(1) | 80.5(2) | O(10) | Mn(2) | O(36) | 91.5(2) |
| O(11) | Mn(2) | O(36) | 93.8(2) | O(3) | Mn(2) | O(36) | 172.8(2) |
| O(4) | Mn(2) | O(36) | 88.7(2) | O(1) | Mn(2) | O(36) | 94.3(2) |
| O(7) | Mn(3) | O(6) | 84.5(2) | O(7) | Mn(3) | O(2) | 93.4(2) |
| O(6) | Mn(3) | O(2) | 88.9(2) | O(7) | Mn(3) | O(1) | 174.6(2) |
| O(6) | Mn(3) | O(1) | 99.9(2) | O(2) | Mn(3) | O(1) | 83.7(2) |
| O(7) | Mn(3) | O(20) | 93.8(2) | O(6) | Mn(3) | O(20) | 92.3(2) |
| O(2) | Mn(3) | O(20) | 172.8(2) | O(1) | Mn(3) | O(20) | 89.1(2) |
| O(7) | Mn(3) | O(4) | 95.9(2) | O(6) | Mn(3) | O(4) | 172.1(2) |
| O(2) | Mn(3) | O(4) | 83.2(2) | O(1) | Mn(3) | O(4) | 79.3(2) |
| O(2) | Mn(3) | O(4) | 95.5(2) | O(8) | Mn(4) | O(9) | 84.8(2) |
| O(8) | Mn(4) | O(4) | 90.8(2) | O(9) | Mn(4) | O(4) | 90.3(2) |
| O(8) | Mn(4) | O(2) | 97.5(2) | O(9) | Mn(4) | O(2) | 173.6(2) |
| O(4) | Mn(4) | O(2) | 83.7(2) | O(8) | Mn(4) | O(3) | 173.4(2) |
| O(9) | Mn(4) | O(3) | 96.8(2) | O(4) | Mn(4) | O(3) | 82.8(2) |
| O(2) | Mn(4) | O(3) | 80.2(2) | O(8) | Mn(4) | O(29) | 90.1(2) |
| O(9) | Mn(4) | O(29) | 96.4(2) | O(4) | Mn(4) | O(29) | 173.3(2) |
| O(2) | Mn(4) | O(29) | 89.7(2) | O(3) | Mn(4) | O(29) | 96.1(2) |
| O(5) | Mn(5) | O(12) | 82.6(2) | O(5) | Mn(5) | O(15) | 95.3(2) |
| O(12) | Mn(5) | O(15) | 174.6(2) | O(5) | Mn(5) | O(42) | 176.8(2) |
| O(12) | Mn(5) | O(42) | 95.3(2) | O(15) | Mn(5) | O(42) | 86.6(2) |
| O(5) | Mn(5) | O(42) | 92.2(2) | O(12) | Mn(5) | O(13) | 93.7(2) |
| O(15) | Mn(5) | O(13) | 91.4(2) | O(42) | Mn(5) | O(13) | 90.4(2) |
| O(5) | Mn(5) | O(44) | 87.1(2) | O(12) | Mn(5) | O(44) | 82.6(2) |
| O(15) | Mn(5) | O(44) | 92.4(2) | O(42) | Mn(5) | O(44) | 90.1(2) |
| O(13) | Mn(5) | O(44) | 176.3(2) | O(7) | Mn(5) | O(6) | 81.6(2) |
| O(7) | Mn(5) | O(23) | 95.3(2) | O(6) | Mn(6) | O(23) | 169.2(2) |
| O(7) | Mn(6) | O(18) | 177.7(2) | O(6) | Mn(6) | O(18) | 96.3(2) |
| O(23) | Mn(6) | O(18) | 86.6(2) | O(7) | Mn(6) | O(21) | 95.4(2) |
| O(6) | Mn(6) | O(21) | 95.9(2) | O(23) | Mn(6) | O(21) | 94.7(2) |
| O(18) | Mn(6) | O(21) | 85.6(2) | O(7) | Mn(6) | O(19) | 88.5(2) |
| O(6) | Mn(6) | O(19) | 84.7(2) | O(23) | Mn(6) | O(19) | 84.9(2) |
| O(18) | Mn(6) | O(19) | 90.4(2) | O(21) | Mn(6) | O(19) | 176.1(2) |
| O(8) | Mn(7) | O(9) | 83.1(2) | O(8) | Mn(7) | O(31) | 173.9(2) |
| O(9) | Mn(7) | O(31) | 94.7(2) | O(8) | Mn(7) | O(26) | 95.8(2) |
| O(9) | Mn(7) | O(26) | 173.1(2) | O(31) | Mn(7) | O(26) | 85.8(2) |
| O(8) | Mn(7) | O(28) | 95.1(2) | O(26) | Mn(7) | I(28) | 94.0(2) |
| O(31) | Mn(7) | O(28) | 90.7(2) | O(26) | Mn(7) | O(28) | 92.9(2) |
| O(8) | Mn(7) | O(39) | 84.6(2) | O(9) | Mn(7) | O(30) | 87.9(2) |
| O(31) | Mn(7) | O(30) | 89.7(2) | O(26) | Mn(7) | O(30) | 85.2(2) |
| O(28) | Mn(7) | O(30) | 178.1(2) | O(10) | Mn(8) | O(11) | 82.9(2) |
| O(10) | Mn(8) | O(34) | 94.7(2) | O(11) | Mn(8) | O(34) | 176.6(2) |
| O(10) | Mn(8) | O(39) | 171.6(2) | O(11) | Mn(8) | O(39) | 95.0(2) |
| O(34) | Mn(8) | O(39) | 87.0(2) | O(10) | Mn(8) | O(37) | 93.9(2) |
| O(11) | Mn(8) | O(37) | 93.9(2) | O(34) | Mn(8) | O(37) | 88.7(2) |
| O(39) | Mn(8) | O(37) | 94.4(2) | O(10) | Mn(8) | O(35) | 84.7(2) |
| O(11) | Mn(8) | O(35) | 88.6(2) | O(34) | Mn(8) | O(35) | 88.8(2) |
| O(39) | Mn(8) | O(35) | 87.0(2) | O(37) | Mn(8) | O(35) | 177.0(2) |
| O(6) | Mn(9) | O(16) | 173.8(2) | O(6) | Mn(9) | O(5) | 95.7(2) |
| O(6) | Mn(9) | O(17) | 90.6(2) | O(5) | Mn(9) | O(16) | 90.3(2) |
| O(16) | Mn(9) | O(17) | 83.3(3) | O(5) | Mn(9) | O(17) | 172.2(3) |
| O(5) | Mn(9) | O(14) | 93.8(2) | O(6) | Mn(9) | O(14) | 91.5(2) |
| O(17) | Mn(9) | O(14) | 90.7(3) | O(16) | Mn(9) | O(14) | 89.6(3) |
| O(5) | Mn(9) | O(45) | 88.3(3) | O(6) | Mn(9) | O(45) | 92.7(2) |
| O(17) | Mn(9) | O(45) | 86.7(4) | O(16) | Mn(9) | O(45) | 85.9(3) |
| O(8) | Mn(10) | O(7) | 94.0(2) | O(14) | Mn(9) | O(45) | 175.1(3) |
| O(7) | Mn(10) | O(24) | 92.0(2) | O(8) | Mn(10) | O(24) | 173.0(2) |
| O(7) | Mn(10) | O(25) | 173.8(2) | O(8) | Mn(10) | O(25) | 91.8(2) |
| O(8) | Mn(10) | O(22) | 92.5(2) | O(24) | Mn(10) | O(25) | 82.4(2) |
| O(24) | Mn(10) | O(22) | 90.7(3) | O(7) | Mn(10) | O(22) | 94.8(2) |
| O(8) | Mn(10) | O(27) | 92.0(2) | O(25) | Mn(10) | O(22) | 82.9(2) |
| O(24) | Mn(10) | O(27) | 84.0(3) | O(7) | Mn(10) | O(27) | 93.6(2) |
| O(22) | Mn(10) | O(27) | 170.2(3) | O(25) | Mn(10) | O(27) | 88.3(3) |
| O(10) | Mn(11) | O(32) | 172.2(3) | O(10) | Mn(11) | O(99) | 95.6(2) |
| O(10) | Mn(11) | O(33) | 90.5(2) | O(9) | Mn(11) | O(32) | 91.8(2) |
| O(32) | Mn(11) | O(33) | 82.2(3) | O(9) | Mn(11) | O(33) | 173.8(2) |
| O(9) | Mn(11) | O(47) | 90.9(2) | O(10) | Mn(11) | O(47) | 92.6(2) |
| O(33) | Mn(11) | O(47) | 87.2(3) | O(32) | Mn(11) | O(47) | 89.8(3) |
| O(9) | Mn(11) | O(46) | 95.6(3) | O(10) | Mn(11) | O(46) | 87.7(30 |
| O(33) | Mn(11) | O946) | 86.1(3) | O(32) | Mn(11) | O(46) | 89.0(3) |
| O912) | Mn(12) | O(11) | 94.6(2) | O(47) | Mn(11) | O(46) | 173.4(2) |
| O(11) | Mn(12) | O(41) | 173.8(3) | O((12) | Mn(12) | O(41) | 90.7(2) |
| O(11) | Mn(12) | O(40) | 92.2(2) | O(12) | Mn(12) | O(40) | 172.2(2) |
| O(12) | Mn(12) | O(38) | 93.4(2) | O(41) | Mn(12) | O(40) | 82.3(3) |
| O(41) | Mn(12) | O(38) | 88.4(2) | O(11) | Mn(12) | O(38) | 94.6(2) |
| O(12) | Mn(12) | O(48) | 90.8(2) | O(40) | Mn(12) | O(38) | 89.7(3) |
| O(41) | Mn(12) | O(48) | 88.6(3) | O(11) | Mn(12) | O(48) | 88.0(2) |
| O(38) | Mn(12) | O(48) | 174.9(3) | O(40) | Mn(12) | O(48) | 85.7(3) |
| Mn(1) | O(1) | Mn(3) | 95.9(2) | Mn(1) | O(1) | Mn(2) | 94.7(2) |
| Mn(3) | O(2) | Mn(4) | 95.7(2) | Mn(2) | O(1) | Mn(3) | 100.1(2) |
| Mn(4) | O(2) | Mn(1) | 99.8(2) | Mn(3) | O(2) | Mn(1) | 95.2(2) |
| Mn(2) | O(3) | Mn(4) | 95.3(2) | Mn(2) | O(3) | Mn(1) | 94.6(2) |
| Mn(2) | O(4) | Mn(4) | 96.4(2) | Mn(1) | O(3) | Mn(4) | 99.7(2) |
| Mn(4) | O(4) | Mn(3) | 95.4(2) | Mn(2) | O(4) | Mn(3) | 100.1(2) |
| Mn(1) | O(5) | Mn(5) | 96.0(2) | Mn(1) | O(5) | Mn(9) | 132.3(3) |
| Mn(3) | O(6) | Mn(9) | 131.8(3) | Mn(9) | O(5) | Mn(5) | 125.5(2) |
| Mn(9) | O(6) | Mn(6) | 129.3(3) | Mn(3) | O(6) | Mn(6) | 95.1(2) |
| Mn(3) | O(7) | Mn(10) | 133.6(3) | Mn(3) | O(7) | Mn(6) | 96.6(2) |
| Mn(4) | O(8) | Mn(10) | 134.1(2) | Mn(6) | O(7) | Mn(10) | 121.6(3) |
| Mn(10) | O98) | Mn(7) | 125.3(3) | Mn(4) | O(8) | Mn(7) | 95.6(2) |
| Mn(4) | O(9) | Mn(7) | 94.5(2) | Mn(4) | O(9) | Mn(11) | 131.9(3) |
| Mn(2) | O(10) | Mn(11) | 131.3(3) | Mn(11) | O(9) | Mn(7) | 126.2(2) |
| Mn(11) | O(10) | Mn(8) | 131.4(3) | Mn(2) | O(10) | Mn(8) | 95.8(2) |
| Mn(12) | O(11) | Mn(8) | 123.6(3) | Mn(12) | O(11) | Mn(2) | 131.8(3) |
| Mn(1) | O(12) | Mn(12) | 133.1(3) | Mn(2) | O(11) | Mn(8) | 95.2(2) |
| Mn(12) | O(12) | Mn(5) | 129.4(3) | Mn(1) | O(12) | Mn(5) | 95.2(2) |

[a]For the [$Mn_{12}O_{12}(H_2O)_4$] core only; a full listing is available in the supporting material.

**Table 11**. Assignment of $^{1}H$ NMR spectra of $[Mn_{12}O_{12}(O_2CPh)_{16}(H_2O)_4]$ (**3**).

| peak | δ-ppm (integral)[b] CDCl$_3$ | δ-ppm (integral)[b] CD$_2$Cl$_2$ | $T_1$[a] (ms) | Assignment |
|---|---|---|---|---|
| A | 13.1 (2) | 13.0 (2) | 10.9 | *m* ax(III...III) |
| B | 11.0 | 11.0 | 4.0 | *o* ax(III...IV) |
| C | 9.7 (2) | 9.6 (2) | 16.6 | *m* ax(III...IV) |
| D | 7.9§ | 7.7 | 8.4 | o ax(III...III) |
| E | 7.3 (1) | 7.3 (1) | 24.0 | *p* ax(III...IV) |
| F | 5.0 (4) | 5.1 (4) | 17.5□ | 2*m* eq(III...III) |
| G | 4.2 (1) | 4.3 (1) | 19.8 | *p* ax(III...III) |
| H | 1.2 | 1.4 | □-□ | 2 *o* eq(III...III) |
| I□ | –0.8 (2) | –0.6 (2) | 26.2 | 2 *p* eq(III...III) |

[a]$T_1$ values measured in CD$_2$Cl$_2$. [b]Integrals not measured for ortho-protons because of broadness.

Table **12**. Assignment of $^1$H NMR spectra of $[Mn_{12}O_{12}(O_2CC_6H_4\text{-}p\text{-}Me)_{16}(H_2O)_4]$ (**6**).

| peak | δ-ppm $CD_2Cl_2$ | $T_1{}^a$ (ms) | Assignment |
|---|---|---|---|
| A | 12.6 | - | *m* ax(III...III) |
| B | 12.3 | 51.7 | 2 Me eq(III...III) |
| C | 10.7 | 1.6 | *o* ax(III...IV) |
| D | 9.3 | 16.0 | *m* ax(III...IV) |
| E | 7.6 | 26.9 | *o* ax(III...III) |
| F | 7.0 | 28.6 | Me ax(III...IV) |
| G | 5.0 | 18.0 | 2*m* eq(III..III) |
| H | 2.9 | 43.7 | Me ax(III...III) |
| I | 1.4 | - | 2*o* eq(III...III) |

$^a T_1$ values measured in $CD_2Cl_2$.

**Figure Captions**

*Figure 1.* Plot of potential energy *vs* magnetization direction for a single $Mn_{12}$ molecule in zero applied field with a S = 10 ground state experiencing a zero-field splitting of $H = D\hat{S}_z^2$, where D < 0.

*Figure 2.* Plots of $\chi_M' \cdot T$ *vs*. temperature (upper) and $\chi_M''$ *vs*. temperature (lower) for the *p*-tert-butylbenzoate complex **8** in an ac field of 1 G oscillating at 50 Hz (●), 250 Hz (▼) and 1000 Hz (■).

*Figure 3.* Plots of $\chi_M' \cdot T$ *vs*. temperature (upper) and $\chi_M''$ *vs*. temperature (lower) for the *p*-chlorobenzoate complex **5** in an ac field of 1 G oscillating at 50 Hz (●), 250 Hz (▼) and 1000 Hz (■).

*Figure 4.* Plots of $\chi_M''$ *vs*. temperature for the *p*-methylbenzoate complexes **6** (upper) and **7** (lower) in an ac field of 1 G oscillating at 50 Hz (●), 250 Hz (▼) or 1000 Hz (■).

*Figure 5.* Plot of $\chi_M'T$ *vs*. temperature for the *p*-methylbenzoate complexes **6** (upper) and **7** (lower) in an ac field of 1 G oscillating at 50 Hz (●), 250 Hz (▼) or 1000 Hz (■).

*Figure 6.* Plots of $\chi_M''$ *vs*. temperature for a microcrystalline sample of the propionate complex (upper) and a frozen solution of the same complex dissolved in 0.5 mL $CD_2Cl_2$:0.5 toluene-$d_8$ (lower) in a 1 G ac field oscillating at 50 Hz (●), 250 Hz (▼) or 1000 Hz (■).

*Figure 7.* ORTEP representation of the $Mn_{12}$ complex in $[Mn_{12}O_{12}(O_2CC_6H_4$-*p*-$Cl)_{16}(H_2O)_4] \cdot 8CH_2Cl_2$ (Complex **5**). Only the $[Mn_{12}O_{12}]$ core atoms are labeled to avoid congestion.

*Figure 8.* ORTEP representation of $[Mn_{12}O_{12}(O_2CEt)_{16}(H_2O)_3]$ (Complex **2b**) with thermal ellipsoids shown at the 50% level. Hydrogen and the methyl carbon atoms of the propionate ligands are omitted and the outer atoms are de-emphasized for clarity (upper). Plotted at the bottom is an ORTEP representation of a unit cell of $[Mn_{12}O_{12}(O_2CEt)_{16}(H_2O)_3]$ (**2b**) with all atoms de-emphasized for clarity.

*Figure 9.* Stereoview of the best molecular fit of the Mn atoms of $[Mn_{12}O_{12}(O_2CEt)_{16}(H_2O)_3]$ (**2b**) with those of $[Mn_{12}O_{12}(O_2CEt)_{16}(H_2O)_3] \cdot 4H_2O \cdot ½\ C_6H_5CH_3$ (**2a**), showing the relative positions of coordinated $H_2O$ molecules on the two cores. Atoms from $[Mn_{12}O_{12}(O_2CEt)_{16}(H_2O)_3]$ (**2b**) are shown as shaded ellipsoids.

*Figure 10.* ORTEP representations of the cores (*i.e.*, without *p*-methylbenzoate ligands) of the $Mn_{12}$ complexes in: (top) $[Mn_{12}O_{12}(O_2CC_6H_4\text{-}p\text{-}Me)_{16}(H_2O)_4] \cdot (HO_2CC_6H_4\text{-}p\text{-}Me)$ (complex **6**) and (bottom) $[Mn_{12}O_{12}(O_2CC_6H_4\text{-}p\text{-}Me)_{16}(H_2O)_4] \cdot 3(H_2O)$ (complex **7**).

*Figure 11.* ORTEP representations of the side views of the cores of (top) $[Mn_{12}O_{12}(O_2CC_6H_4\text{-}p\text{-}Me)_{16}(H_2O)_4] \cdot 3(H_2O)$ (complex **7**) and (bottom) $[Mn_{12}O_{12}(O_2CC_6H_4\text{-}p\text{-}Me)_{16}(H_2O)_4] \cdot (HO_2CC_6H_4\text{-}p\text{-}Me)$ (complex **6**). The coordination geometries about each Mn atom are shown. Each of the eight $Mn^{III}$ ions show a tetragonally elongated Jahn-Teller distortion. In the case of complex **7** (top) these JT elongation axes are indicated as solid lines. For complex **6** the JT elongation axis dashed lines is pointed at an $O^{2-}$ ion and is unusual. There are two dashed lines because the molecule has a crystallographic $C_2$ axis disorder.

*Figure 12.* COSY and 1-D $^1H$ NMR spectra of $[Mn_{12}O_{12}(O_2CPh)_{16}(H_2O)_4]$ (**3**) obtained at 499.81 MHz. The solvent was referenced to the residual protic solvent signal at 5.33 ppm. This spectrum was obtained with an acquisition time of 88 ms and 2048 data points were collected in F1 and 1024 in F2. Unshifted sine-bell weighting and line broadening functions were applied prior to the Fourier transform. Zero filling of F2 to 2048 was applied and the data were symmetrized.

*Figure 13.* TOCSY and 1-D $^1$H NMR spectra of $[Mn_{12}O_{12}(O_2CPh)_{16}(H_2O)_4]$ (**3**) obtained at 499.81 MHz. The solvent was referenced to the residual protic solvent signal at 5.33 ppm. This spectrum was obtained with an acquisition time of 88 ms and a mixing time of 20 ms; 2048 data points were collected in F1 and 1024 in F2. Unshifted sine-bell weighting and line broadening functions were applied prior to the Fourier transform. Zero filling of F2 to 2048 was applied and the data were symmetrized. Only the positive contours are shown.

*Figure 14.* 1-D $^1$H NMR spectra of $[Mn_{12}O_{12}(O_2CPh\text{-}p\text{-}Me)_{16}(H_2O)_4]$ obtained at 499.81 MHz. The solvent was referenced to the residual protic solvent signal at 5.33 ppm.

*Figure 15.* Plots of magnetization *versus* external magnetic field for ) $[Mn_{12}O_{12}(O_2CC_6H_4\text{-}p\text{-}Me)_{16}(H_2O)_4] \cdot 3(H_2O)$ (complex **7**) at five temperatures in the 1.72-2.50 K range. Five small crystals (1.2 mg) were oriented in a frozen eicosane matrix so that the magnetic field is parallel to the principal axis of magnetization.

*Figure 16.* Plots of magnetization *versus* external magnetic field for $[Mn_{12}O_{12}(O_2CC_6H_4\text{-}p\text{-}Me)_{16}(H_2O)_4] \cdot (HO_2CC_6H_4\text{-}p\text{-}Me)$ (complex **6**) at five temperatures in the 1.72-2.20 K range. Six small crystals (2.2 mg) were oriented in a frozen eicosane matrix so that the magnetic field is parallel to the principal axis of magnetization.

*Figure 17.* Plot of magnetic induction B *versus* temperature for $[Mn_{12}O_{12}(O_2CC_6H_4\text{-}p\text{-}Me)_{16}(H_2O)_4] \cdot (HO_2CC_6H_4\text{-}p\text{-}Me)$ (complex **6**). Two magnetic induction (B) values were calculated from $B = H + 4\pi M$, with the external field (H) values obtained from the peak positions determined in the first-derivative magnetization plots.

*Figure 18*. Arrhenius plot of the natural logarithm of the relaxation rate *versus* the inverse absolute temperature for $[Mn_{12}O_{12}(O_2CC_6H_4\text{-}p\text{-}Me)_{16}(H_2O)_4] \cdot 3(H_2O)$ (complex **7**). The straight line represents a least-squares fit of the data to eq. 1.

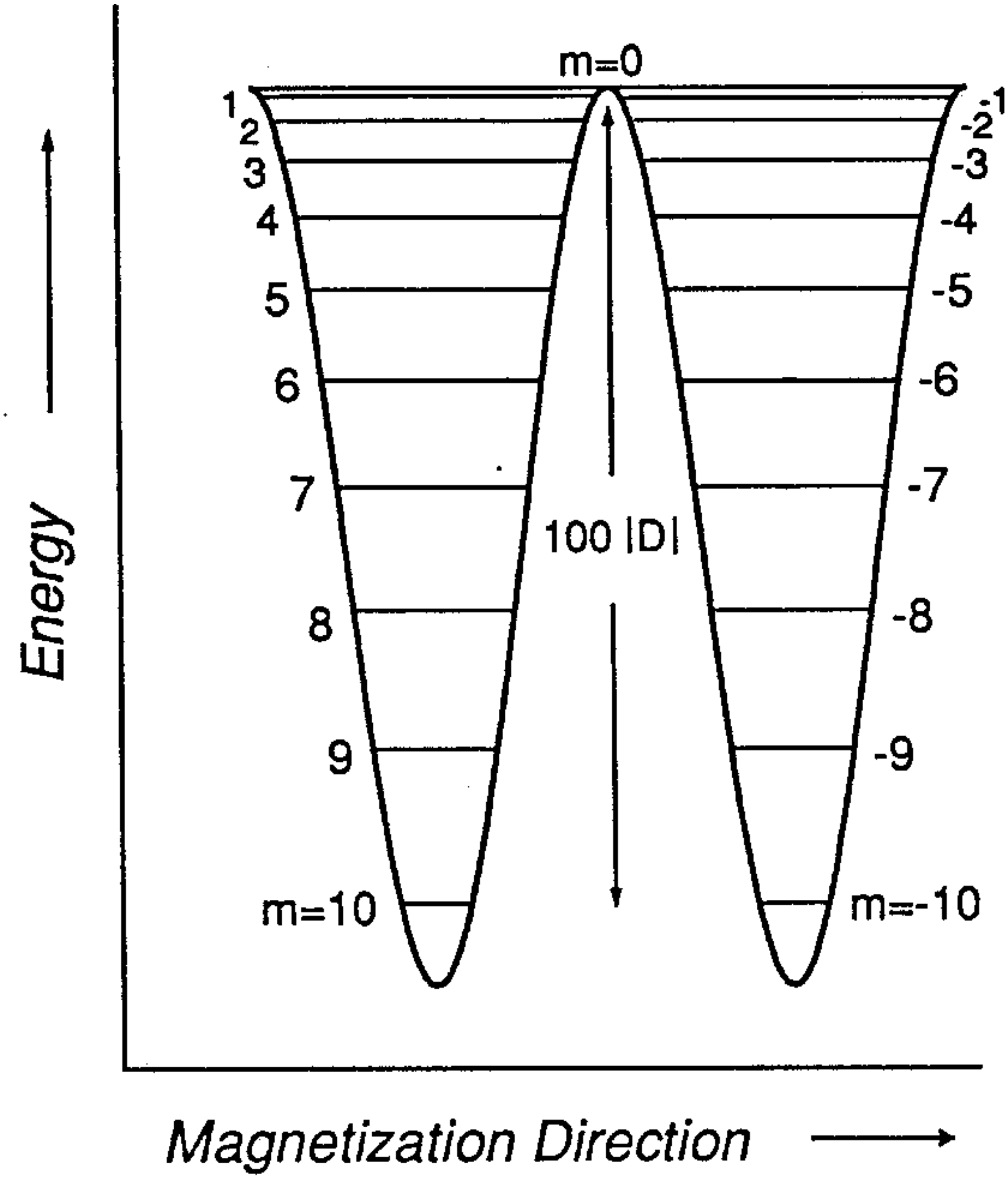

Figure 1.

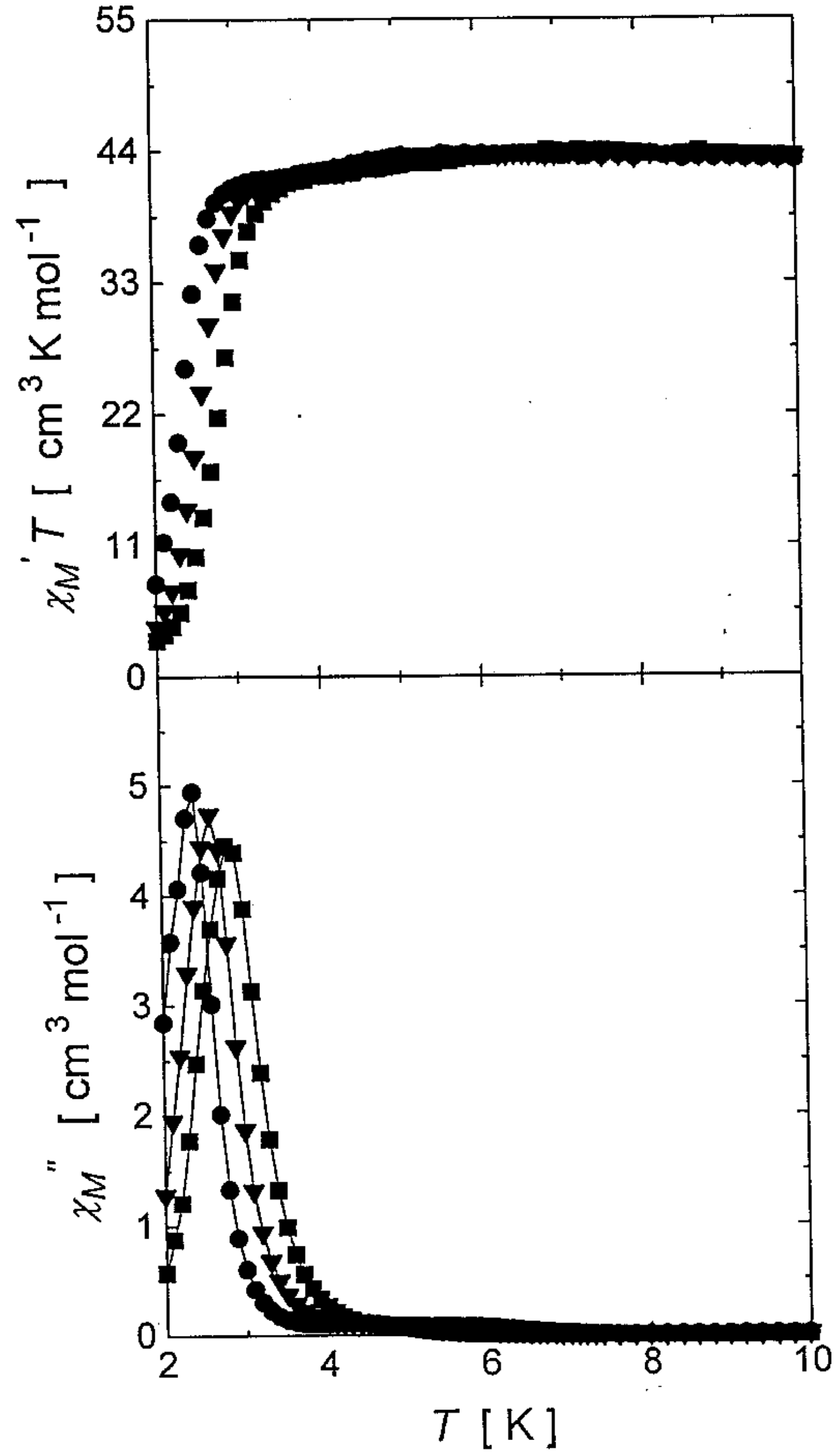

Figure 2.

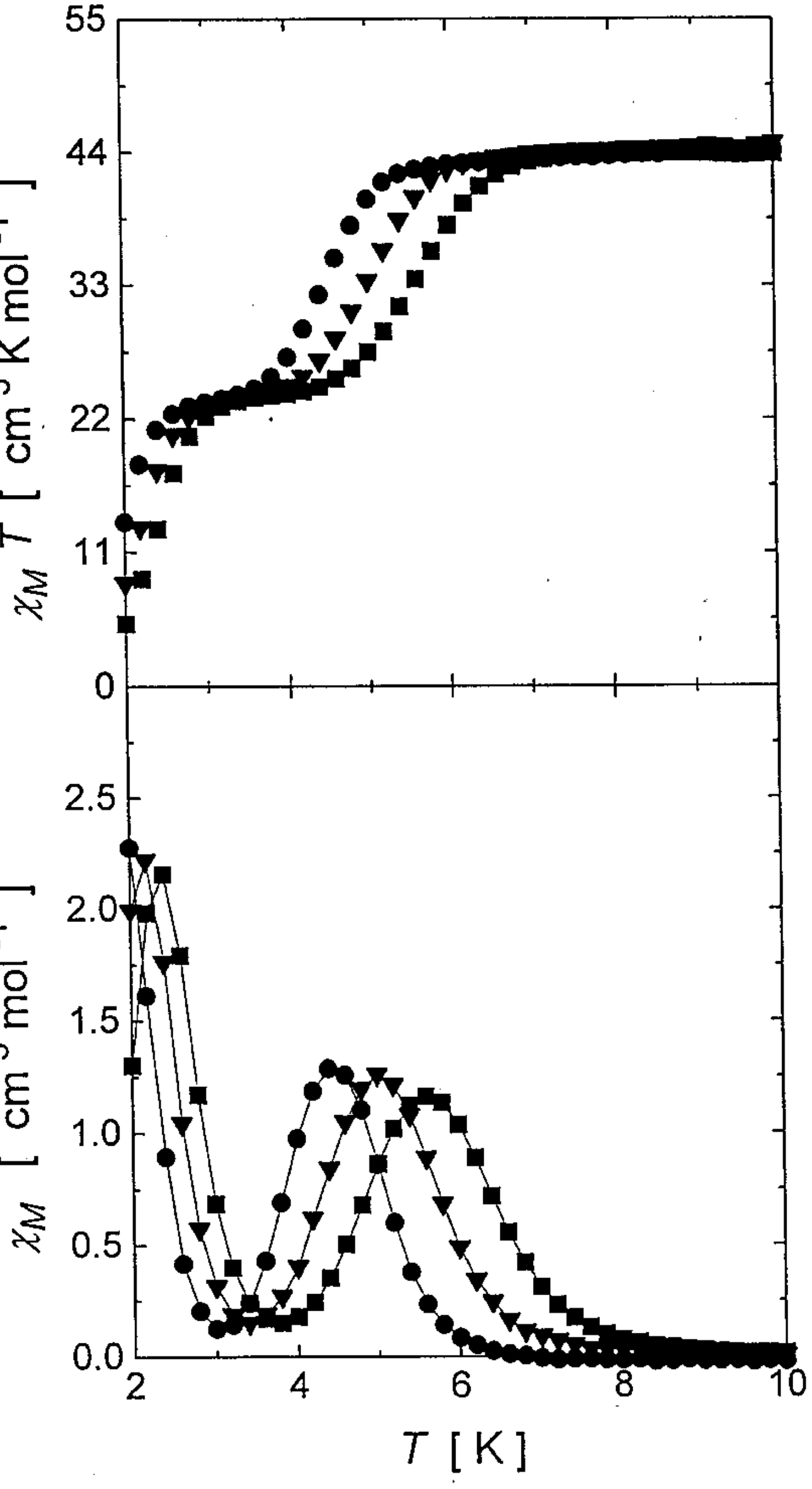

55
44
33
22
11
0
χ_M' T [ cm 3 K mol -1 ]
2.5
2.0
1.5
1.0
0.5
0.0
χ_M'' [ cm 3 mol -1 ]
2
4
6
8
10
T [ K ]

Figure 3.

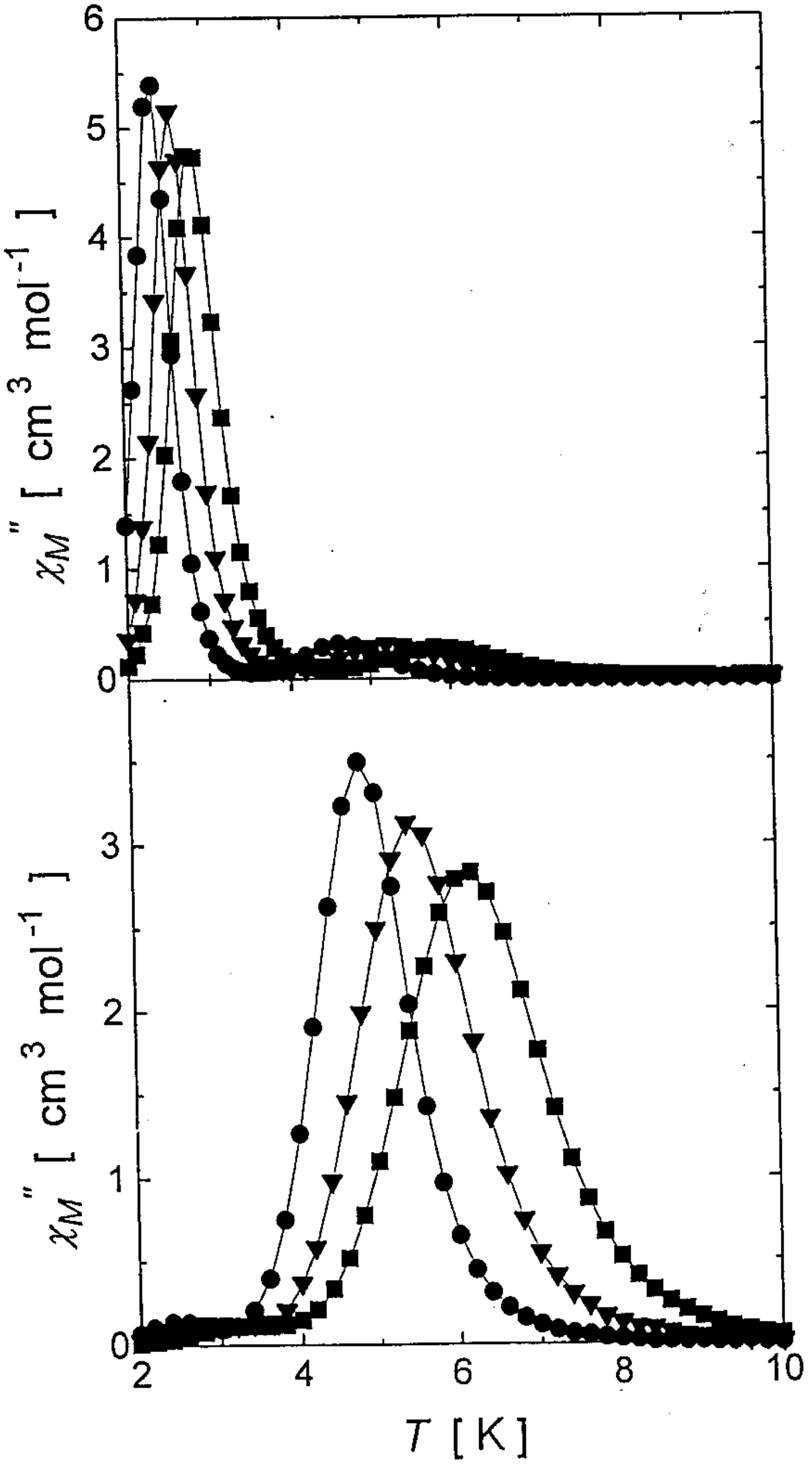

Figure 4.

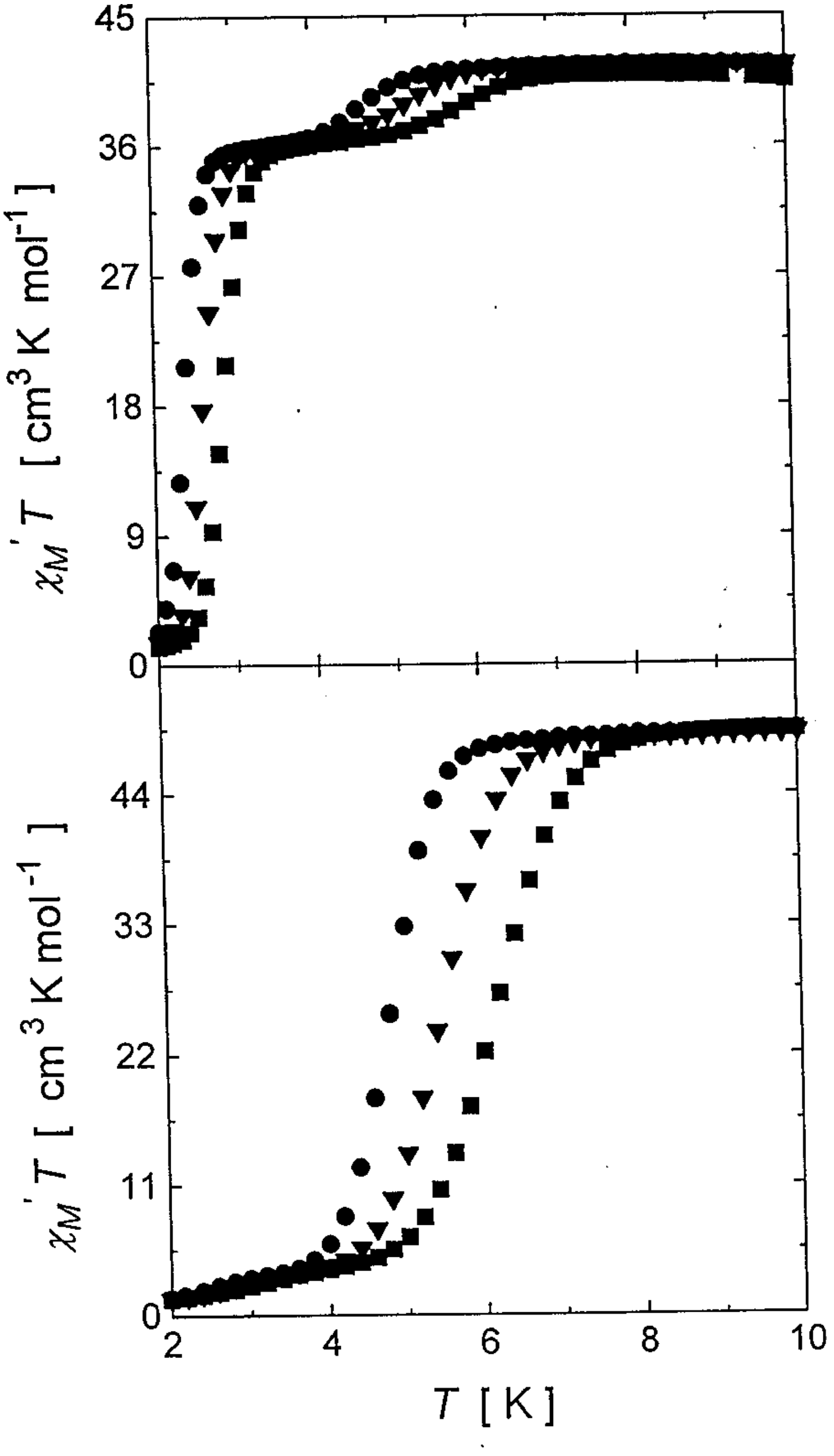

Figure 5.

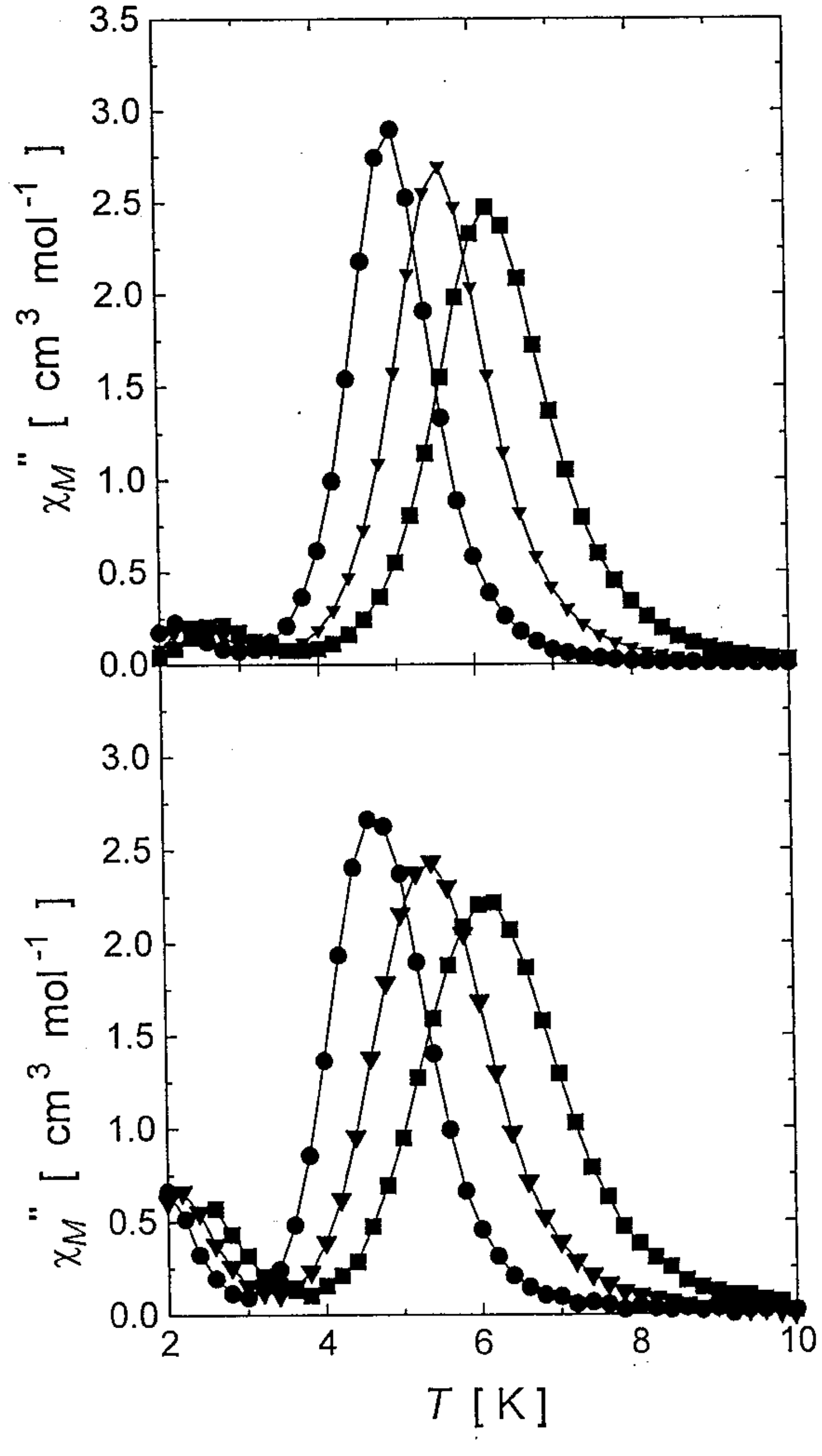

Figure 6.

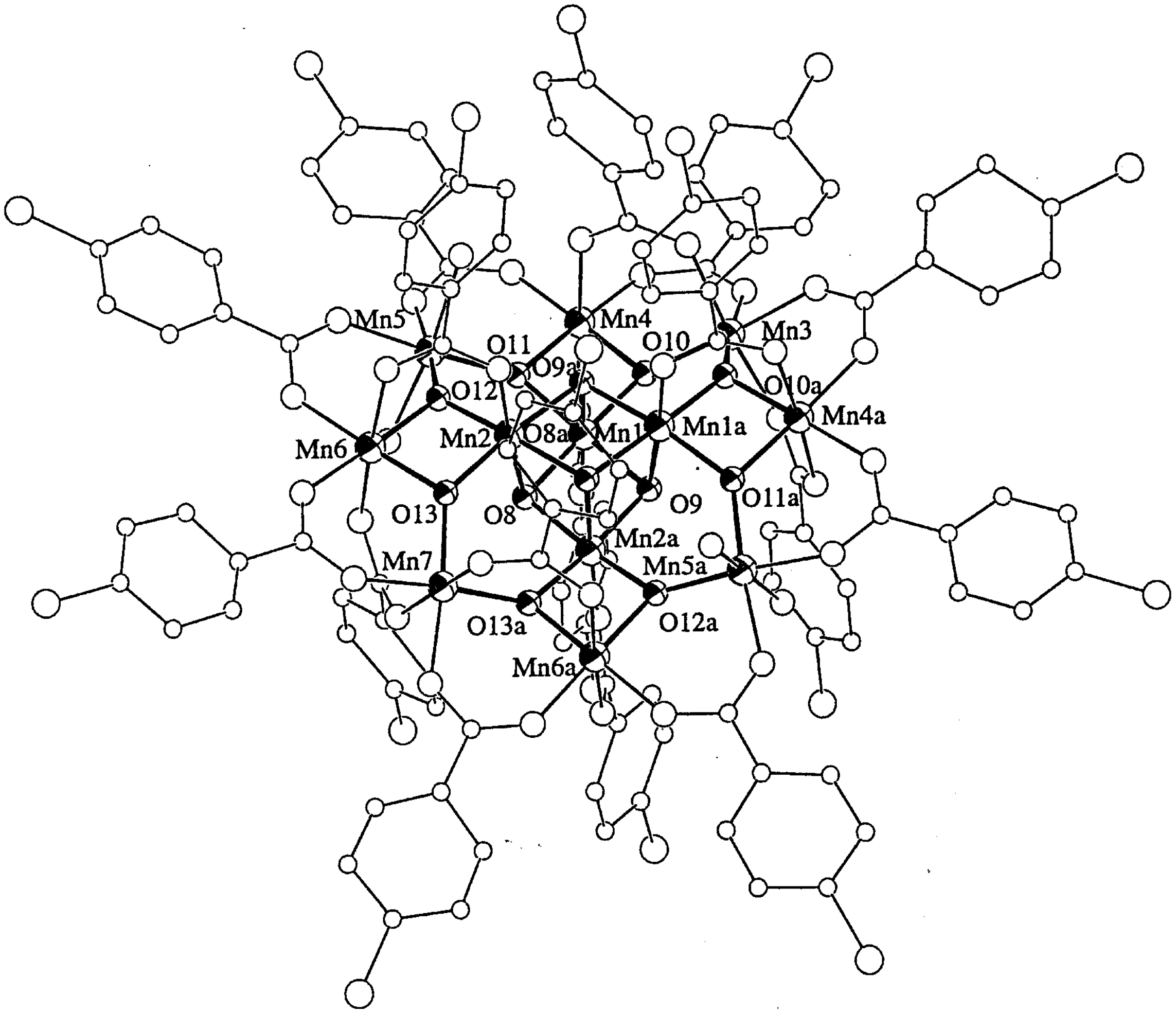

Figure 7.

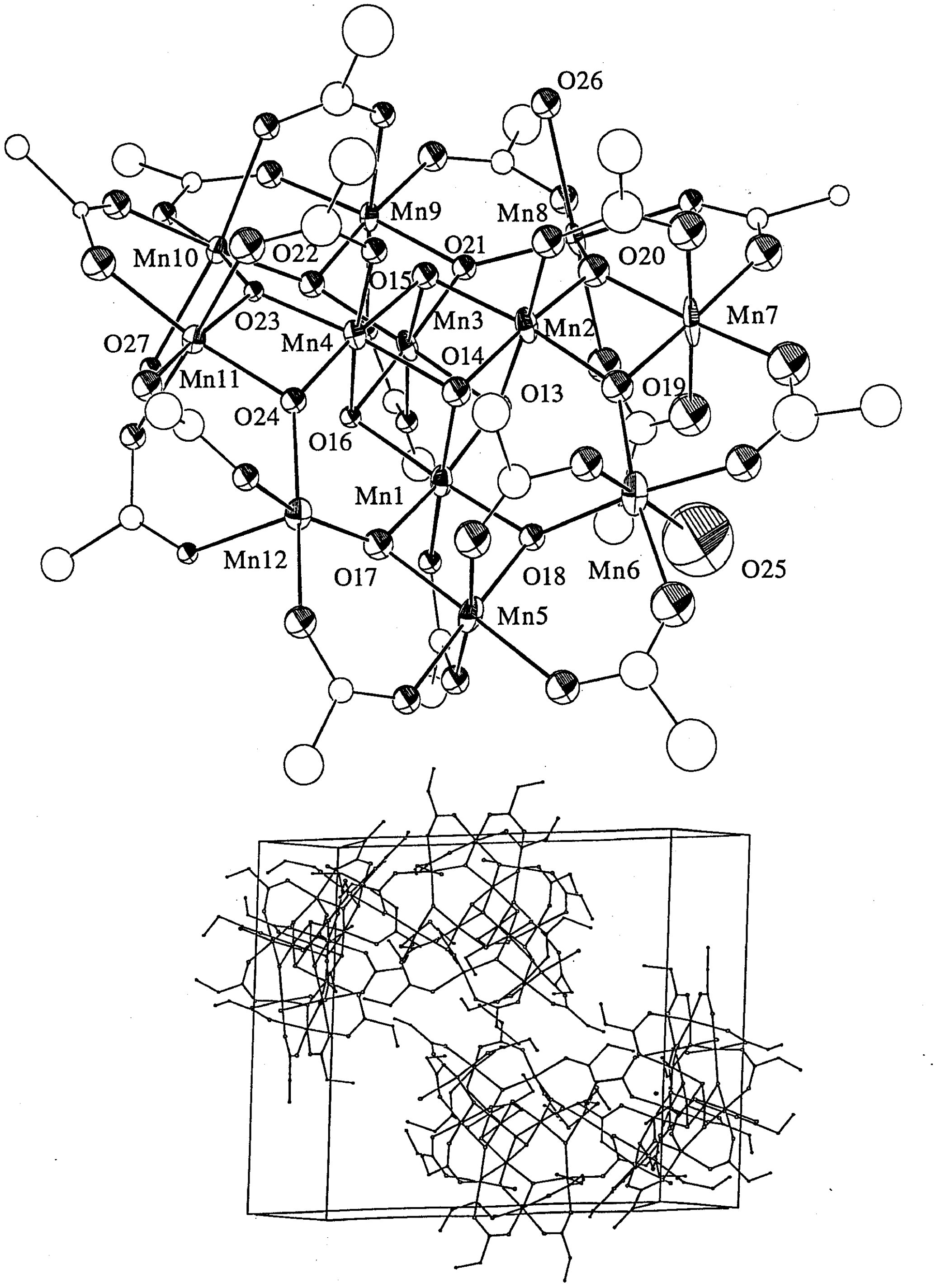

Figure 8.

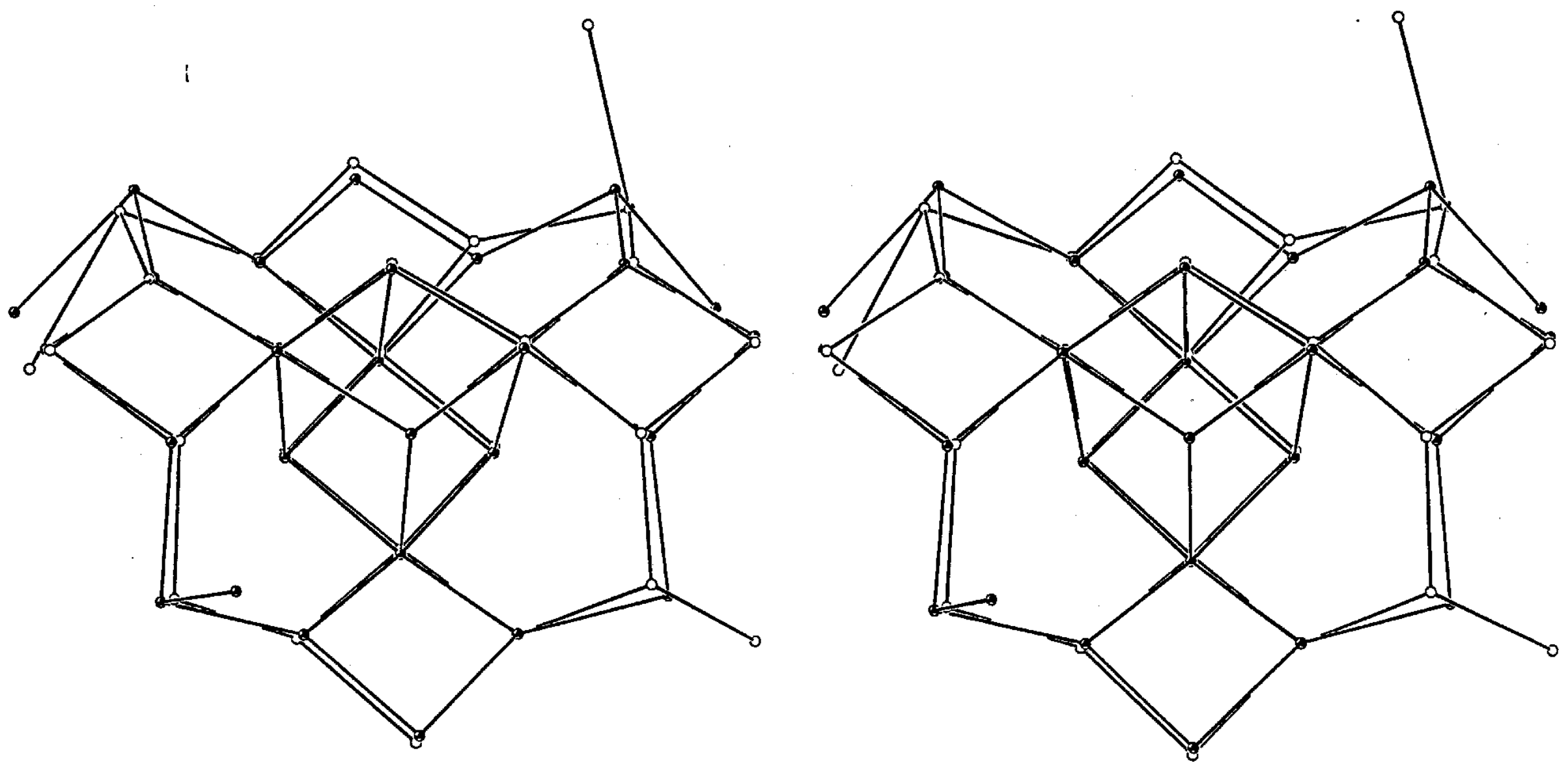

Figure 9.

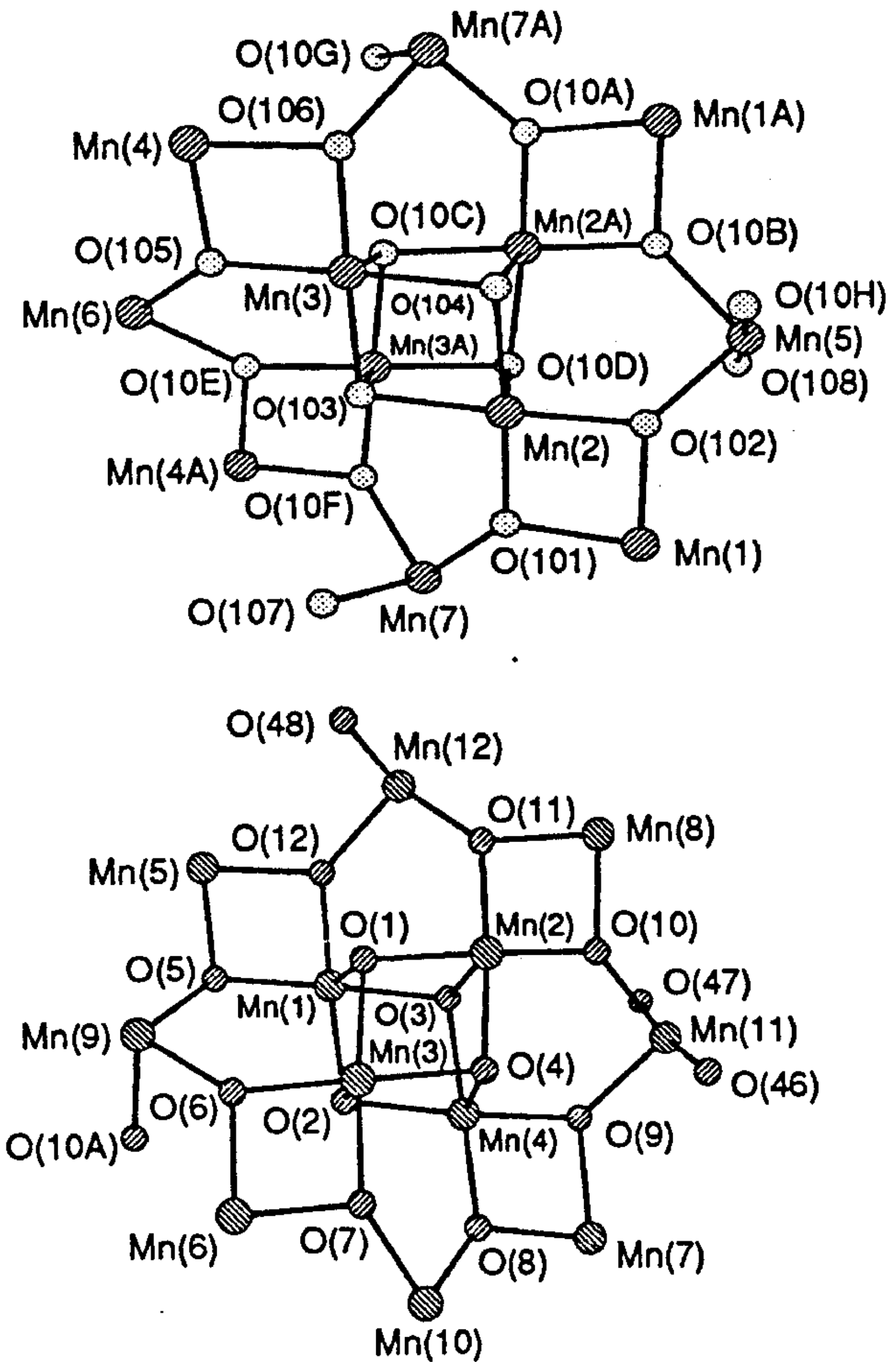

Figure 10.

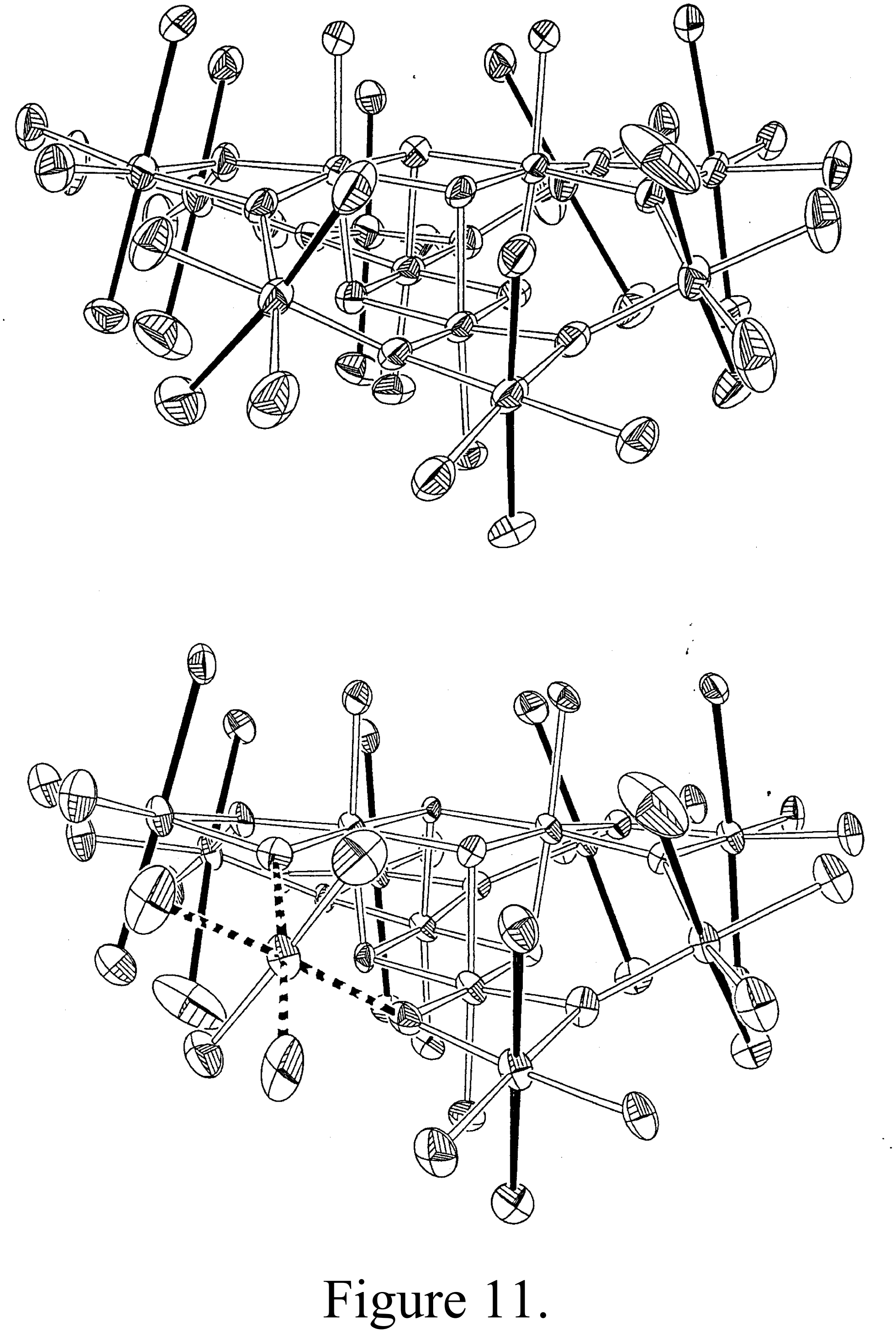

Figure 11.

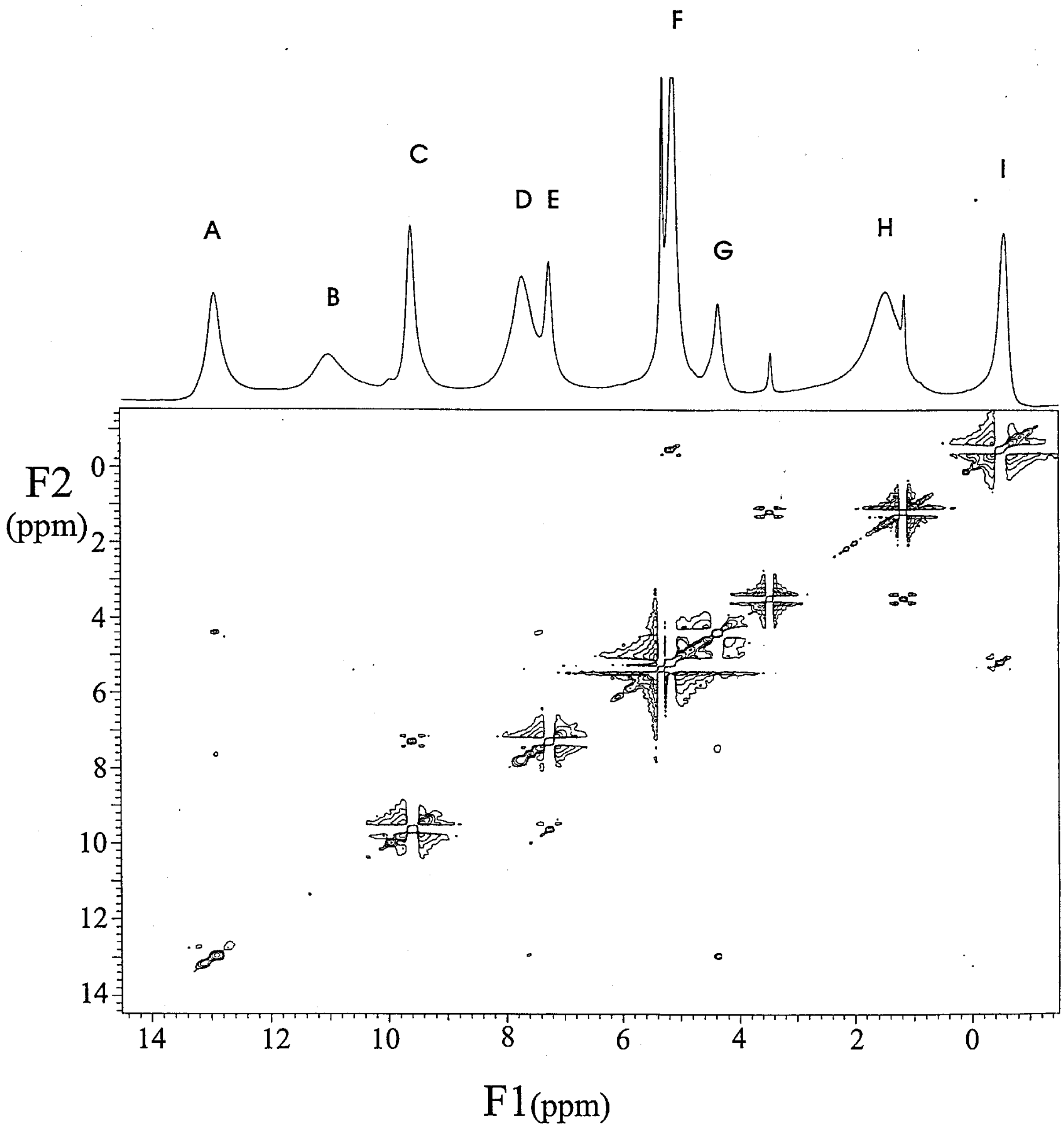

Figure 12.

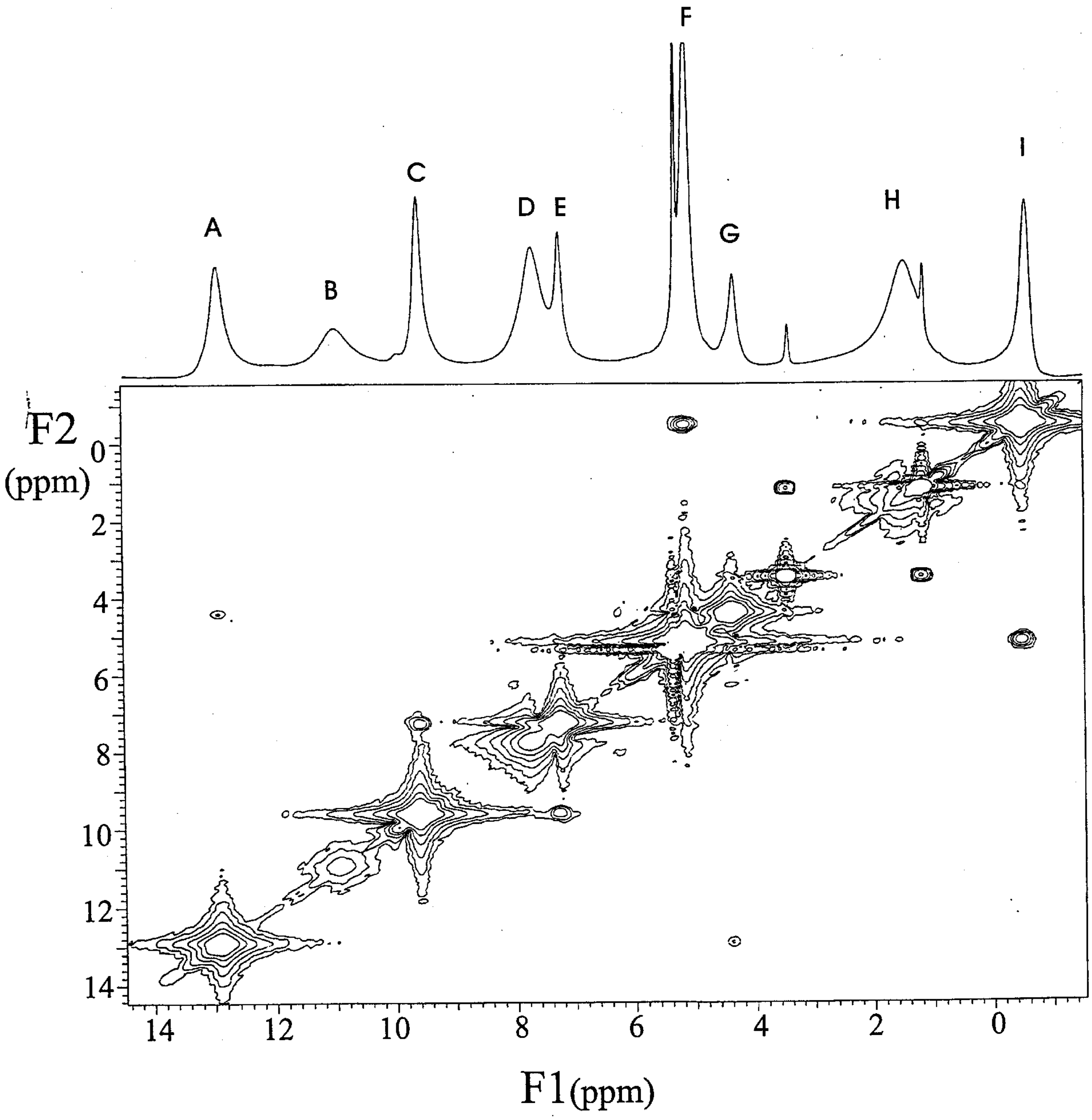

Figure 13.

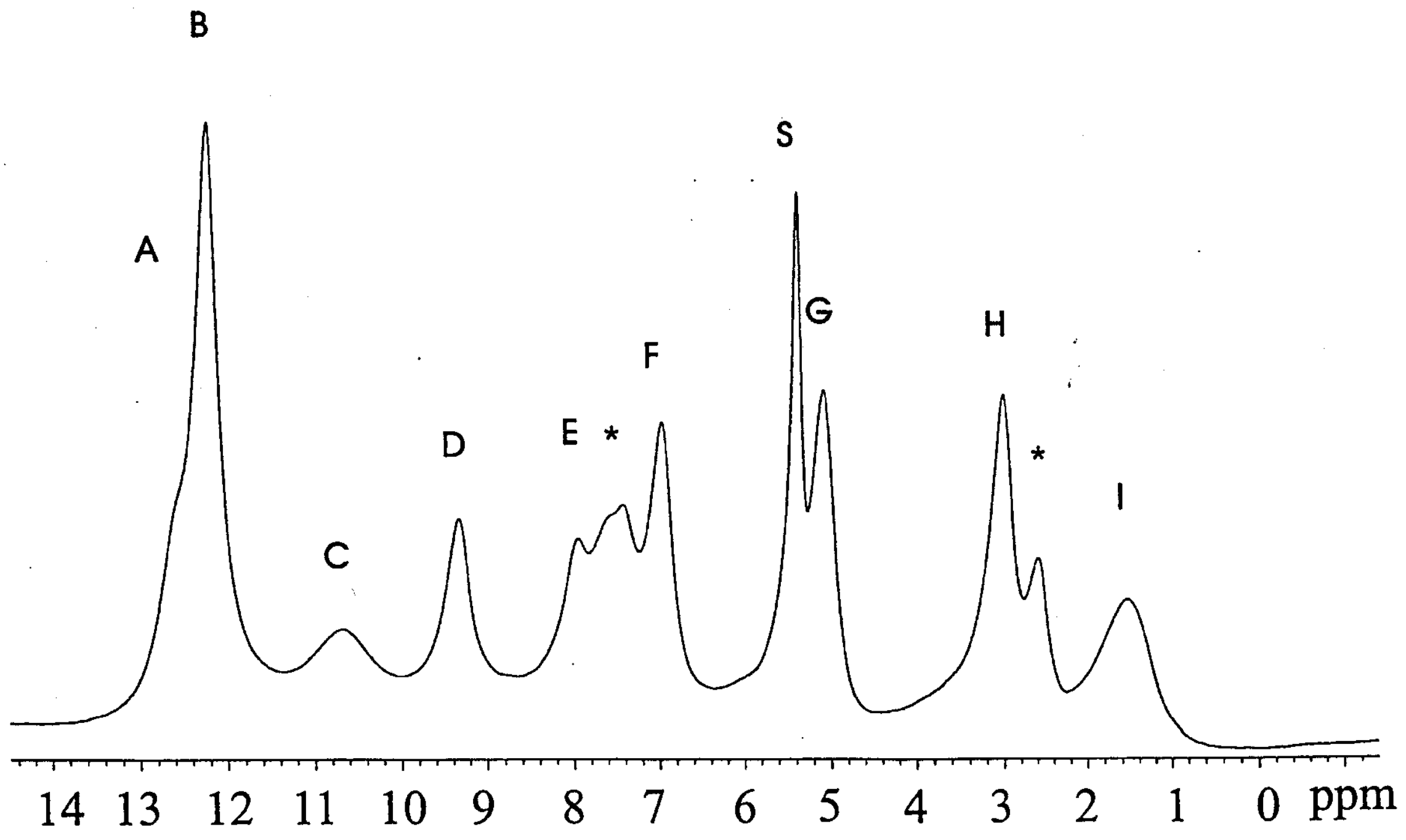

Figure 14.

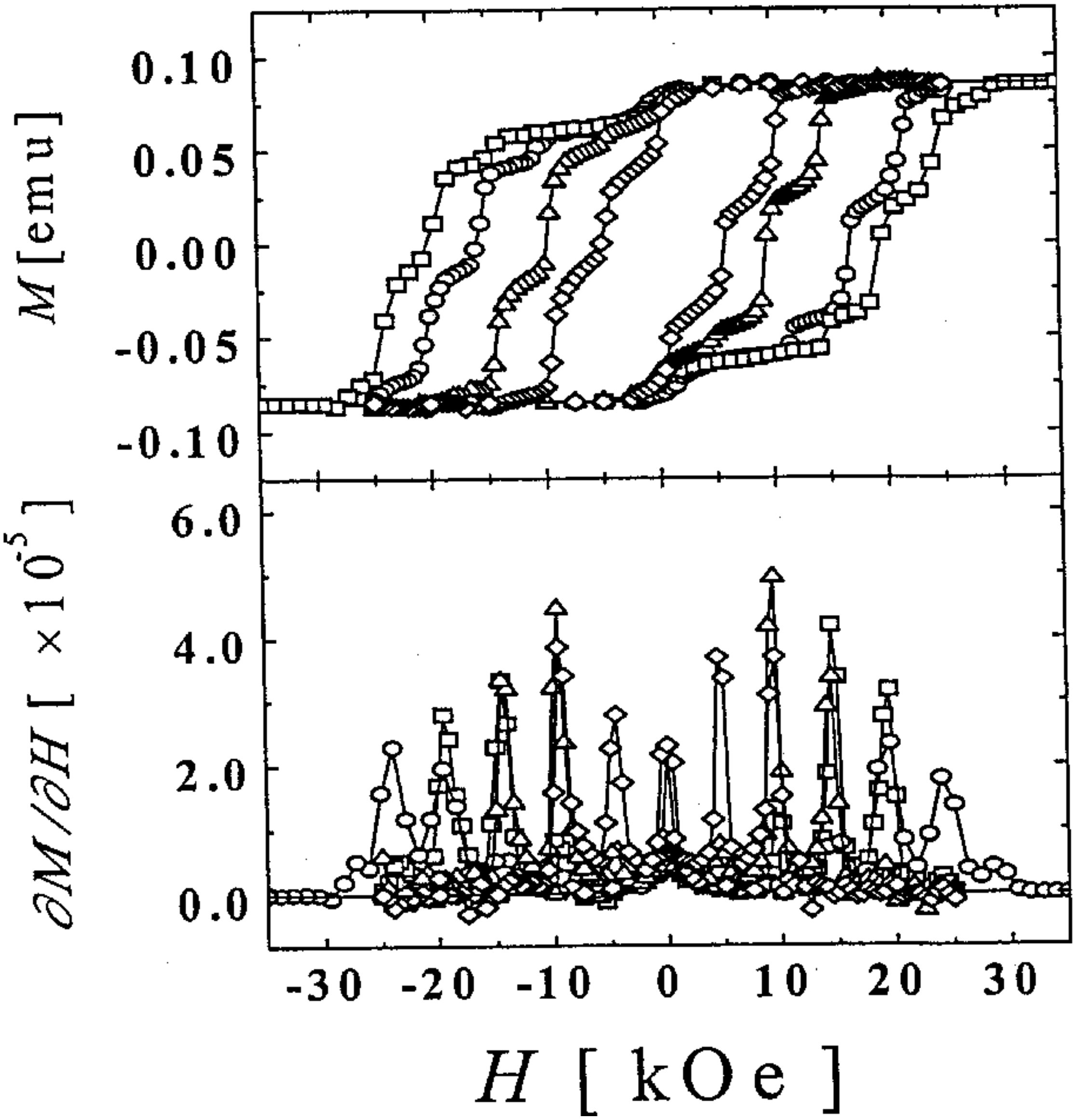

Figure 15.

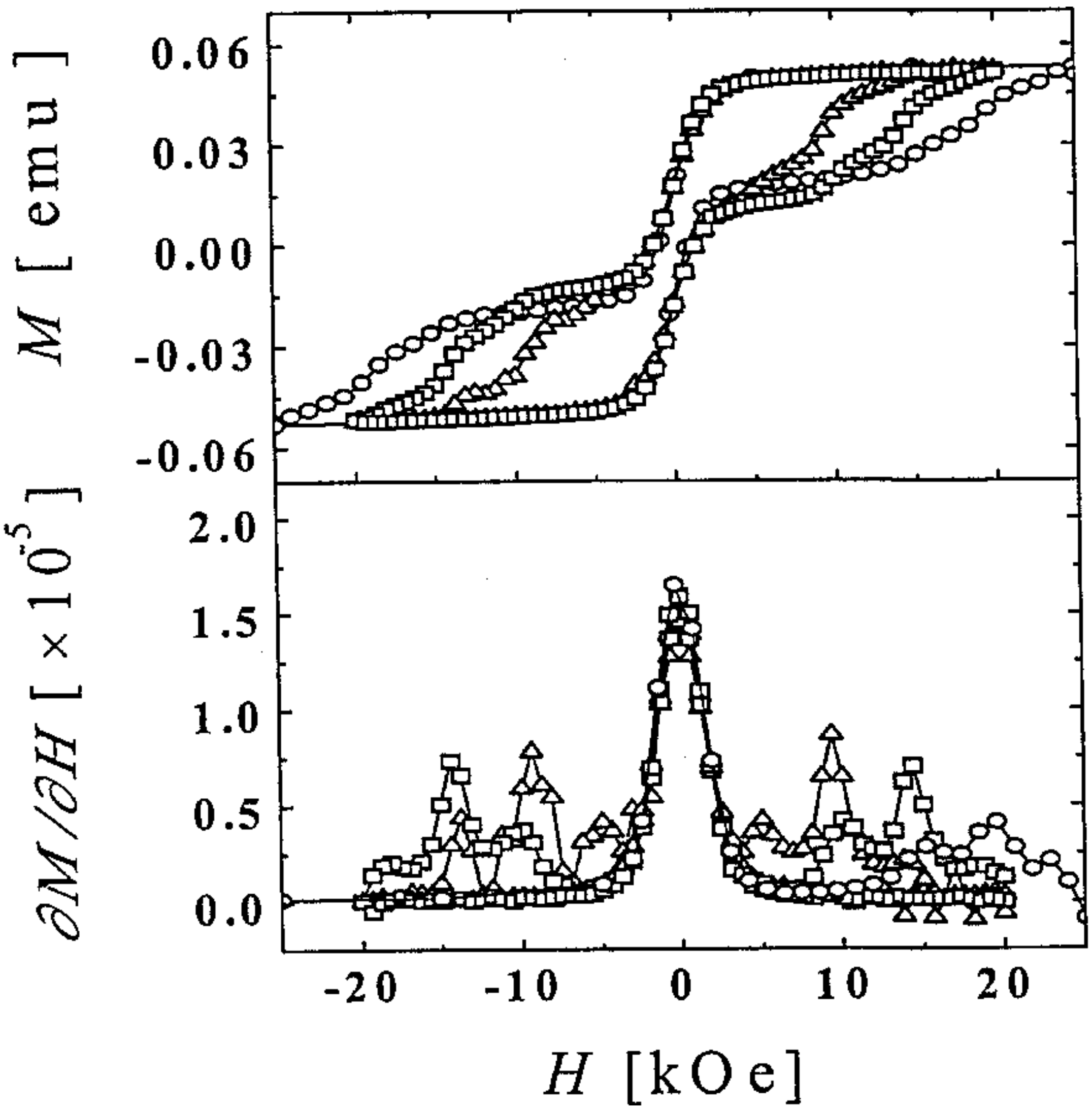

Figure 16.

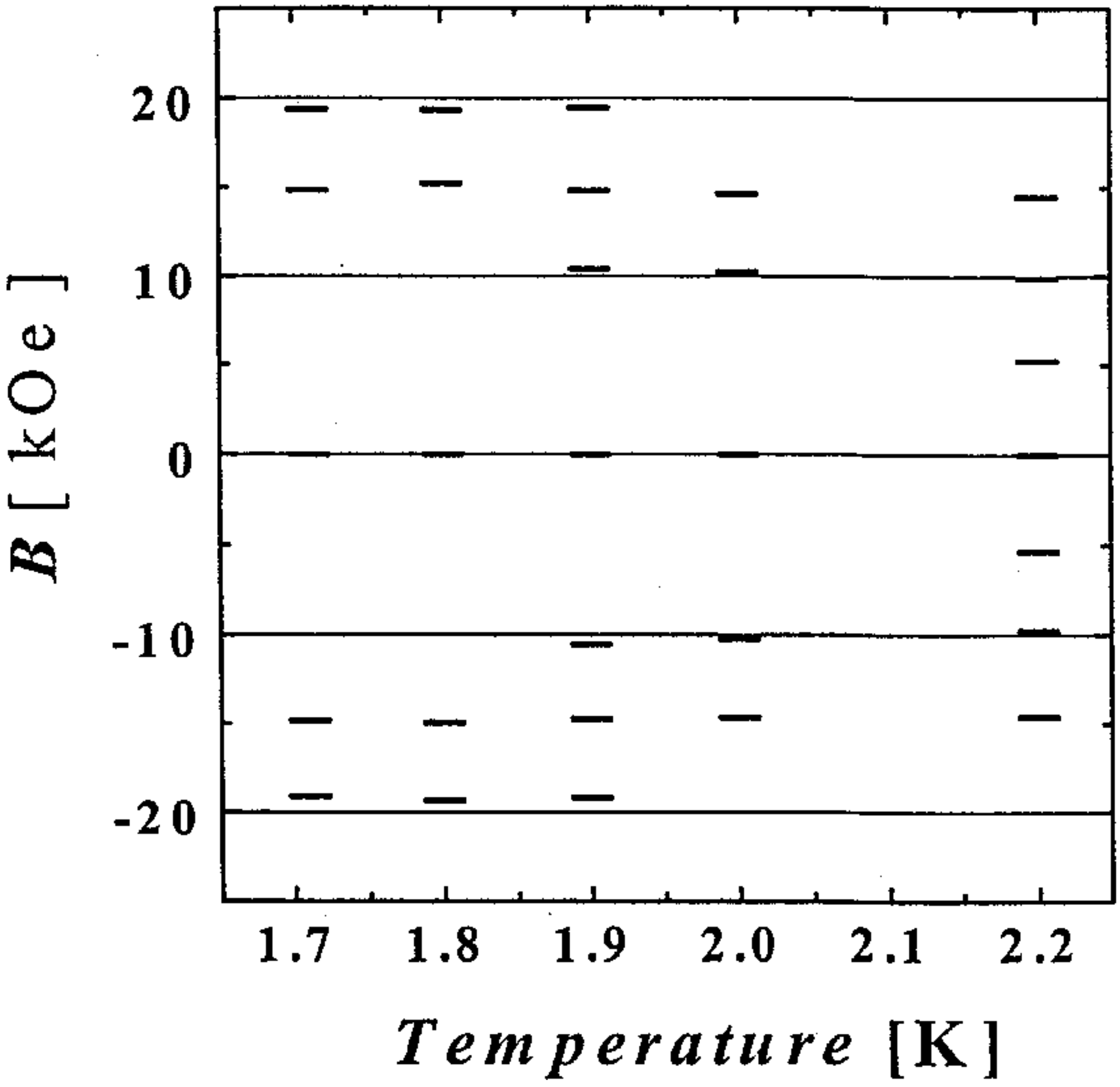

Figure 17.

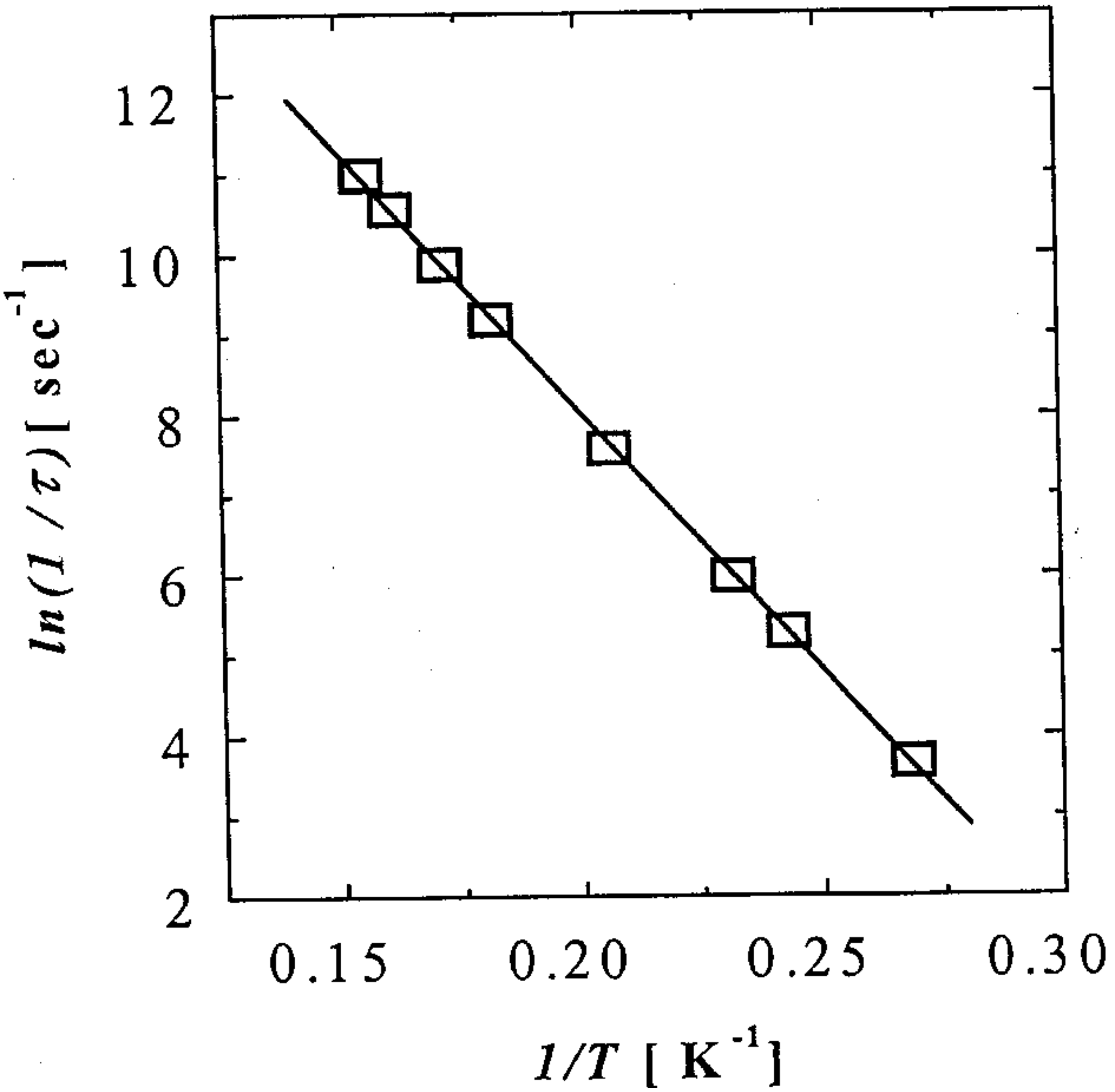

Figure 18.